\documentclass[reprint,superscriptaddress,amsmath,amssymb,aps]{revtex4-1}
 
\usepackage{graphicx}
\usepackage{dcolumn}
\usepackage{bm}
\usepackage{url} 
\usepackage{hyperref} 
\usepackage{wasysym} 
\usepackage{stmaryrd} 

\begin{document}


\title[Analytic Description of Zinc-Blende Nanowires]{Analytic Description of Cross Sections for Regular Forms of Zincblende- and Diamond-Lattice Nanowires}

\author{Dirk K{\"o}nig}
\email{solidstatedirk@gmail.com}
\affiliation{Integrated Materials Design Centre (IMDC), UNSW Node, University of New South Wales, NSW 2052, Australia}
\altaffiliation[Also at ]{Institute of Semiconductor Electronics (IHT), RWTH Aachen University, 52074 Aachen, Germany}


\begin{abstract}
Semiconductor nanowires (NWires) experience stress and charge transfer by their environment and impurity atoms. In return, the environment of NWires experiences a NWire stress response which may lead to propagated strain and change in shape and size of NWire cross sections. We deduce geometrical number series for zinc-blende- (zb-) and diamond-lattice NWires of diameter $d_{\mathrm{Wire}}$ to obtain the number of NWire atoms $N_{\mathrm{Wire}}(d_{\mathrm{Wire}}[i])$, bonds between NWire atoms $N\mathrm{_{bnd}}(d_{\mathrm{Wire}}[i])$ and interface bonds $N\mathrm{_{IF}}(d_{\mathrm{Wire}}[i])$ for six high symmetry zb NWires with low-index faceting frequently ocurring in bottom-up and top-down approaches of NWire processing. Along with these primary parameters, we present specific length of interface facets, cross section widths and heights as well as the cross section area. The fundamental insights into NWire structures revealed here offer a universal gauge and thus enable major advancements in data interpretation and understanding of all zb- and diamond-lattice based NWires. We underpin this statement with results from the literature on cross-section images from III-V core-shell NWire growth and on Si-NWires undergoing self-limiting oxidation and etching. The massive breakdown of impurity doping due to self-purification is shown to occur for both, Si NWires and Si nanocrystals (NCs) for a ratio of $N\mathrm{_{bnd}}/N_{\mathrm{Wire}}=1.94\pm0.01$ using published experimental data.
\end{abstract}

\maketitle

\section{\label{intro}Introduction}
It is well known that the electronic structure and optical response of nanoscale systems such as NWires or NCs is a function of 
lattice strain which is routinely measured by Raman- and Fourier-Transformation InfraRed (FT-IR) spectroscopy to probe phononic spectra \cite{Jac66,Ana70,Nak81,Boy82,Boy87}. Such spectra are sensitive to lattice strain which is a function of the material via Young's modulus \cite{Ell98}. Changes in compressive or expansive stress were shown to modify the optical response of NWires by photoluminescence (PL) \cite{Tomi11,Joyc11,Treu15}. The growth of monolithic NWires depends critically on balanced stress to avoid stacking faults which deteriotate electronic NWire properties. Attempts to place phosphorus atoms as donors onto lattice sites in free-standing Si NCs were shown to fail increasingly with shrinking NC diameter \cite{Steg08a,Steg09}. These findings were confirmed recently in theory and experiment for SiO$_2$-embedded Si NCs \cite{Koe15,Gnas14}. As structural cause, self-purification was indentified \cite{Dalp06,Dalp08,Chan08,Ossi05}. While no detailed analog investigations were carried out on Si NWires yet, it is apparent from the self-purification mechanism that a structural limit also exists for Si NWires and -- to a lesser extent -- for III-V NWires. All these structures are key to future electronic devices such as radial light emitting diodes \cite{Tomi11} or next generation very-large scale integration (VLSI) Si field effect transistors (FETs) \cite{Webe17}.

The goal of this work is to provide an analytic metrology for zb- and diamond-lattice NWire cross sections by analytic number series down to the individual atom/bond (see e.g. Figure \ref{fig13} and Table \ref{tab_4}). These number series can deliver crucial input to interpret experimental and theoretical data of such NWire systems.  A high accuracy of predicting  NWire cross sections requires adequate experimental data such as exact lattice positions or interface-specific defect ratios. Alternatively, iterative techniques can be used \cite{Koe07}.

With a universal gauge for stress such phenomena could be scaled as a function of NWire diameter $d_{\mathrm{Wire}}$. Principal parameters are the number of atoms forming the zb-NWire $N_{\mathrm{Wire}}(d_{\mathrm{Wire}})$, the bonds between such atoms $N\mathrm{_{bnd}}(d_{\mathrm{Wire}})$ and the number of bonds $N\mathrm{_{IF}}(d_{\mathrm{Wire}})$ terminating the NWire interface. As parameters for size and shape identification, characteristic interface lengths $d_{abc\rm{-IF}}(d_{\mathrm{Wire}})$ per interface orientation \{$abc$\} serve as identification tool for assigning experimental data to the respective analytic NWire cross section. The total cross section area $A$ is a key parameter in particular for electronic devices where current densities can be used instead of absolute currents to interprete device behavior. The ratio $N\mathrm{_{bnd}}(d_{\mathrm{Wire}})/N_{\mathrm{Wire}}(d_{\mathrm{Wire}})$ yields the bonds per atom within zb-NWires as a gauge for the response to external stress, while the ratio  $N\mathrm{_{IF}}(d_{\mathrm{Wire}})/N\mathrm{_{bnd}}(d_{\mathrm{Wire}})$ describes the ability of embedding materials or ligands to exert stress onto NWires. The impact of a highly polar surface termination on the zb-NWire electronic structure is assessed by  $N\mathrm{_{IF}}(d_{\mathrm{Wire}})/N_{\mathrm{Wire}}(d_{\mathrm{Wire}})$ which provides a gauge to interface charge transfer \cite{Koe14,Koe18a}.

\begin{figure}[t!]
\begin{center}
\includegraphics[width=0.3642\textheight]{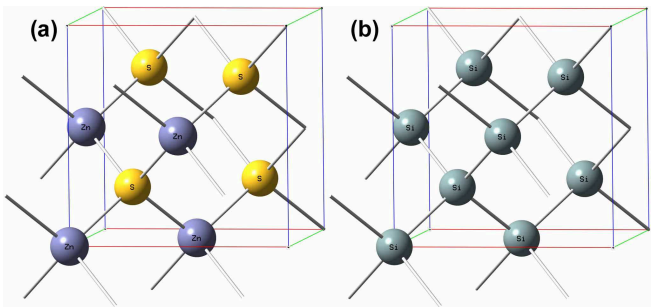}
\end{center}
\caption{\label{fig01}Examples of periodic unit cells of space group F$\bar{4}$3m (zb) and Fd$\bar{3}$m (diamond) covered in this work \cite{Koe16}. Zinc sulfide (\emph{Zinkblende}) ZnS (a), standing for many II-VI and III-V semiconductors obeying zb-lattice symmetry. Main group IV semiconductors such as diamond, germanium or silicon (b) obey diamond-lattice symmetry. Both space groups share the same crystal symmetry apart from their primitive basis which is A--A (Si--Si) for diamond lattices and A--B (Zn--S) for zb-lattices.}
\end{figure}

We illustrate results on Si NWires (diamond lattice), using one color for NWire atoms without interface bonds. This facilitates the color coding of outermost NC atoms in accord with their number of interface bonds, see figure \ref{fig03}. Analytical number series introduced below also hold for zb-NWires due to straightforward symmetry arguments, \emph{cf.} Figure \ref{fig01}. 
Hence, analytical descriptions below cover all NWires with zb symmetry in addition to Si and Ge. Material properties resulting from differences in the base cell -- \emph{cf.} Figure \ref{fig01} -- are \emph{not} considered here. This constraint has no impact on the applicability of the analytics of our work unless data of zb-type NWires are directly compared with diamond-type NWires. 

Due to the complexity of symmetry arguments, zb-NWires with index faceting beyond \{$1\bar{3}1$\} are not considered in our work. For Si, the \{111\} (\{001\}) facets have the lowest (second-lowest) experimental values of surface free energy, \emph{cf.} Table \ref{tab_1}, followed by \{110\} facets. Regarding surface bond density, \{111\}-oriented facets have the lowest surface bond density for all facets up to \{433\} orientation \cite{Hes93}. These findings also hold for other diamond- and zb-NWires due to symmetry arguments.  

The number series follow a run index $i$ which relates to $d_{\mathrm{Wire}}[i]$ via the respective NWire cross section area $A[i]$. Since the NWires considered here differ in their cross section shape, we use $d_{\mathrm{Wire}}$ of a spherical NWire which allows to compare the same parameter of different NWire shapes as function of $d_{\rm{Wire}}^{\,\rm{sphr}}$:
\begin{eqnarray}\label{eqn-01}
  d_{\rm{Wire}}^{\,\rm{sphr}}[i]&=&\sqrt{\frac{4}{\pi}\times A[i]}
\end{eqnarray}  
\begin{table}[t!]
\caption{ \label{tab_1}Bond densities and free energies per square for low index Si facets. Bond density values taken from \cite{Hes93}, experimental surface energy values taken from \cite{Eagl93}.}\vspace*{0.1cm}
\renewcommand{\baselinestretch}{1.2}\small\normalsize
  \begin{tabular}{c|c|c}
    \hline
    facet&surface bond&surface free\\ 
    orientation&density [cm$^{-2}$]&energy [Jm$^{-2}$]\\ \hline 
  \{001\}&$1.36\times 10^{15}$&1.36\\ 
  \{110\}&$0.96\times 10^{15}$&1.43\\ 
  \{111\}&$0.78\times 10^{15}$&1.23\\ \hline
  \end{tabular}
\renewcommand{\baselinestretch}{1.0}\small\normalsize
\end{table}

\section{\label{nomiWire}Analytical Number Series of Nanowire Cross Sections}
Below, we provide a general procedure to pick the right NWire cross section to use for interpreting experimental data as a function of NWire structure. For direct imaging, e.g. by high resolution transmission electron microscopy (HR-TEM, see Figure \ref{fig13}) or scanning TEM (STEM, see Figure \ref{fig12}), the ratio of interface lengths $d_{\rm{IF}}$, widths $w$ and heights $h$ of NWire cross sections can be readily extracted. If atomic information is available as is the case in Figure \ref{fig13}, it helps further to pick the right symmetry center of which two exist for NWires with hexagonal cross section. This is reflected in an \emph{even} and an \emph{odd} set of number series per cross section type. Other spectroscopic methods such as electron paramagnetic resonance (EPR) or Raman spectroscopy are sensitive to certain interface defects or modifications which either exist exclusively or with a very high probability at specific interface orientations. The partitions per interface-specific defect can then be used to determine ratios of side lengths for specifying the NWire cross section. An example with EPR is given at the end of Section \ref{Outlook}.

For the description of zb- and diamond-lattice NWire cross sections, 
we consider slabs with a thickness $d_{\rm{slab}}$ of one unit cell thickness $a_{\rm{uc}}$ multiplied with the respective factor in the direction of NWire growth to achieve periodicity; values are summarized in Table \ref{tab_2}. In addition, Table \ref{tab_2} lists the amount of atoms and bonds per column (i.e. per atom or bond visible) as a function of NWire axis orientation for a top view onto the cross section, thereby allowing to count atoms and bonds. Respective images are provided for all NWire cross section types presented here. 
\begin{table}[h!]
\caption{ \label{tab_2}Slab thickness $d_{\rm{slab}}$ of NWire cross sections as function of growth axis orientation given in unit cell lengths $a_{\rm{uc}}$ per growth orientation to achieve periodicity. Numbers of atoms and of bonds per column as described per feature seen in cross section top view are given to enable the counting of atoms and NWire-internal bonds.}\vspace*{0.1cm}
\renewcommand{\baselinestretch}{1.2}\small\normalsize
  \begin{tabular}{c|c|c|c}
    \hline
    growth axis&$d_{\rm{slab}}$&atoms&bonds\\ \hline 
    &[$a_{\rm{uc}}$]&\multicolumn{2}{c}{per column, in top view}\\ \hline
  001&1&1&1 per $/$ and $\backslash$\\ 
  110&$\sqrt{2}$&2&2 per $/$ and $\backslash$, 4 per $|$\\ 
  111&$\sqrt{3}$&2&1 per atom column,\\
  &&&1 between atom columns\\ 
  11$\bar{2}$&$\sqrt{6}$&1&2 per $/\!\!\!\backslash$, 1 per --- and -- \footnote{bond symbols must be turned by 90$^{\circ}$ to align with graphs in Figures \ref{fig06} and \ref{fig07}}\\ \hline
  \end{tabular}
\renewcommand{\baselinestretch}{1.0}\small\normalsize
\end{table}
Table \ref{tab_2a} shows all parameters described by number series for each cross section type.
\begin{table}[h!]
\caption{ \label{tab_2a}Parameter list for each NWire cross section. All $N_{\nu}$ are calculated as per NWire slab.}\vspace*{0.1cm}
\renewcommand{\baselinestretch}{1.2}\small\normalsize
  \begin{tabular}{c|c}
    \hline
    parameter&description\\ \hline 
  $N_{\rm{Wire}}$&number of atoms forming NWire\\
  $N_{\rm{bnd}}$&number of bonds within NWire\\
  $N_{\rm{IF}}$&number of interface bonds of NWire\\
  $N_{abc\rm{-IF}}$&number of bonds per IF type $\{abc\}$\\
  $d_{abc\rm{-IF}}$&length of IF with orientation $\{abc\}$\\
  $w$&max. width of NWire cross section\\
  $h$&max. height of NWire cross section\\
  $A$&cross section area\\ \hline
  \end{tabular}
\renewcommand{\baselinestretch}{1.0}\small\normalsize
\end{table}
The indexing of NWire cross section type is given as a superscript with its shape and growth direction, see Table \ref{tab_2b}.
\begin{table}[h!]
\caption{ \label{tab_2b}List of NWire shape indices -- cross section, growth direction and side interfaces (where necessary) -- added to all parameters as a superscript.}\vspace*{0.1cm}
\renewcommand{\baselinestretch}{1.2}\small\normalsize
  \begin{tabular}{l|c|c|c}
    \hline
    superscript&growth axis&cross section&side interfaces\footnote{only when required to distinguish cross sections}\\ \hline 
  $001-\boxempty$&001&square&\\
  $110-\bigbox$&110&rectangular&\\
  $110-\hexagon$&110&hexagon&\\
  $11\bar{2}-\hexagon$&$11\bar{2}$&hexagon&\\
  $111-\hexagon|110$&111&hexagon&111\\
  $111-\hexagon|11\bar{2}$&111&hexagon&$11\bar{2}$\\ \hline
  \end{tabular}
\renewcommand{\baselinestretch}{1.0}\small\normalsize
\end{table} 

For all equations in this section, the limits of run index is $1\leq i < \infty$ unless noted otherwise. On details of the crystallographic algorithm to obtain the number series below, we refer the reader to our previous work \cite{Koe16}.
 
\subsection{\label{AnaNomiSq001} NWires growing along [001] Direction with Square Cross Section and Four \{001\} Interfaces}
\begin{eqnarray}\label{eqn-02}
N_{\rm{Wire}}^{001-\boxempty}[i]&=&8\left(i+1\right)^2
\end{eqnarray}
\begin{eqnarray}\label{eqn-03}
N_{\rm{bnd}}^{001-\boxempty}[i]&=&\left(4i+3\right)^2
\end{eqnarray}
\begin{eqnarray}\label{eqn-04}
N_{\rm{IF}}^{001-\boxempty}[i]&=&2\left(8i+7\right)
\end{eqnarray}
\begin{eqnarray}\label{eqn-05}
d_{\rm{001-IF}}^{\,001-\boxempty}[i]&=&a_{\rm{uc}}\frac{4}{3}\left(i+\frac{3}{4}\right)
\end{eqnarray}
The square shape of the cross section results in $w[i]\equiv h[i]\equiv d_{\rm{001-IF}}^{\,001-\boxempty}[i]$.
\begin{eqnarray}\label{eqn-06}
A^{\,001-\boxempty}[i]&=&\big(a_{\rm{uc}}\big)^2\left(\frac{4}{3}\right)^2\left(i+\frac{3}{4}\right)^2
\end{eqnarray}

\begin{figure}[h!]
\begin{center}
\includegraphics[totalheight=0.1025\textheight]{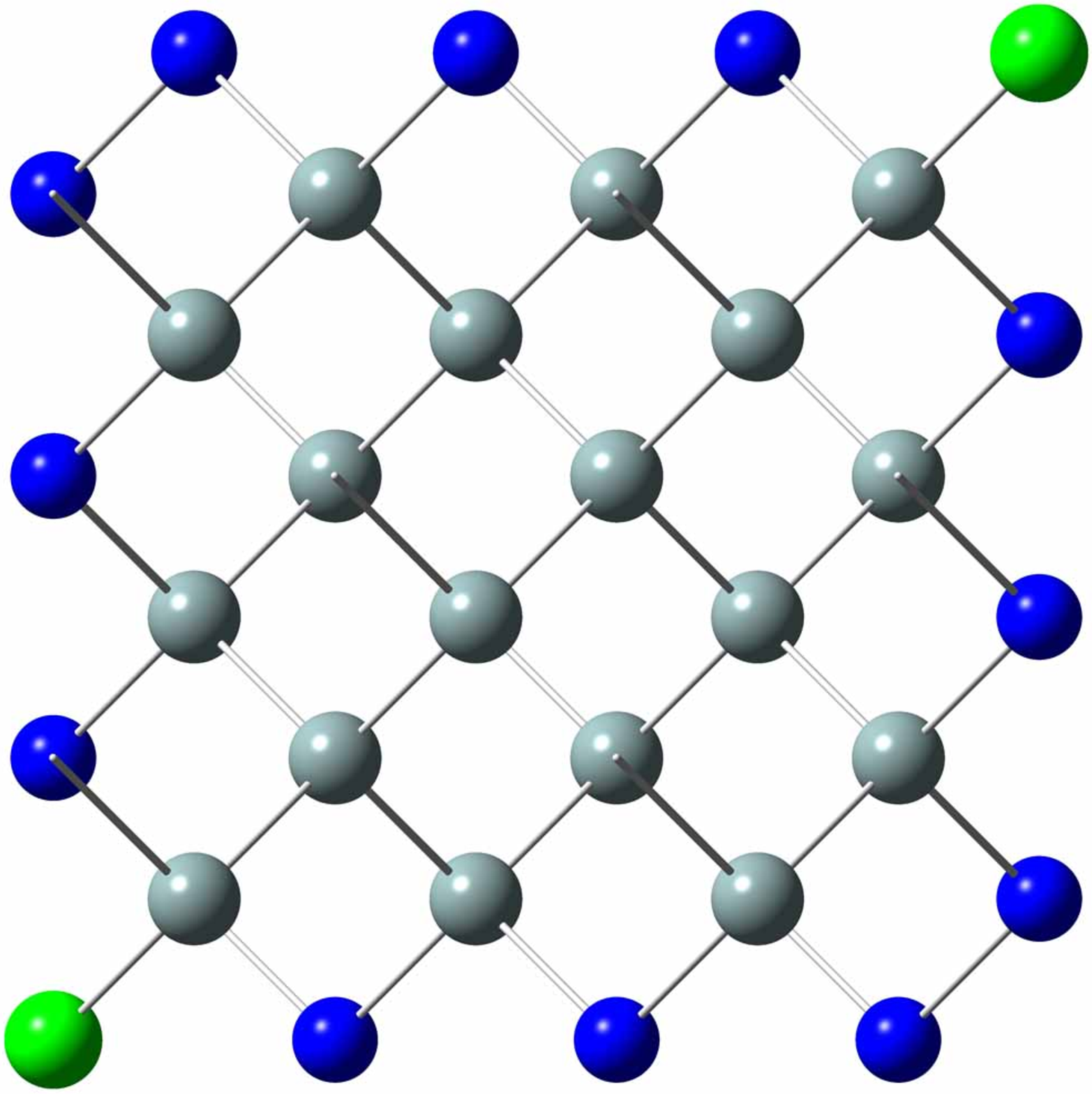}
\includegraphics[totalheight=0.1025\textheight]{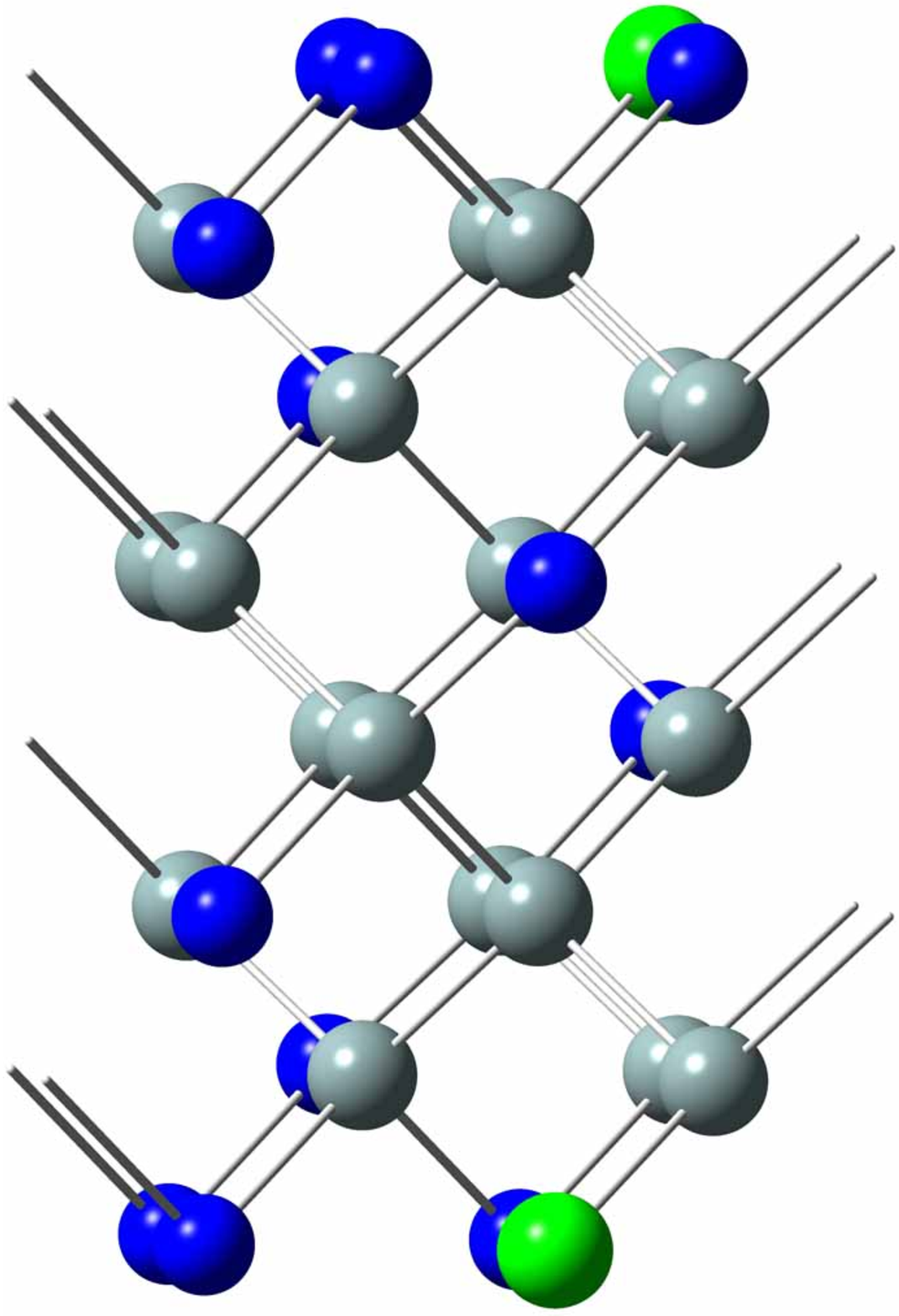}\hfill
\includegraphics[totalheight=0.116\textheight]{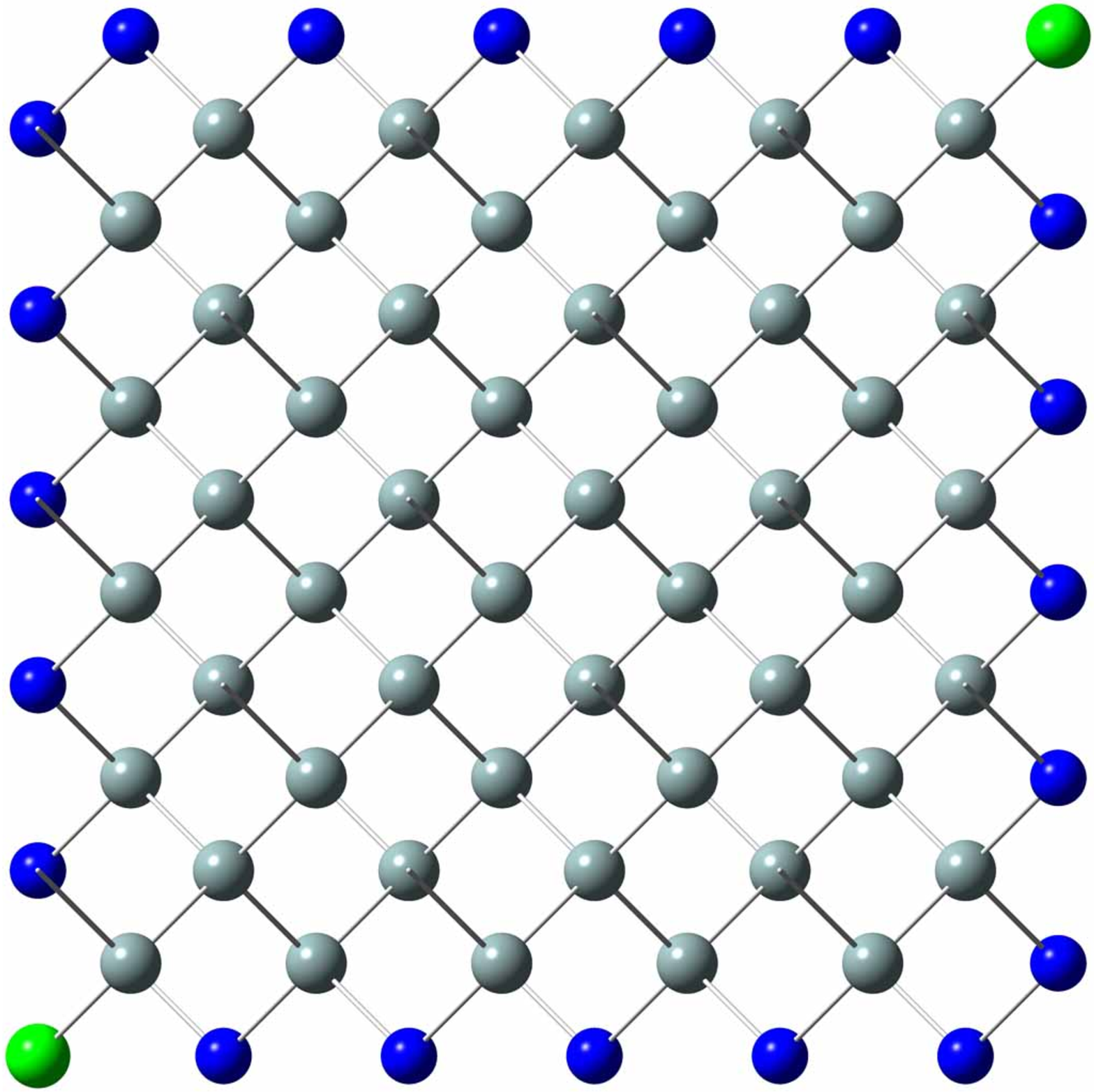}
\includegraphics[totalheight=0.116\textheight]{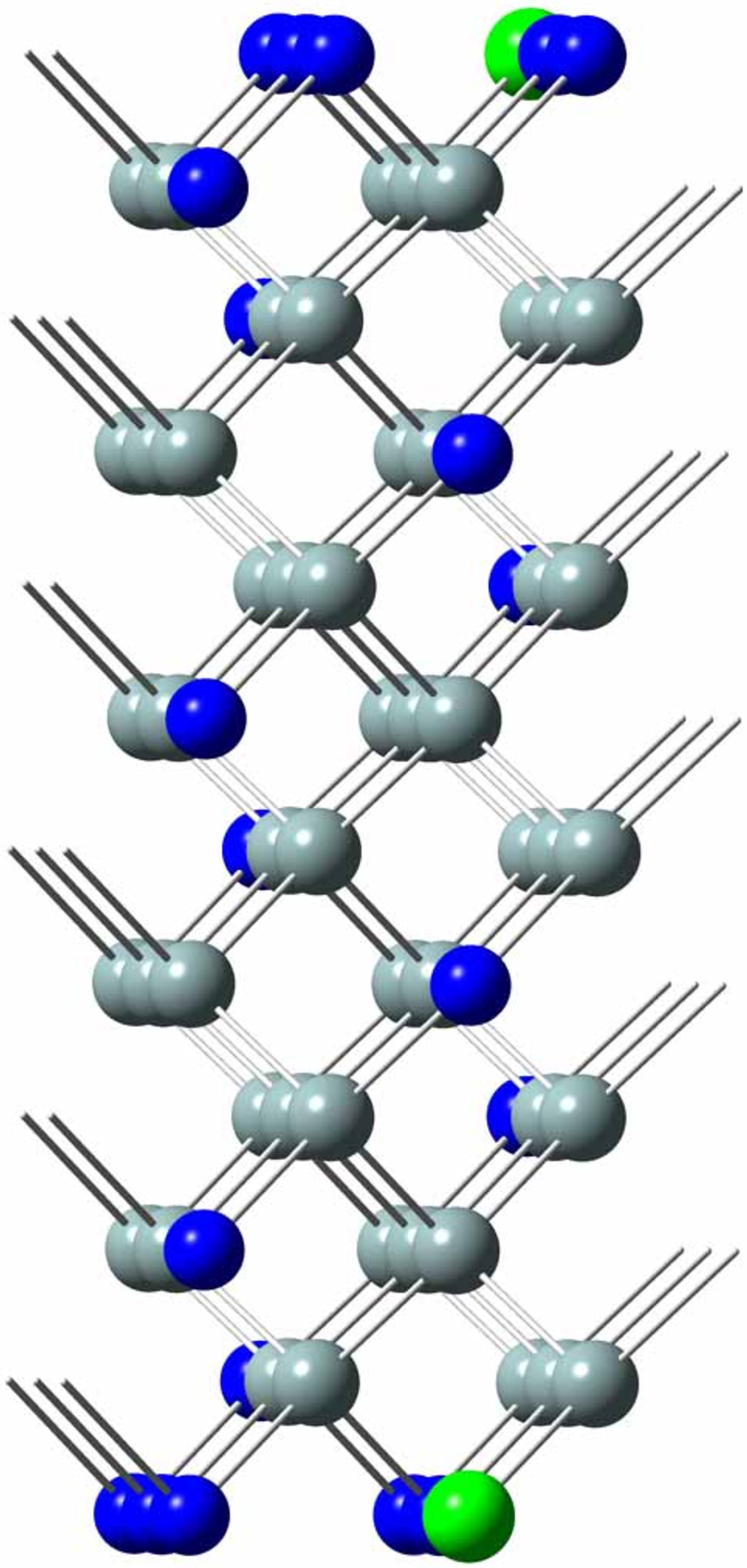}\\[-0.1cm]
\begin{picture}(0,0)
\hspace{-2.45cm}{\bf(a)}\hspace{3.91cm}{\bf (b)}
\end{picture}\\[0.2cm]
\includegraphics[totalheight=0.123\textheight]{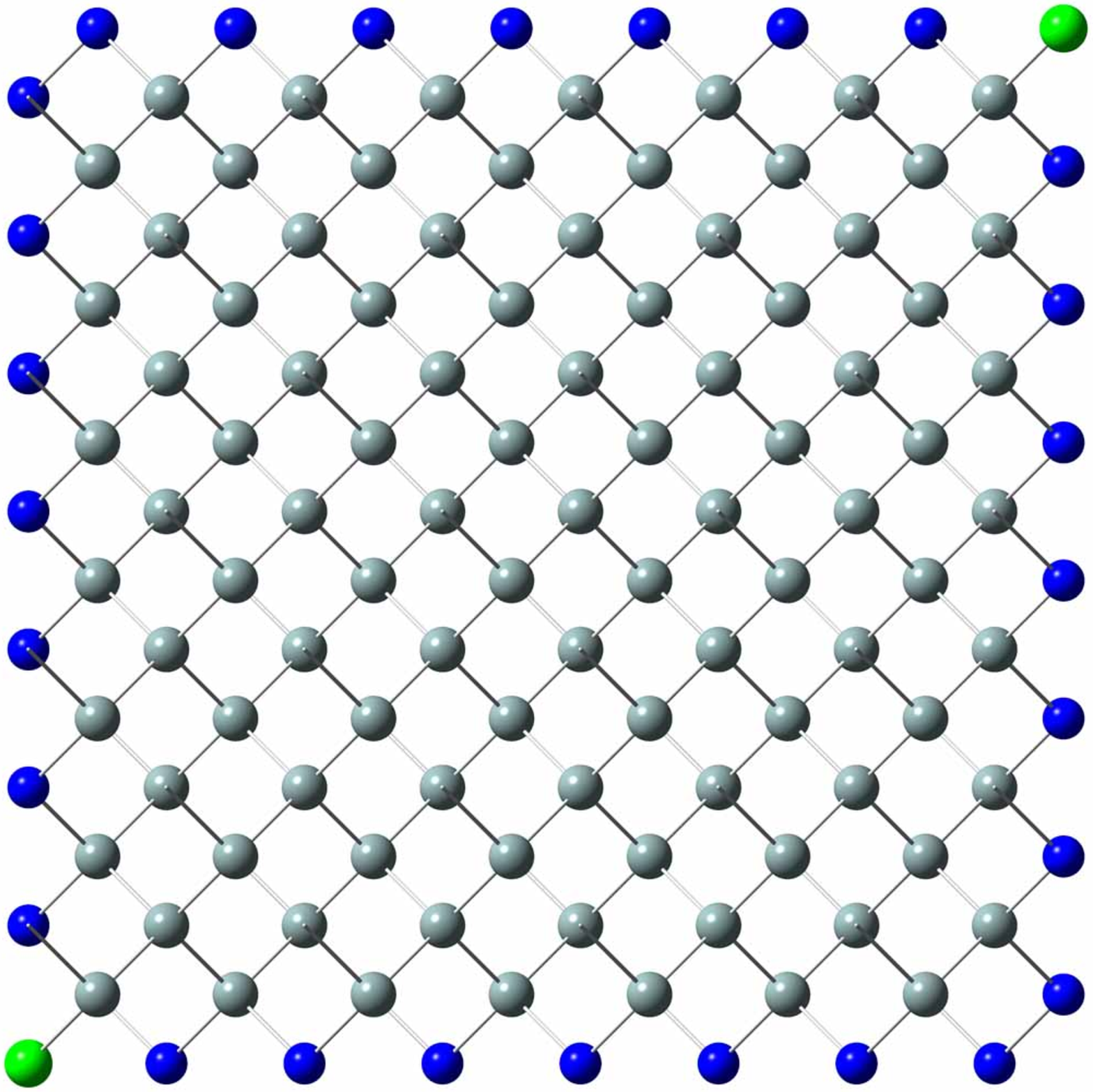}
\includegraphics[totalheight=0.123\textheight]{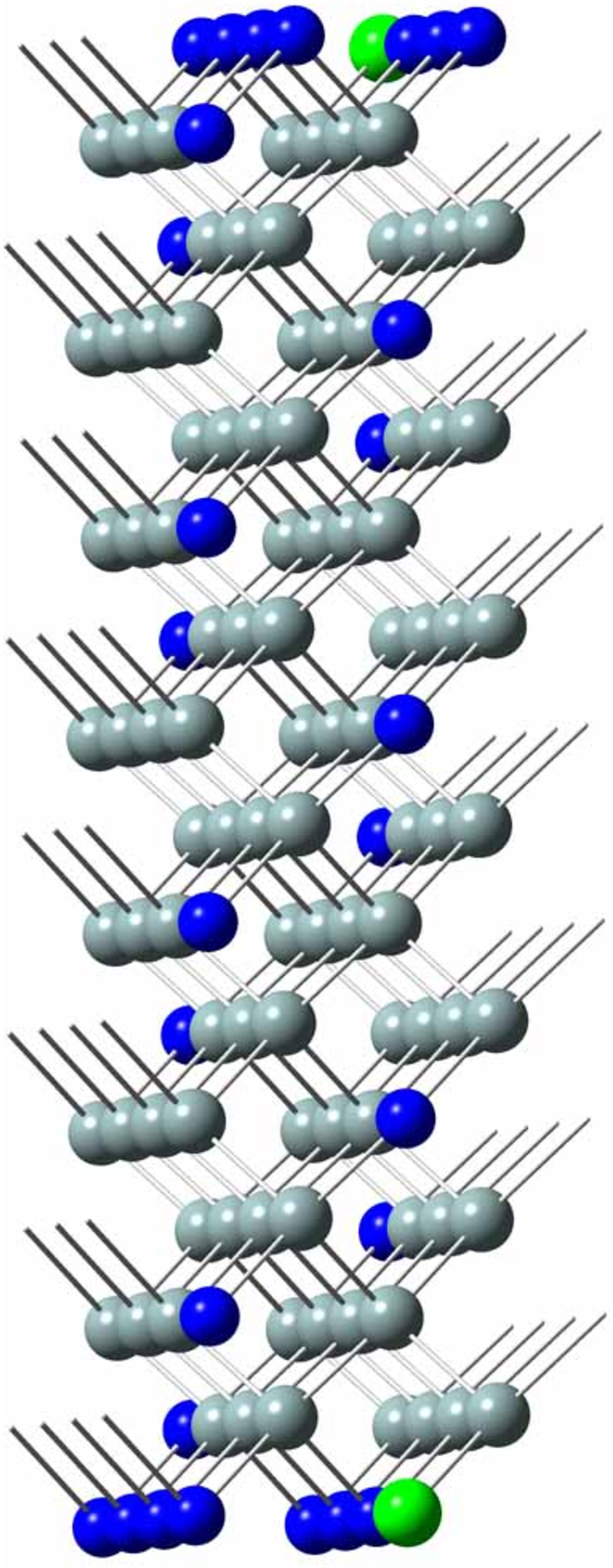}\hfill
\\[-0.1cm]
\begin{picture}(0,0)
\hspace{-2.45cm}{\bf(c)}
\end{picture}
\end{center}
\caption{Cross section and side view of zb-/diamond-lattice NWires growing along [001] axis with square cross section and four \{001\} interfaces for run index $i=1$ to 3: X$_{32}$ (a), X$_{72}$ (b), X$_{128}$ (c). Interior atoms are gray, atoms with two interface bonds are blue, atoms with three interface bonds are green.}
\label{fig02}
\end{figure}

Figure \ref{fig02} shows the the first four members of the square NWire cross sections with growth along [001] direction and four \{001\} interfaces.

\subsection{\label{AnaNomiRect110} NWires growing along [110] Direction with Rectangular Cross Section and Two \{001\} Plus Two \{110\} Interfaces}
\begin{figure}[h!]
\begin{center}
\includegraphics[totalheight=0.095\textheight]{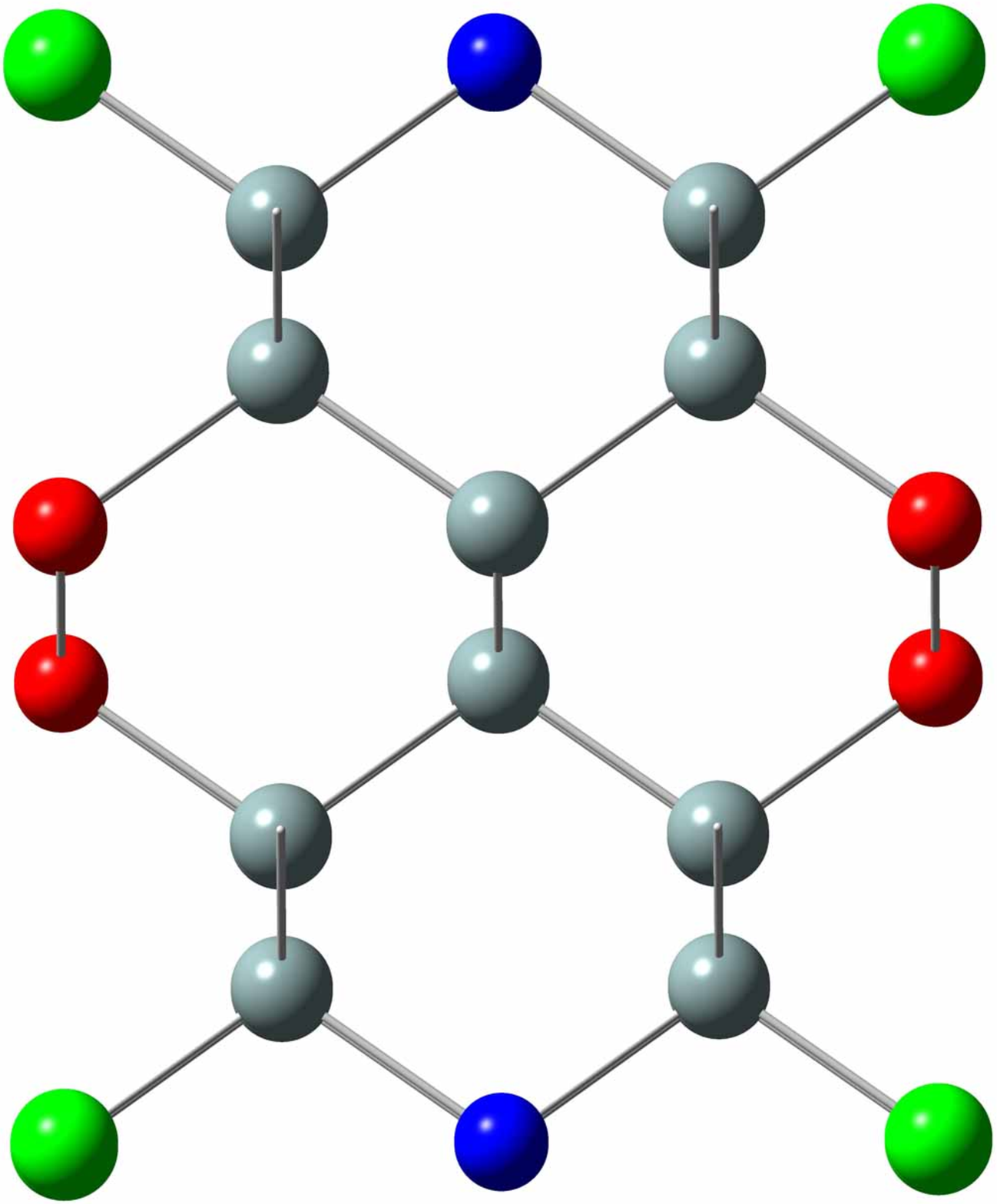}
\includegraphics[totalheight=0.095\textheight]{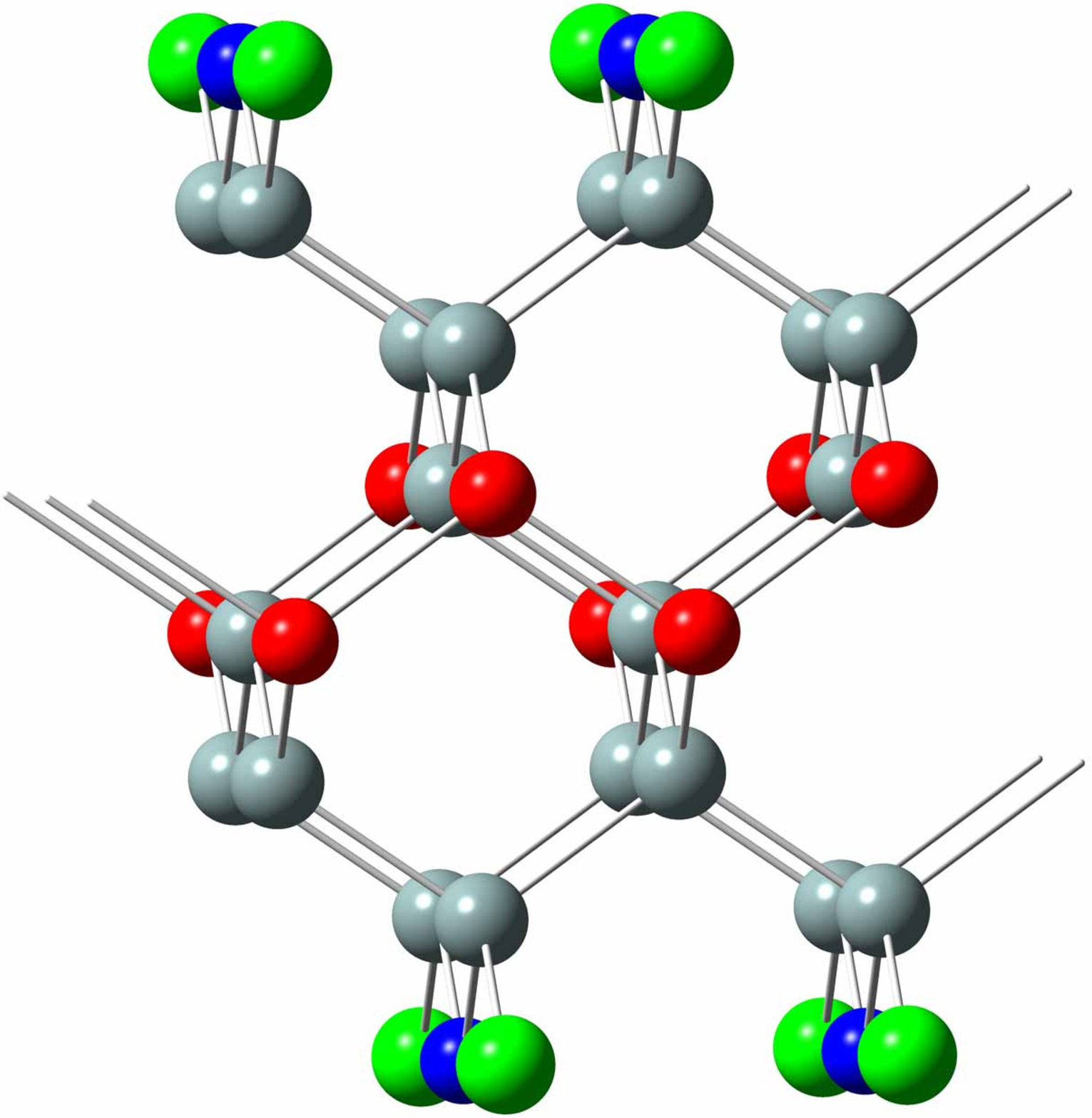}\hfill
\includegraphics[totalheight=0.1135\textheight]{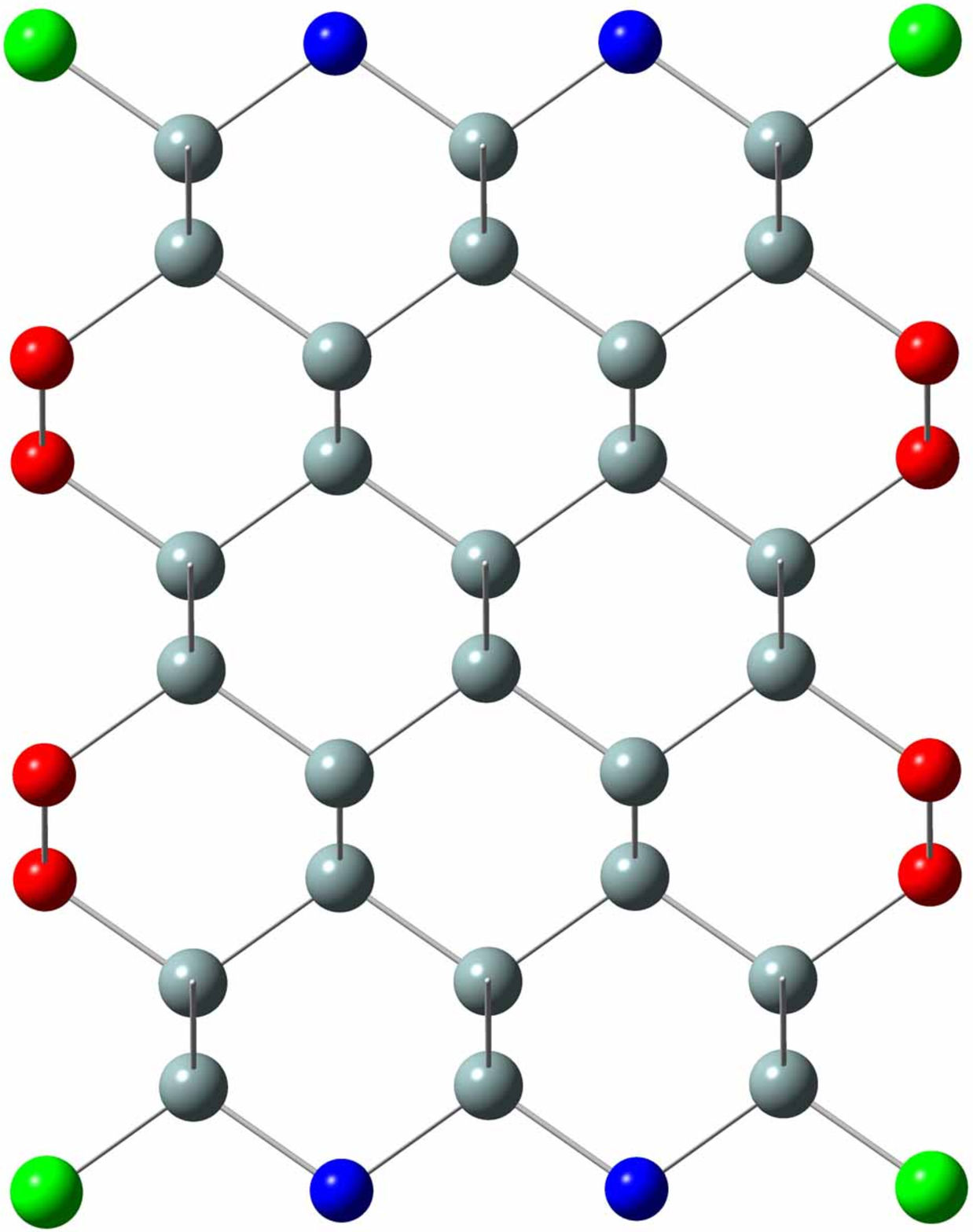}
\includegraphics[totalheight=0.1135\textheight]{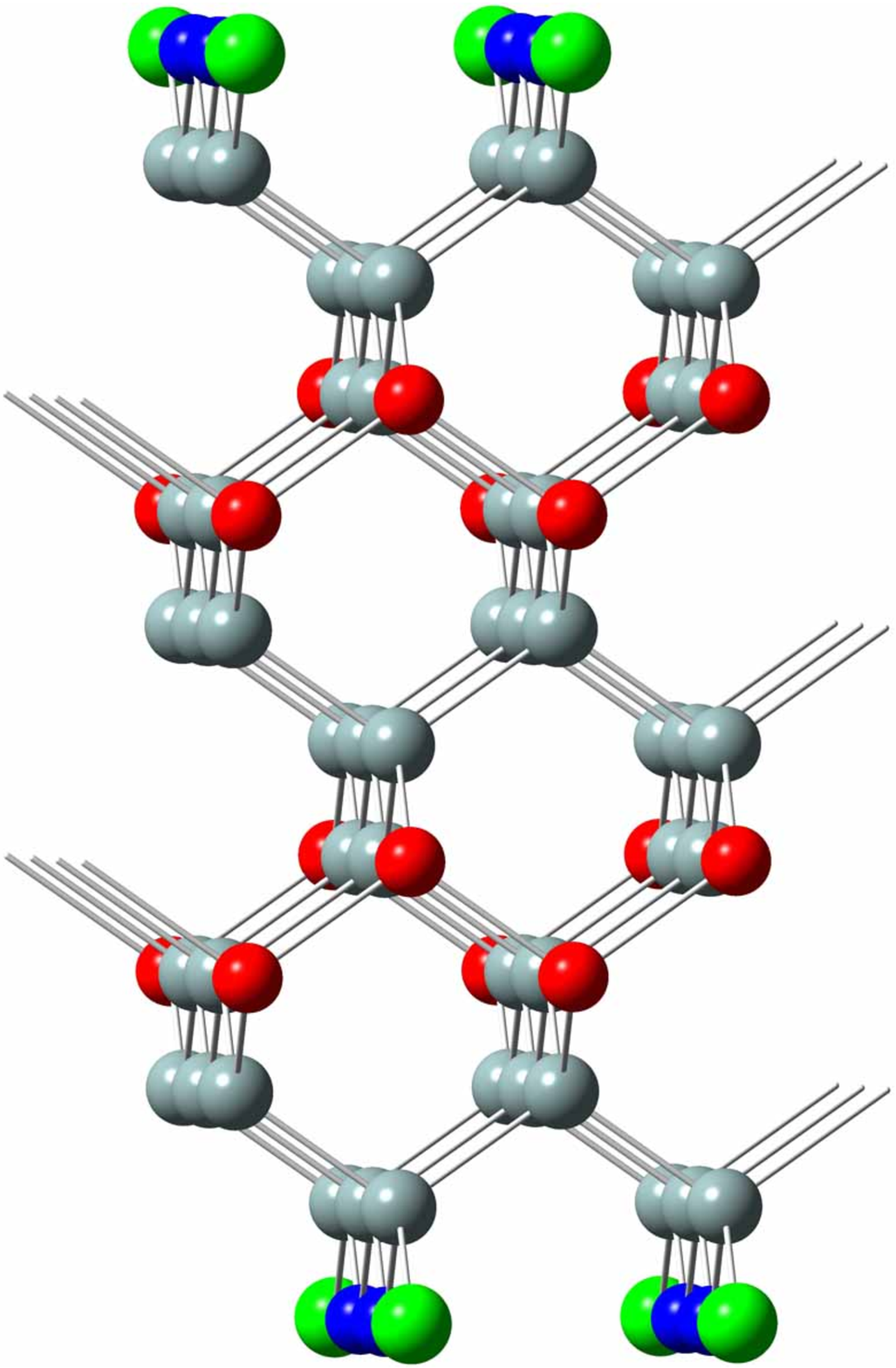}\\[-0.1cm]
\begin{picture}(0,0)
\hspace{-2.45cm}{\bf(a)}\hspace{3.91cm}{\bf (b)}
\end{picture}\\[0.2cm]
\includegraphics[totalheight=0.1305\textheight]{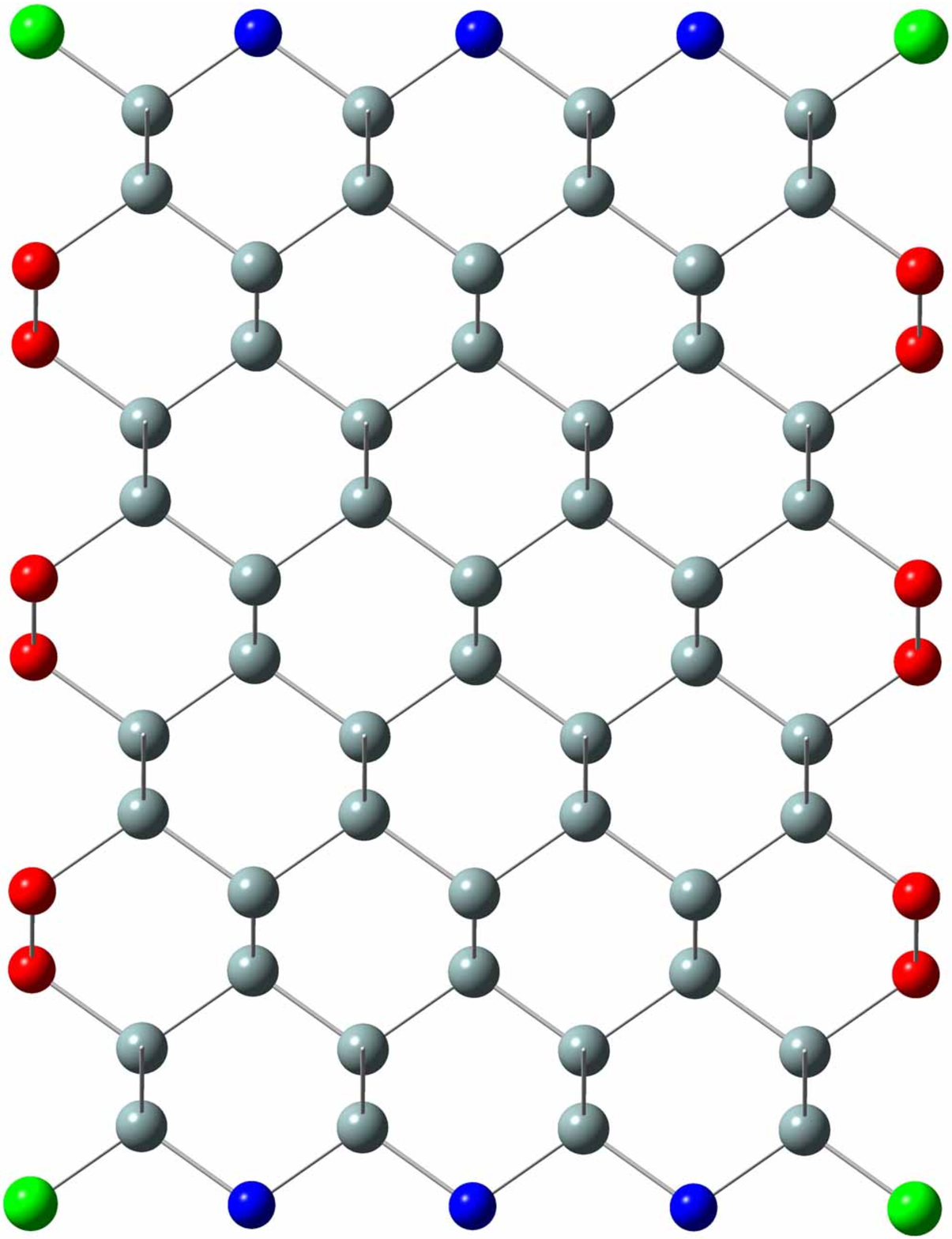}
\includegraphics[totalheight=0.1305\textheight]{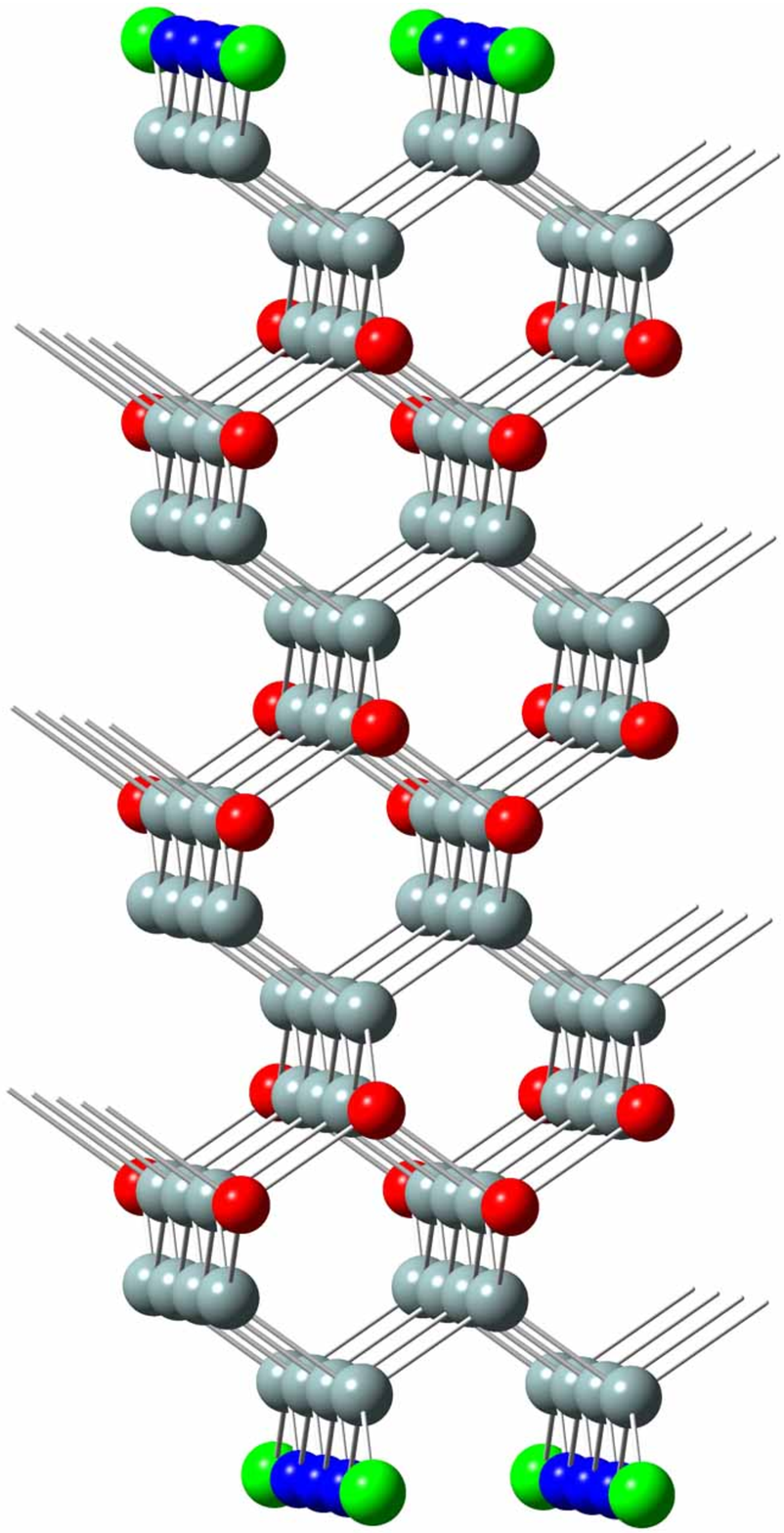}\hfill
\\[-0.1cm]
\begin{picture}(0,0)
\hspace{-2.45cm}{\bf(c)}
\end{picture}
\end{center}
\caption{Cross section and side view of zb-/diamond-lattice NWires growing along [110] axis with rectangular cross section, two \{111\} and two \{001\} interfaces for run index $i=1$ to 3: X$_{40}$ (a), X$_{84}$ (b), X$_{144}$ (c).  Interior atoms are gray, atoms with one/two/three interface bonds are red/blue/green, respectively.}
\label{fig03}
\end{figure}

\begin{eqnarray}\label{eqn-07}
N_{\rm{Wire}}^{110-\bigbox}[i]&=&8(i+1)\left(i+\frac{3}{2}\right)
\end{eqnarray}
\begin{eqnarray}\label{eqn-08}
N_{\rm{bnd}}^{110-\bigbox}[i]&=&4\left(i+1\right)\left(4i+3\right)
\end{eqnarray}
\begin{eqnarray}\label{eqn-09}
N_{\rm{IF}}^{110-\bigbox}[i]&=&8\left(2i+3\right)%
\end{eqnarray}
\begin{eqnarray}\label{eqn-10}
\frac{N_{\rm{110-IF}}^{110-\bigbox}[i]}{N_{\rm{001-IF}}^{110-\bigbox}[i]}&=&
\frac{8i}{8(i+3)}\ =\ \frac{i}{i+3}
\end{eqnarray}
The center expression shows both number series in their explicit form, while the right side expression presents the simplied result of their ratio.
\begin{eqnarray}\label{eqn-11}
d_{\rm{001-IF}}^{\,110-\bigbox}[i]&=
&a_{\rm{uc}}\left(\frac{i+1}{\sqrt{2}}\right)
\end{eqnarray}
The width of the rectangular cross section follows straightforward from  $w[i]\equiv d_{\rm{001-IF}}^{\,110-\bigbox}[i]$.
\begin{eqnarray}\label{eqn-12}
d_{\rm{110-IF}}^{\,110-\bigbox}[i]&=&a_{\rm{uc}}\left(i+\frac{3}{4}\right)
\end{eqnarray}
The height of the rectangular cross section follows straightforward from  $h[i]\equiv d_{\rm{110-IF}}^{\,110-\bigbox}[i]$.
\begin{eqnarray}\label{eqn-13}
A^{\,110-\bigbox}[i]&=&\big(a_{\rm{uc}}\big)^2\frac{(i+1)(4i+3)}{4\sqrt{2}}
\end{eqnarray}
The cross section of this NWire type is shown in Figure \ref{fig03} for $i=1$ to 3.

\subsection{\label{AnaNomiHexa110} NWires growing along [110] Direction with Hexagonal Cross Section and Four \{111\} Plus Two \{001\} Interfaces}
\begin{figure}[h!]
\begin{center}
\includegraphics[totalheight=0.226\textheight]{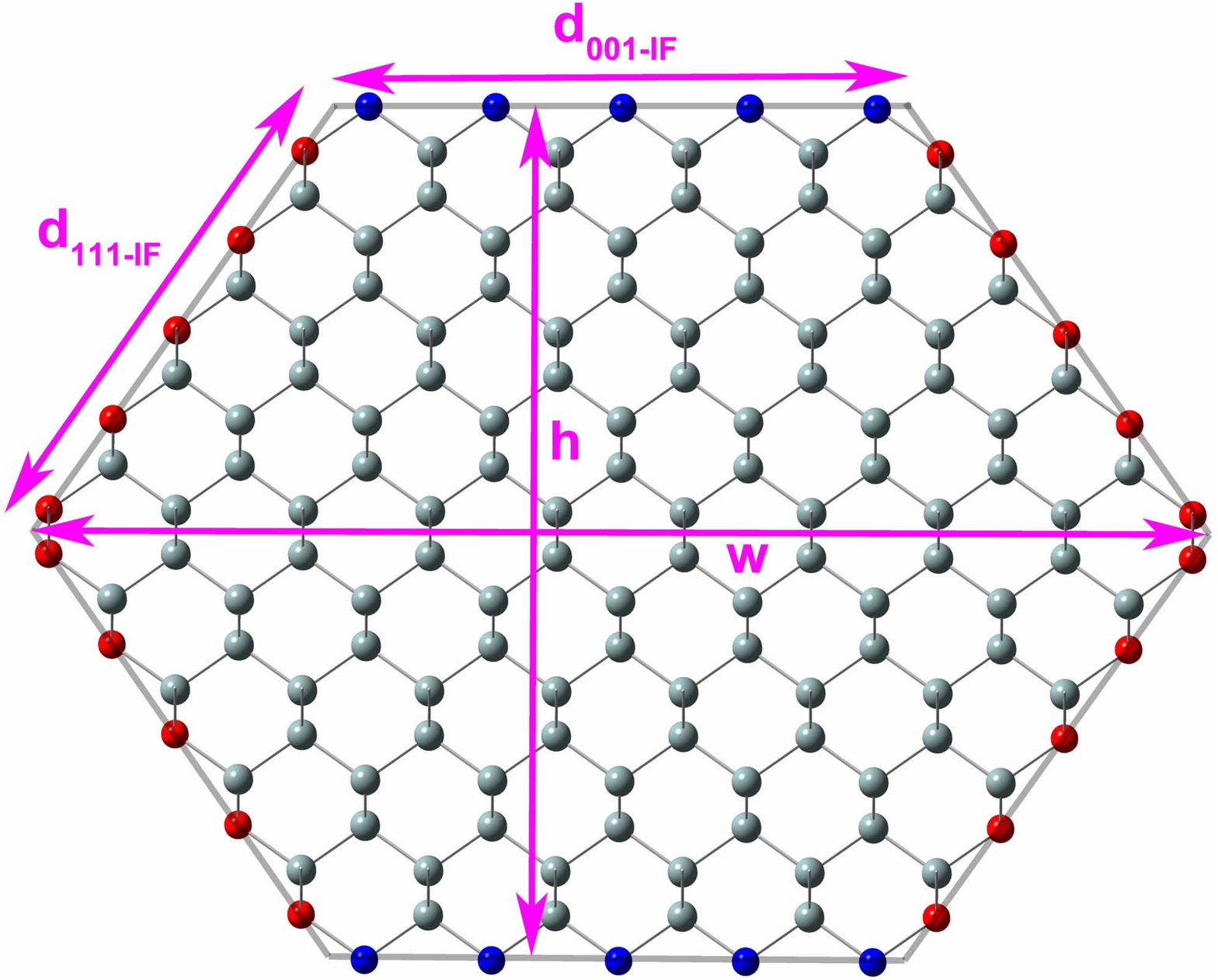}\\[-0.1cm]
\begin{picture}(0,0)
{\bf(a)}
\end{picture}\\[0.2cm]
\includegraphics[totalheight=0.0613\textheight]{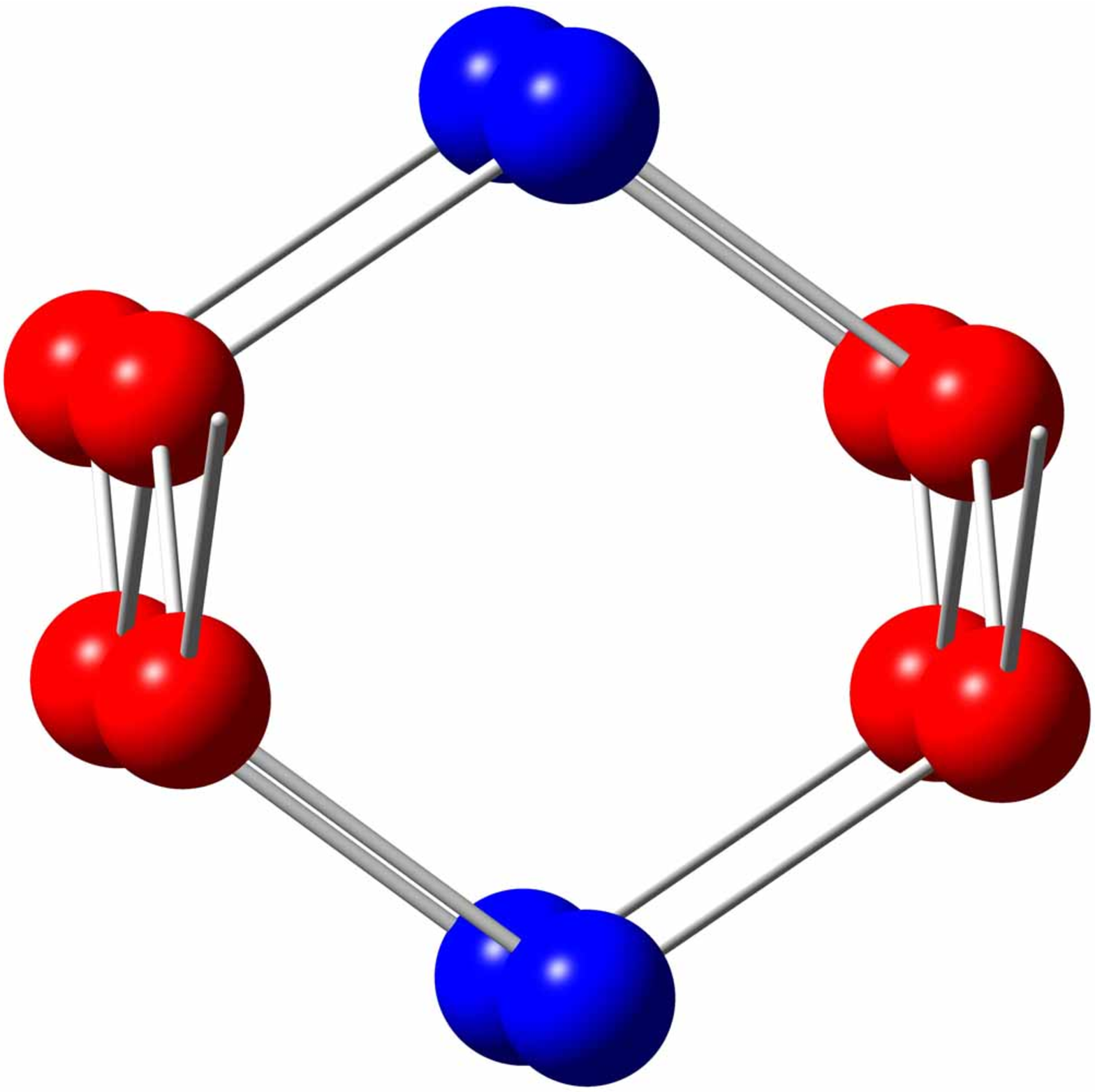}
\includegraphics[totalheight=0.0613\textheight]{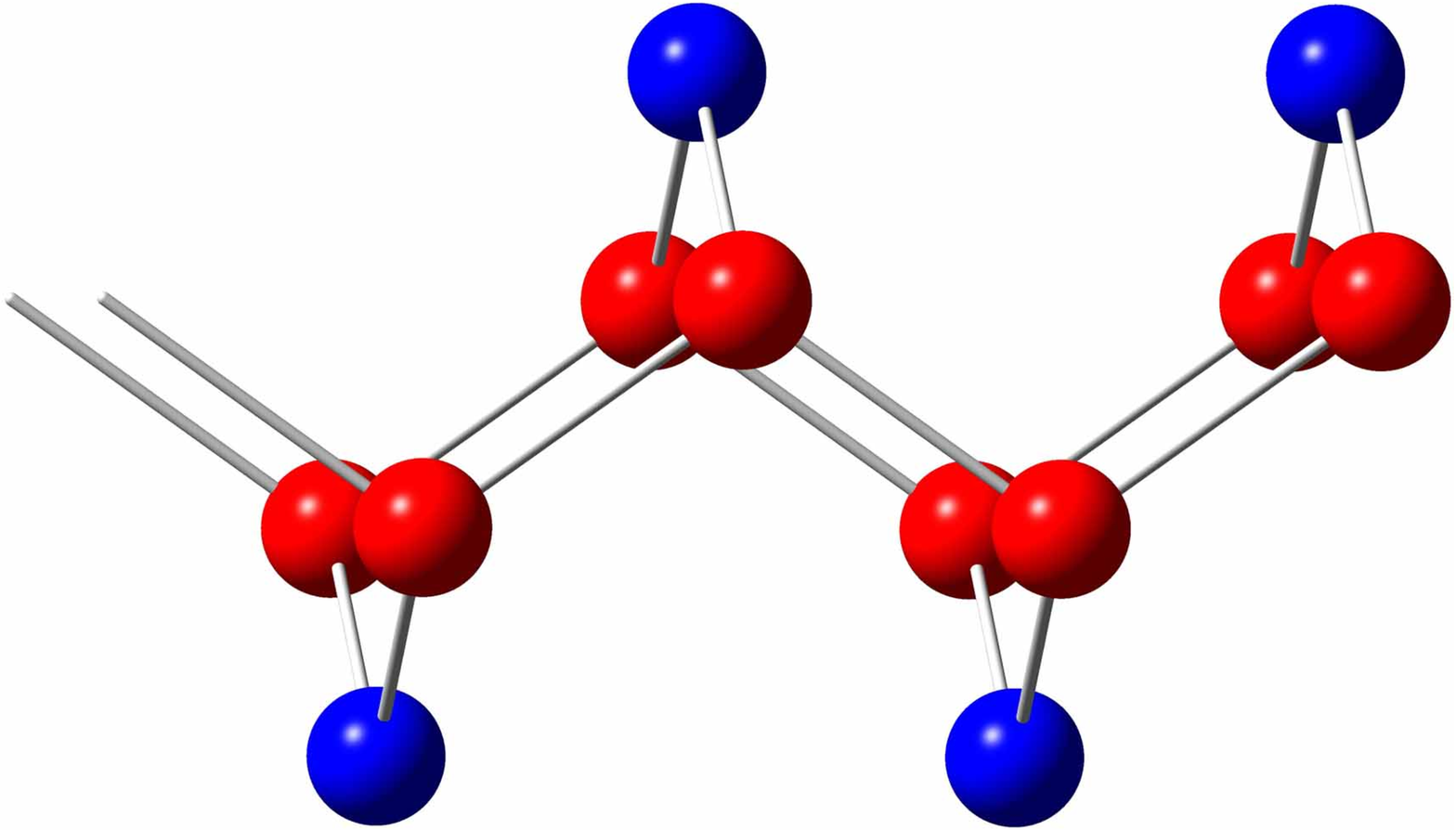}\hfill
\includegraphics[totalheight=0.0763\textheight]{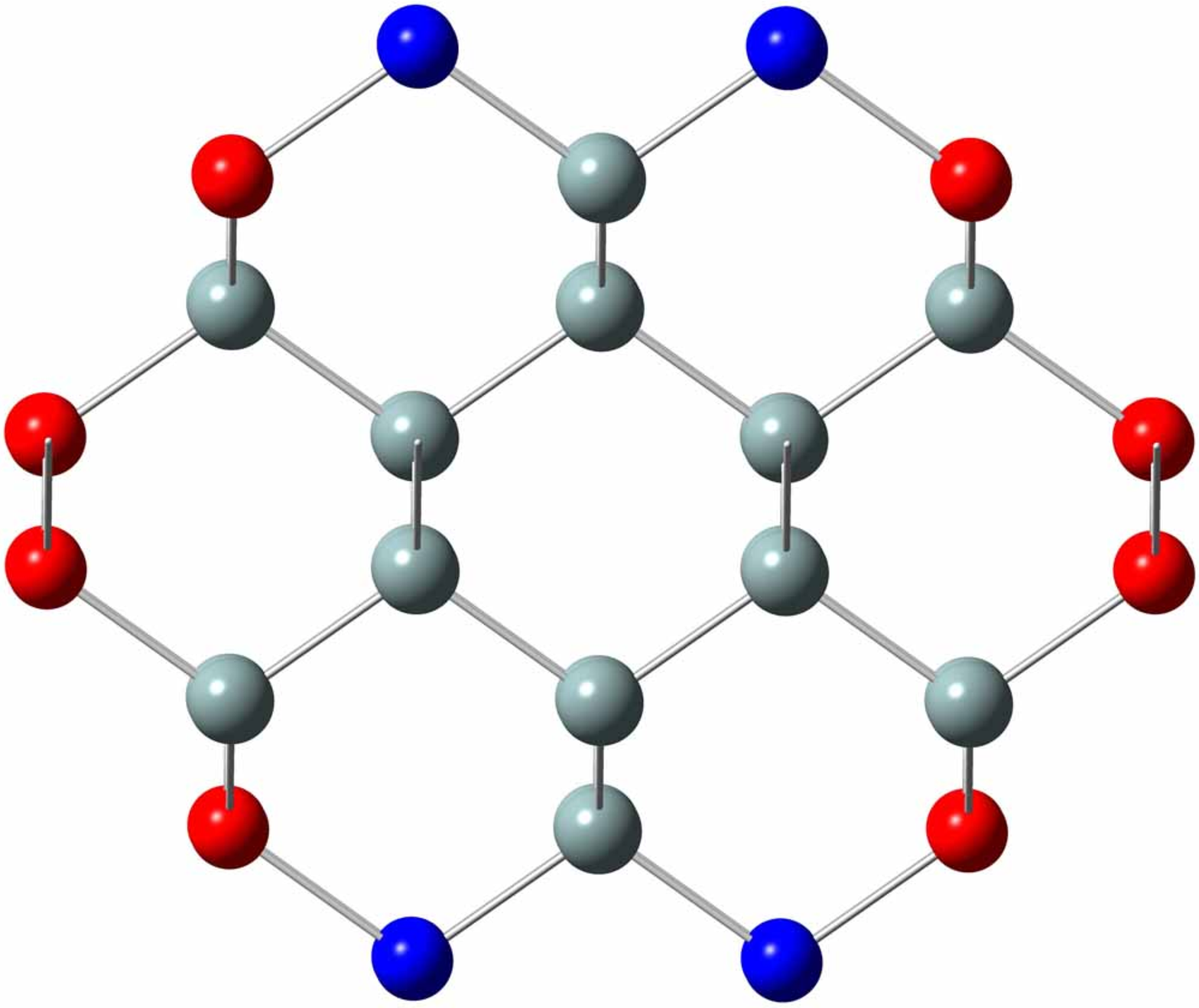}
\includegraphics[totalheight=0.0763\textheight]{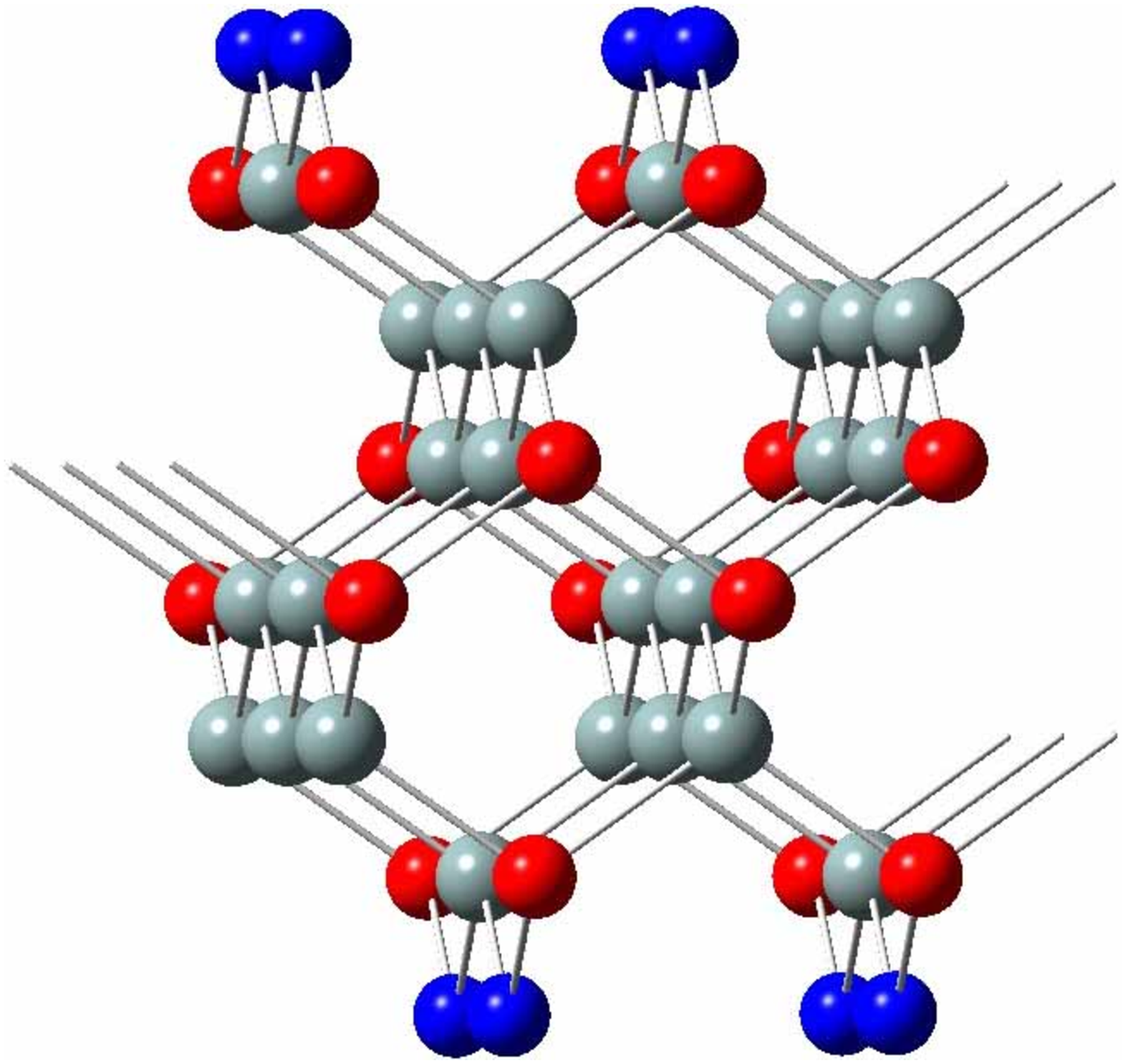}\\[-0.1cm]
\begin{picture}(0,0)
\hspace{-2.45cm}{\bf(b)}\hspace{3.91cm}{\bf (c)}
\end{picture}\\[0.2cm]
\includegraphics[totalheight=0.0863\textheight]{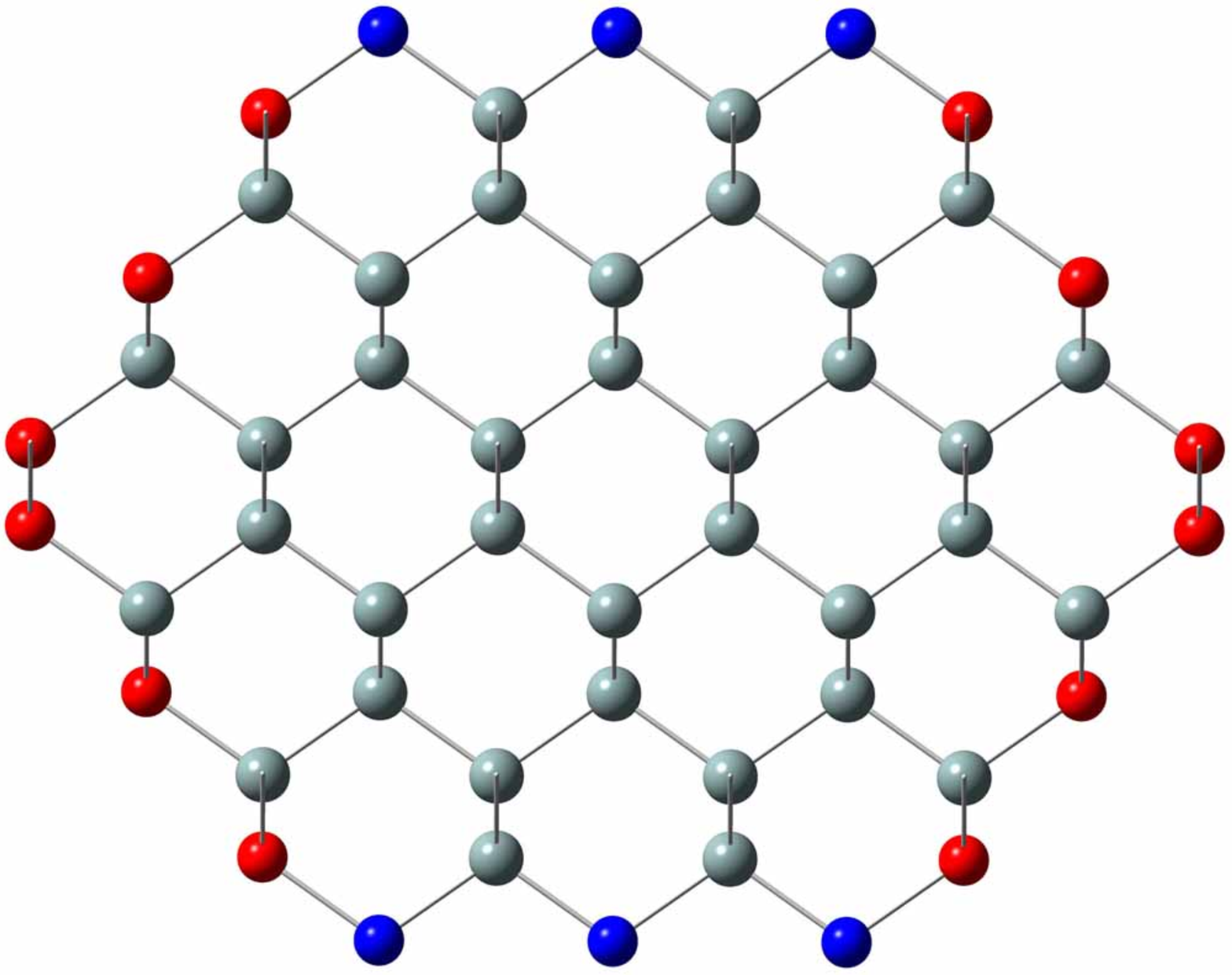}
\includegraphics[totalheight=0.0863\textheight]{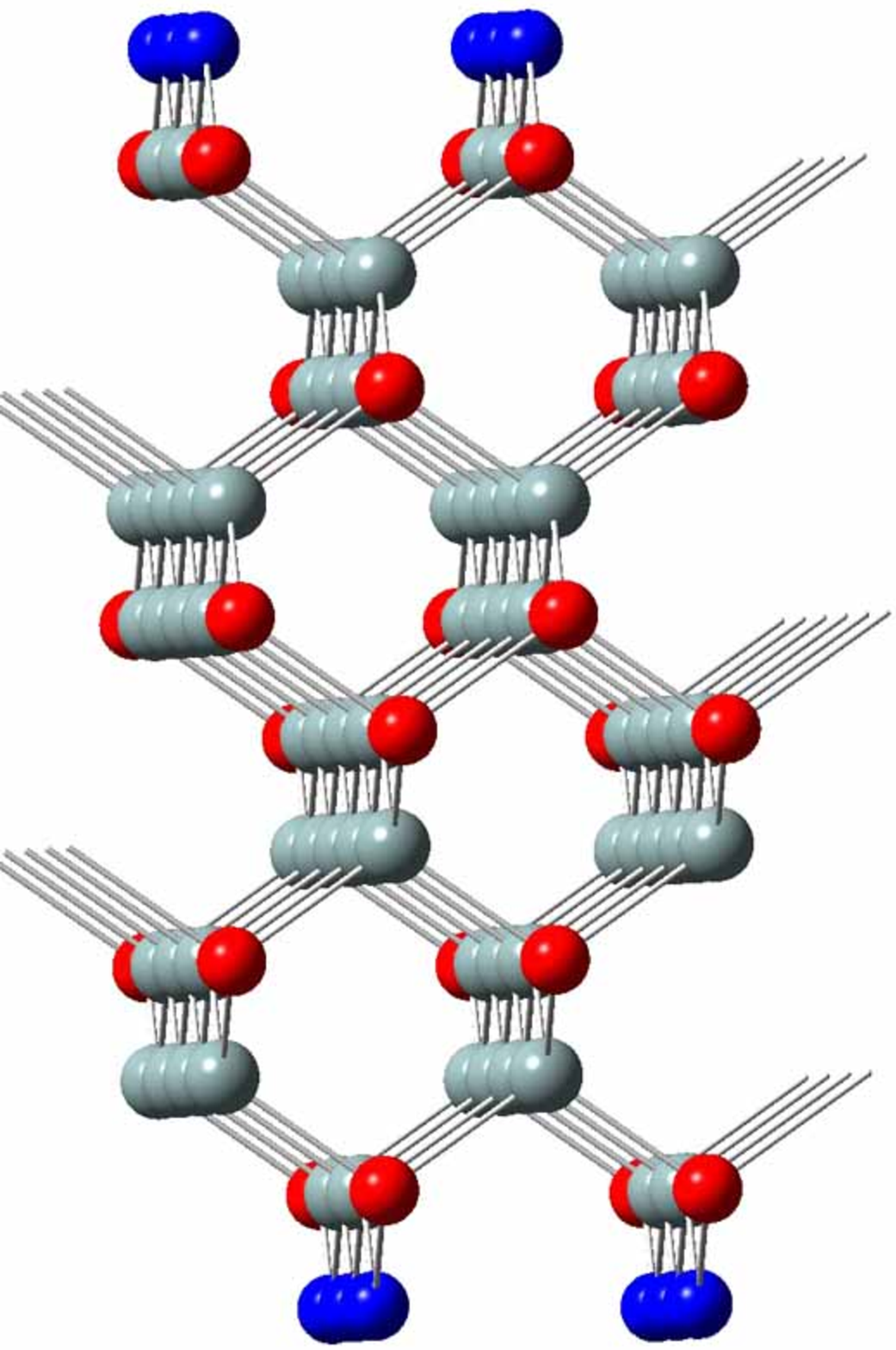}\hfill
\\[-0.1cm]
\begin{picture}(0,0)
\hspace{-2.45cm}{\bf(d)}
\end{picture}
\end{center}
\caption{Definition of characteristic lengths for the zb-/diamond-lattice NWires growing along [110] axis with hexagonal cross section, two \{001\} and four \{111\} interfaces which are shown by translucent black lines. Top and side view of the first three members ($i=1$ to 3) of the \emph{even} series: X$_{12}$ (b), X$_{48}$ (c), X$_{108}$ (d). For atom colours see Figure \ref{fig03}.}
\label{fig04}
\end{figure}

The remaining four NWire types to be investigated have all a hexagonal cross section which has a more complex geometry. Owing to the fact that each hexagonal cross section type has two different configurations -- namely a different configuration at the centerpoint -- we present an \emph{even} and an \emph{odd} series per NWire type. We will see in section \ref{Example_Si-Ox-Etch} that there is a real need for such detailed description when evaluating HR-TEM images.
\begin{eqnarray}\label{eqn-14}
N_{\rm{Wire,even}}^{110-\hexagon}[i]&=&12i^2
\end{eqnarray}
\begin{eqnarray}\label{eqn-15}
N_{\rm{bnd,even}}^{110-\hexagon}[i]&=&8i\left(3i-1\right)
\end{eqnarray}
\begin{eqnarray}\label{eqn-16}
N_{\rm{IF,even}}^{110-\hexagon}[i]&=&16i
\end{eqnarray}
\begin{eqnarray}\label{eqn-17}
\frac{N_{\rm{111-IF,even}}^{110-\hexagon}[i]}{N_{\rm{001-IF,even}}^{110-\hexagon}[i]}&=&1\ \ \forall i
\end{eqnarray}
From Equations \ref{eqn-16} and \ref{eqn-17}, it follows straight that $N_{\rm{111-IF,even}}^{110-\hexagon}[i]=N_{\rm{001-IF,even}}^{110-\hexagon}[i]=16i$.
\begin{eqnarray}\label{eqn-18}
d_{\rm{001-IF,even}}^{110-\hexagon}[i]&=&
a_{\rm{uc}}\sqrt{\frac{1}{8}}\,\big(2i-1\big)
\end{eqnarray}
\begin{eqnarray}\label{eqn-19}
d_{\rm{111-IF,even}}^{110-\hexagon}[i]&=&
a_{\rm{uc}}\sqrt{\frac{3}{8}}\left(i-\frac{1}{4}\right)
\end{eqnarray}
\begin{eqnarray}\label{eqn-20}
w_{\rm{even}}^{110-\hexagon}[i]&=&
a_{\rm{uc}}\sqrt{2}\left(i-\frac{3}{8}\right)
\end{eqnarray}
\begin{eqnarray}\label{eqn-21}
h_{\rm{even}}^{110-\hexagon}[i]&=&
a_{\rm{uc}}\left(i-\frac{1}{4}\right)
\end{eqnarray}
\begin{eqnarray}\label{eqn-22}
A^{110-\hexagon}_{\rm{even}}[i]&=&\big(a_{\rm{uc}}\big)^2\sqrt{\frac{1}{8}} \left(3i-\frac{5}{4}\right)\left(i-\frac{1}{4}\right)
\end{eqnarray}
The cross section of this \emph{even} NWire type is shown in Figure \ref{fig04} for $i=1$ to 3 together with the definition of characteristic lengths.

\begin{figure}[h!]
\begin{center}
\includegraphics[totalheight=0.0908\textheight]{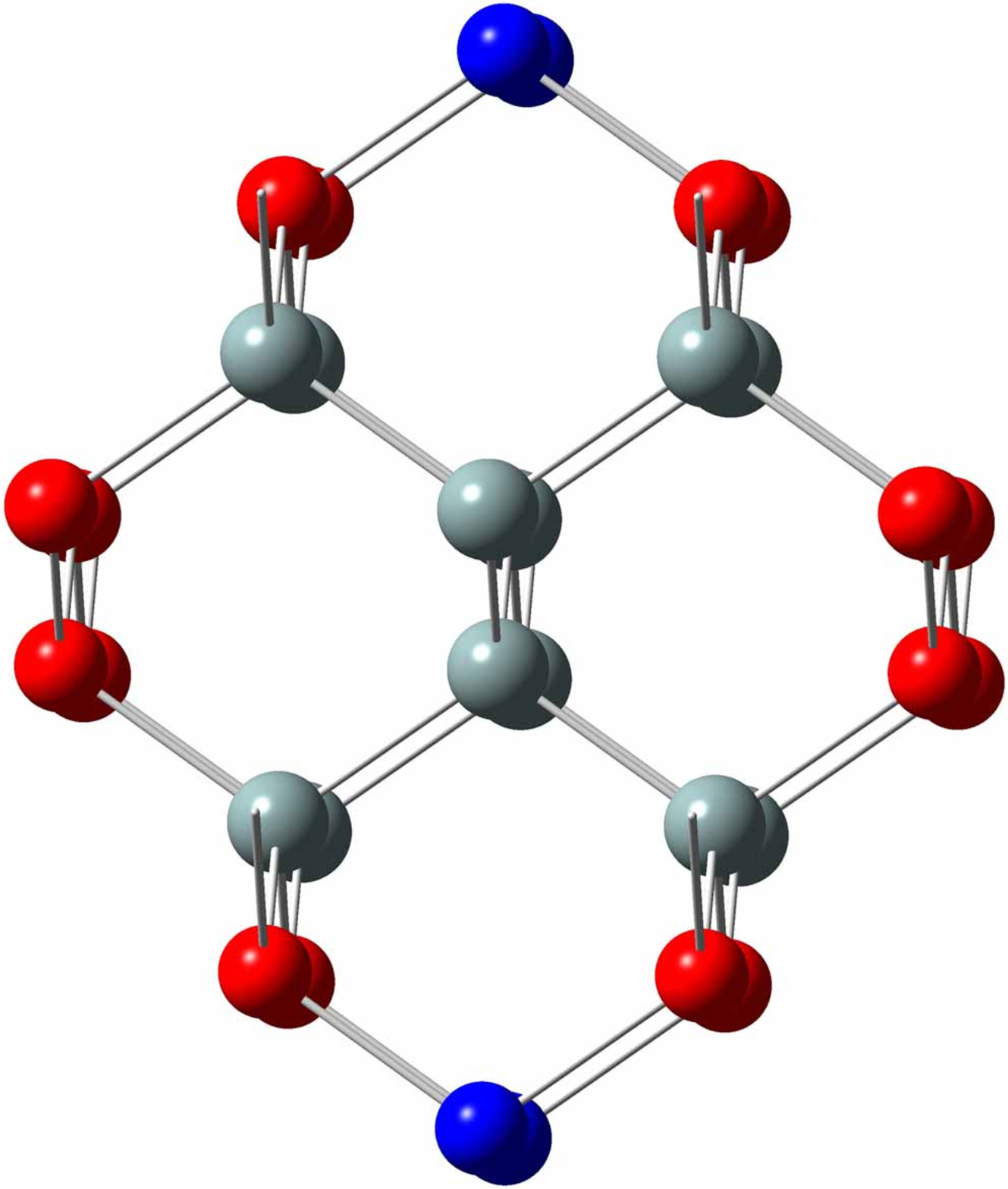}
\includegraphics[totalheight=0.0908\textheight]{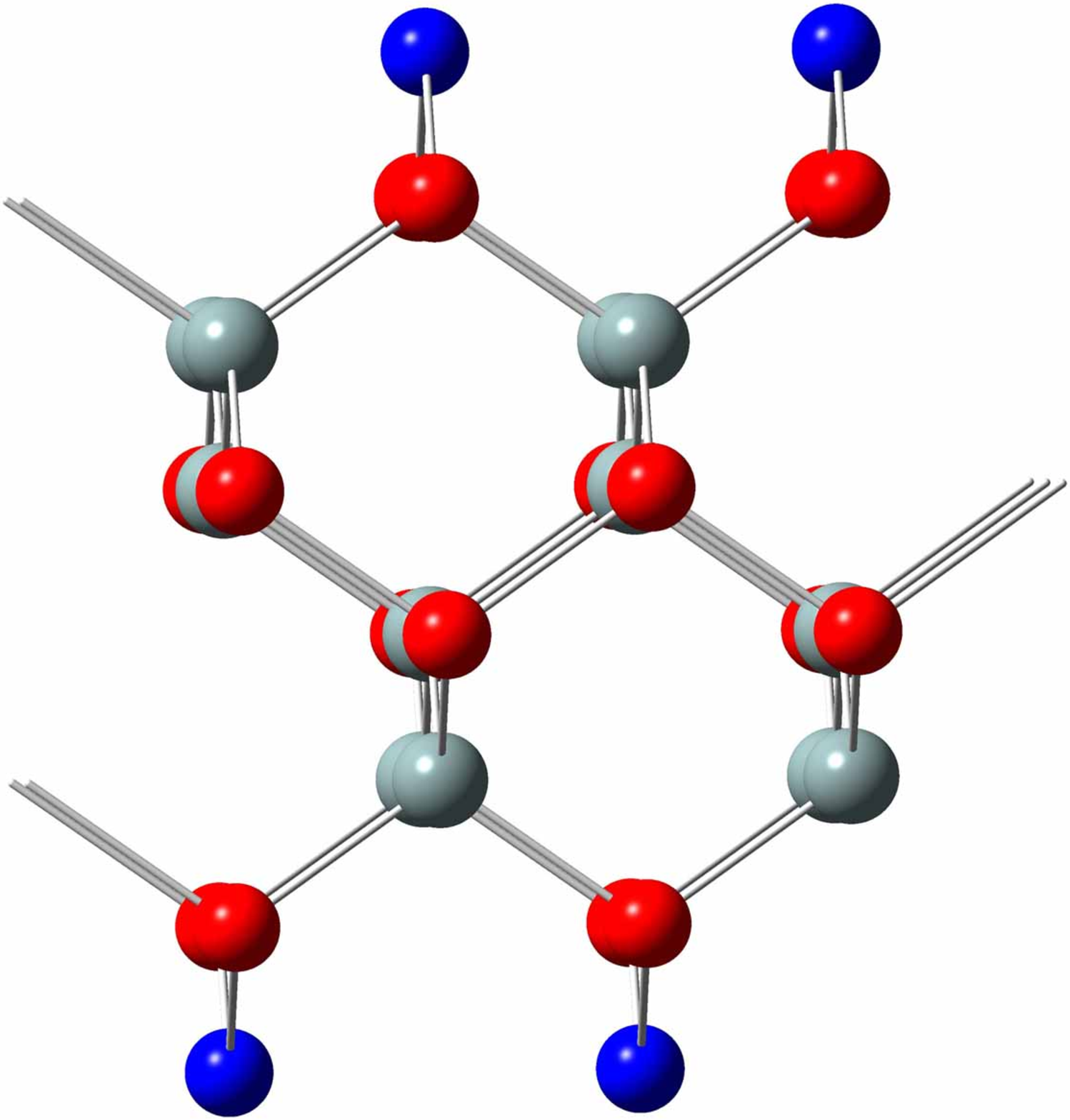}\hfill
\includegraphics[totalheight=0.1014\textheight]{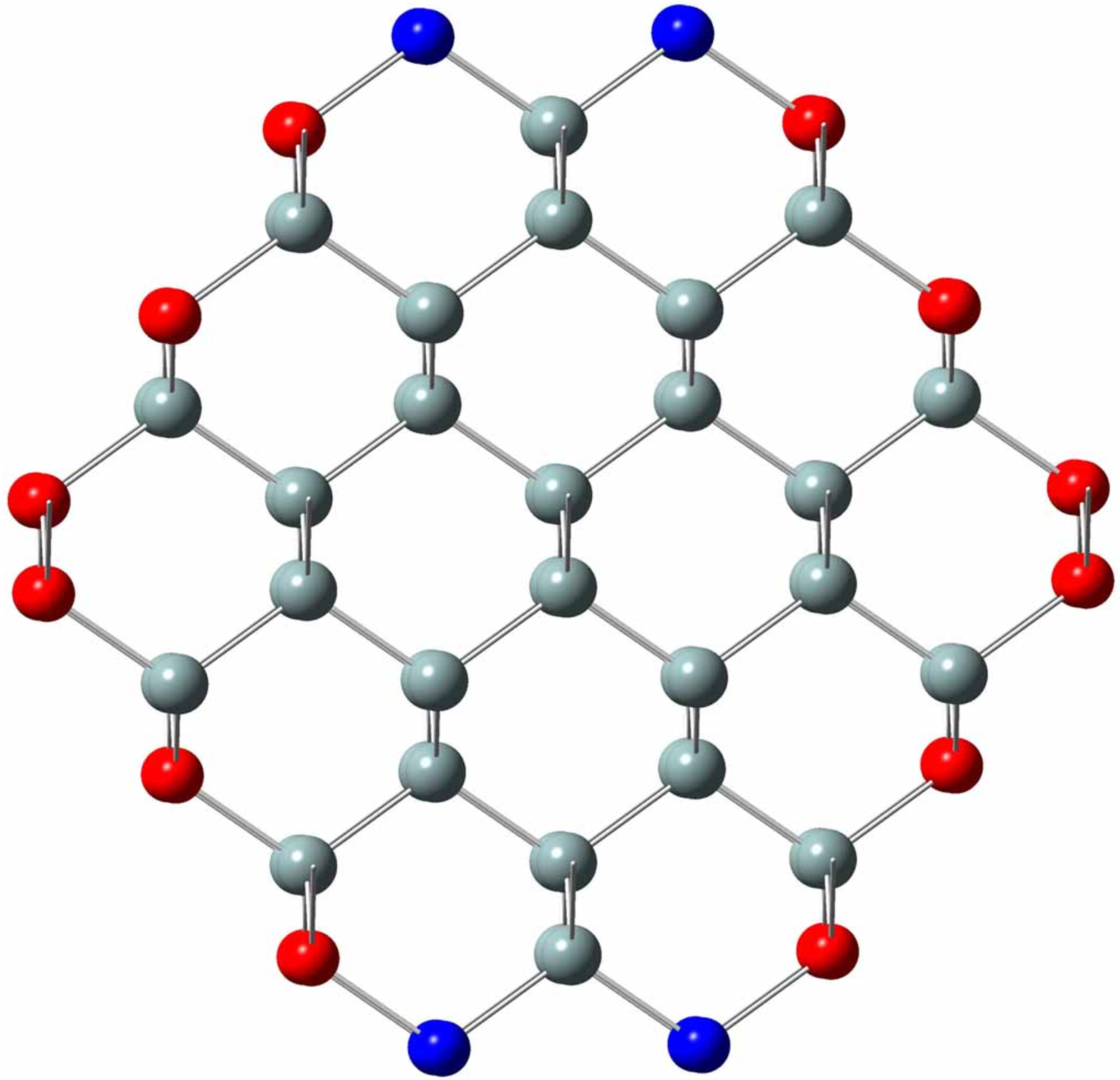}
\includegraphics[totalheight=0.1014\textheight]{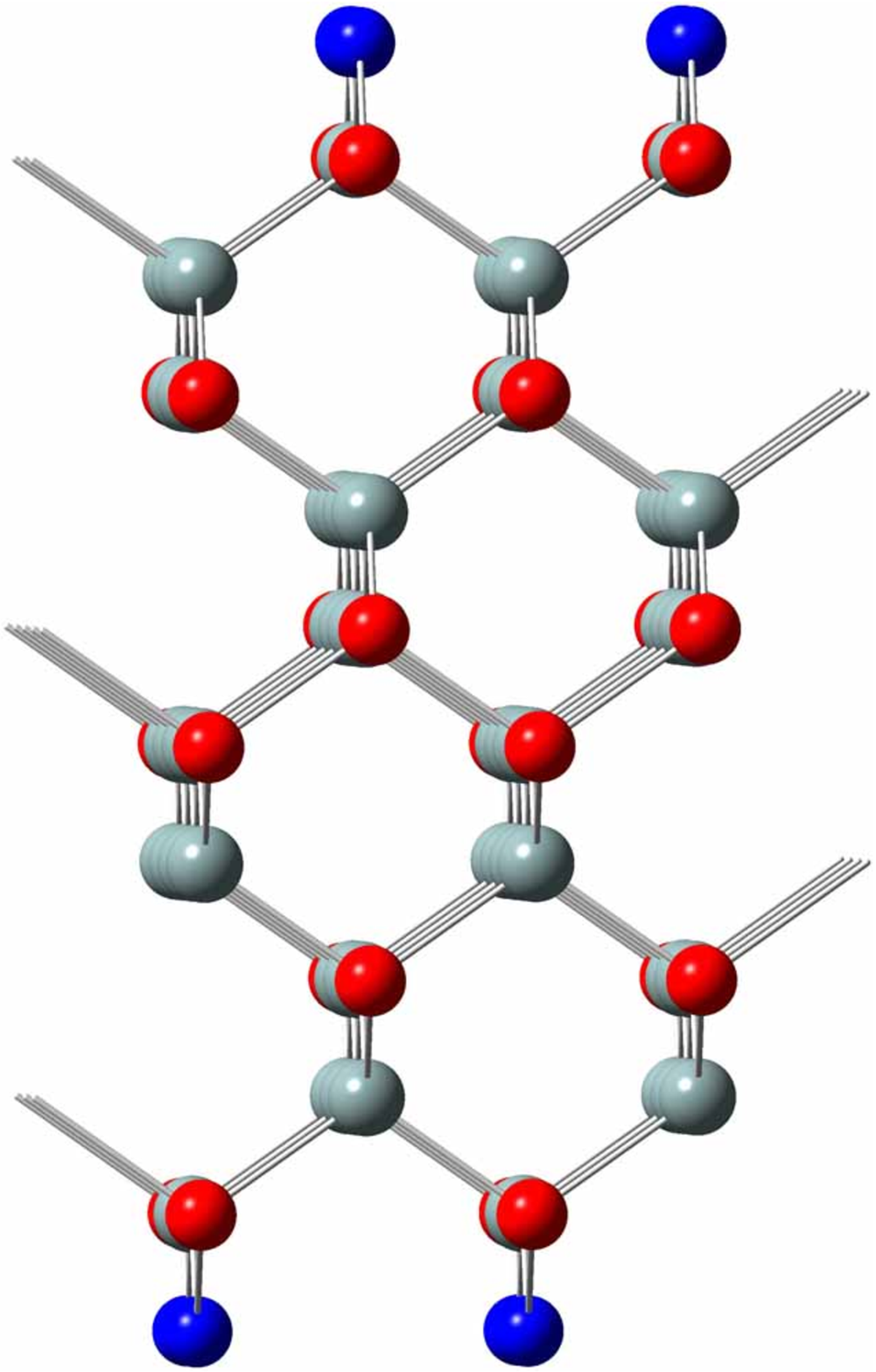}\\[-0.1cm]
\begin{picture}(0,0)
\hspace{-2.45cm}{\bf(a)}\hspace{3.91cm}{\bf (b)}
\end{picture}\\[0.2cm]
\includegraphics[totalheight=0.1052\textheight]{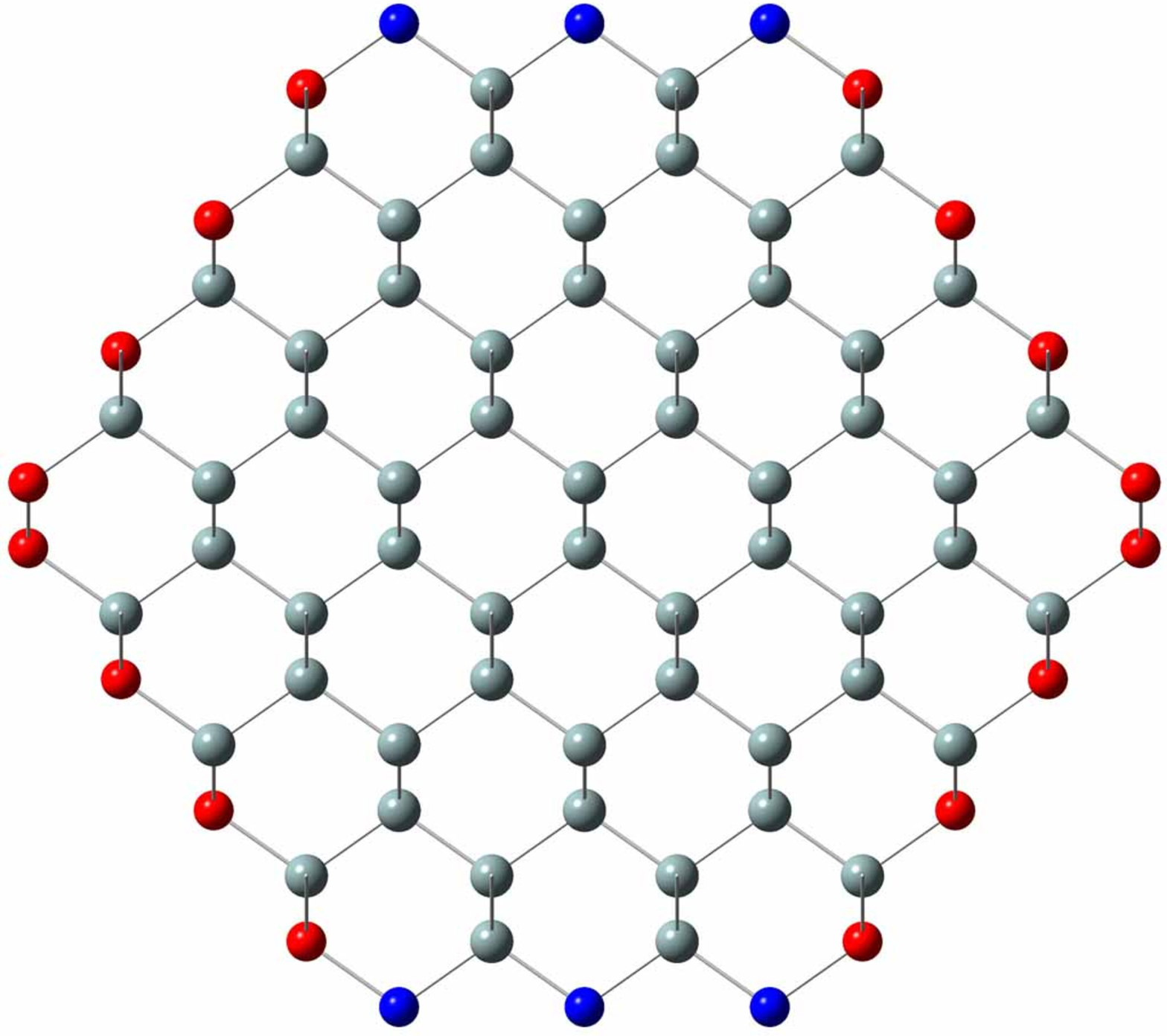}
\includegraphics[totalheight=0.1052\textheight]{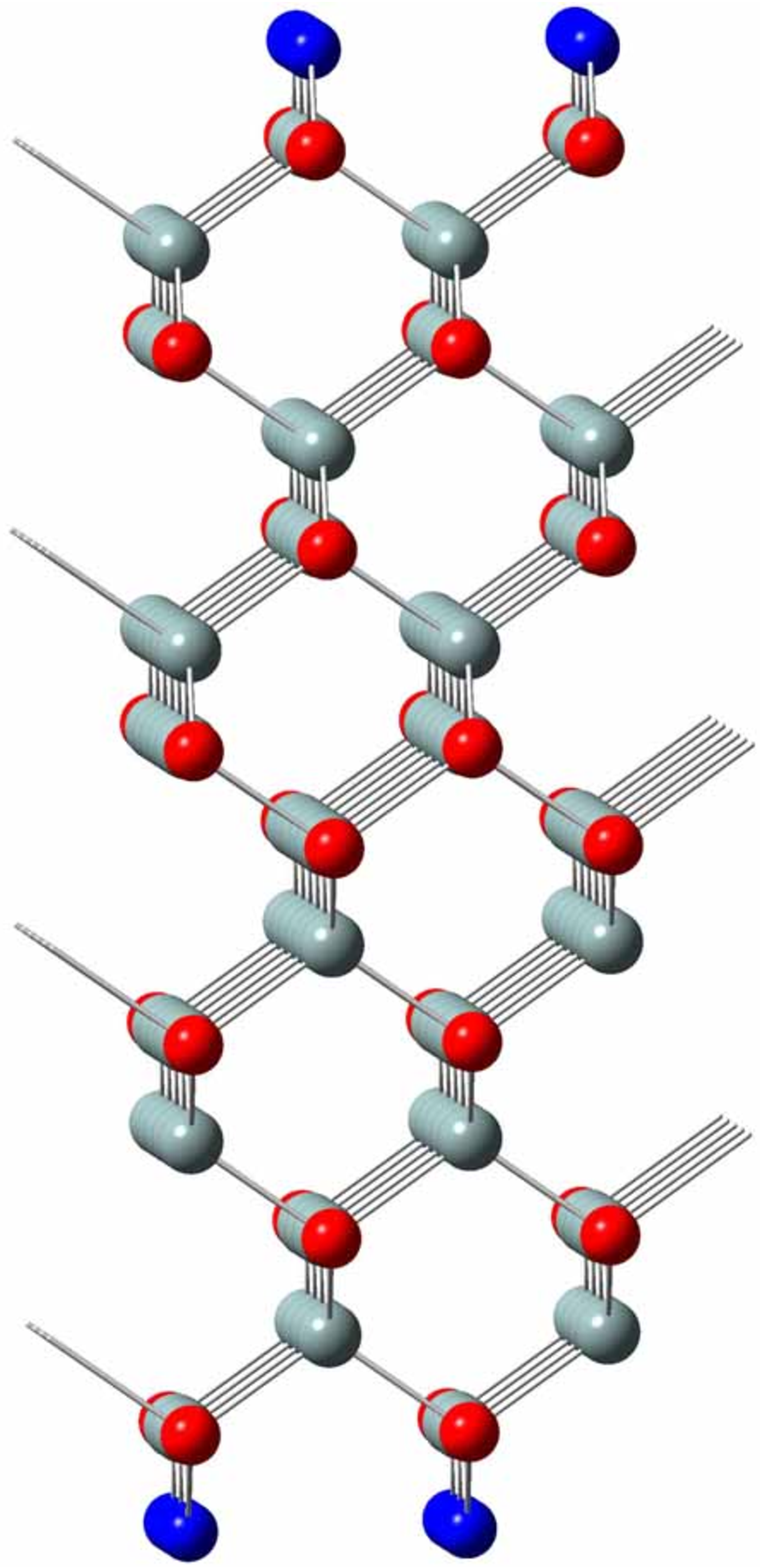}\hfill
\\[-0.1cm]
\begin{picture}(0,0)
\hspace{-2.45cm}{\bf(c)}
\end{picture}
\end{center}
\caption{Cross section and side view of zb-/diamond-lattice NWires growing along [110] axis with hexagonal cross section, two \{001\} and four \{111\} interfaces, \emph{odd} series, for run index $i=1$ to 3: X$_{32}$ (a), X$_{84}$ (b), X$_{160}$ (c). For atom colours and definition of characteristic lengths, see Figure \ref{fig03} and Figure \ref{fig04}, respectively.}
\label{fig05}
\end{figure}

For the odd series we get
\begin{eqnarray}\label{eqn-23}
N_{\rm{Wire,odd}}^{110-\hexagon}[i]&=&4\big(i[3i+4]+1\big)
\end{eqnarray}
\begin{eqnarray}\label{eqn-24}
N_{\rm{bnd,odd}}^{110-\hexagon}[i]&=&4\left(6i[i+1]+1\right)
\end{eqnarray}
\begin{eqnarray}\label{eqn-25}
N_{\rm{IF,odd}}^{110-\hexagon}[i]&=&8(2i+1)
\end{eqnarray}
whereby
\begin{eqnarray}\label{eqn-26}
\frac{N_{\rm{111-IF,odd}}^{110-\hexagon}[i]}{N_{\rm{001-IF,odd}}^{110-\hexagon}[i]}&=&\frac{8(i+1)}{8i}\ =\ 1+\frac{1}{i}
\end{eqnarray}
Again, the center expression shows both number series in their explicit form, while the right side expression presents the simplied result of their ratio.
\begin{eqnarray}\label{eqn-27}
d_{\rm{001-IF,odd}}^{110-\hexagon}[i]&=
&a_{\rm{uc}}\sqrt{\frac{1}{8}}\,\big(2i-1\big)
\end{eqnarray}
Comparing this result with equation \ref{eqn-18}, we see that  $d_{\rm{001-IF,odd}}^{110-\hexagon}[i]=d_{\rm{001-IF,even}}^{110-\hexagon}[i]$. 
\begin{eqnarray}\label{eqn-28}
d_{\rm{111-IF,odd}}^{110-\hexagon}[i]&=
&a_{\rm{uc}}\sqrt{\frac{3}{8}}\left(i+\frac{3}{4}\right)
\end{eqnarray}
Comparing this result with equation \ref{eqn-19}, we see that  $d_{\rm{111-IF,odd}}^{110-\hexagon}[i]=d_{\rm{111-IF,even}}^{110-\hexagon}[i+1]$.
\begin{eqnarray}\label{eqn-29}
w_{\rm{odd}}^{110-\hexagon}[i]&=&a_{\rm{uc}}\sqrt{2}\left(i+\frac{1}{8}\right)
\end{eqnarray}
The relation to the \emph{even} series is $w_{\rm{odd}}^{110-\hexagon}[i]=w_{\rm{111-IF,even}}^{110-\hexagon}[i+\frac{1}{2}]$.
\begin{eqnarray}\label{eqn-30}
h_{\rm{odd}}^{110-\hexagon}[i]&=&a_{\rm{uc}}\left(i+\frac{3}{4}\right)
\end{eqnarray}
This result differs to the \emph{even} series by one $a_{\rm{uc}}$;  $h_{\rm{odd}}^{110-\hexagon}[i]=h_{\rm{111-IF,even}}^{110-\hexagon}[i+1]$.
\begin{eqnarray}\label{eqn-31}
A^{110-\hexagon}_{\rm{odd}}[i]&=&\big(a_{\rm{uc}}\big)^2\sqrt{\frac{1}{8}} \left(3i-\frac{1}{4}\right)\left(i+\frac{3}{4}\right)
\end{eqnarray}
The cross section of this \emph{odd} NWire type is shown in Figure \ref{fig05} for $i=1$ to 3.

\subsection{\label{AnaNomiHexa11-2} NWires growing along [11$\mathbf{\bar{2}}$] Direction with Hexagonal Cross Section and Four \{1$\mathbf{\bar{3}}$1\} Plus Two \{111\} Interfaces}
\begin{figure}[h!]
\begin{center}
\includegraphics[totalheight=0.213\textheight]{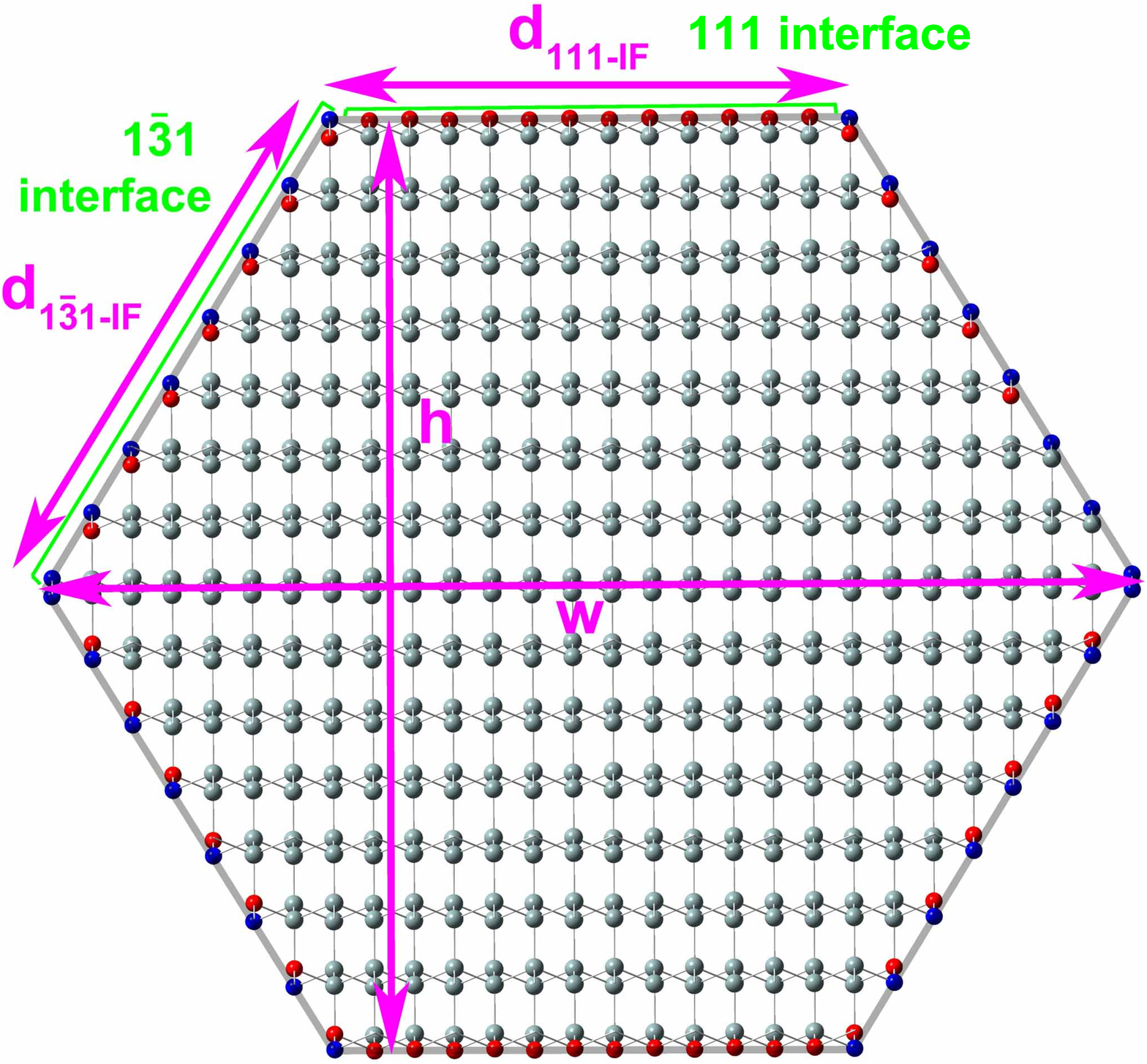}\\[-0.1cm]
\begin{picture}(0,0)
{\bf(a)}
\end{picture}\\[0.2cm]
\includegraphics[totalheight=0.087\textheight]{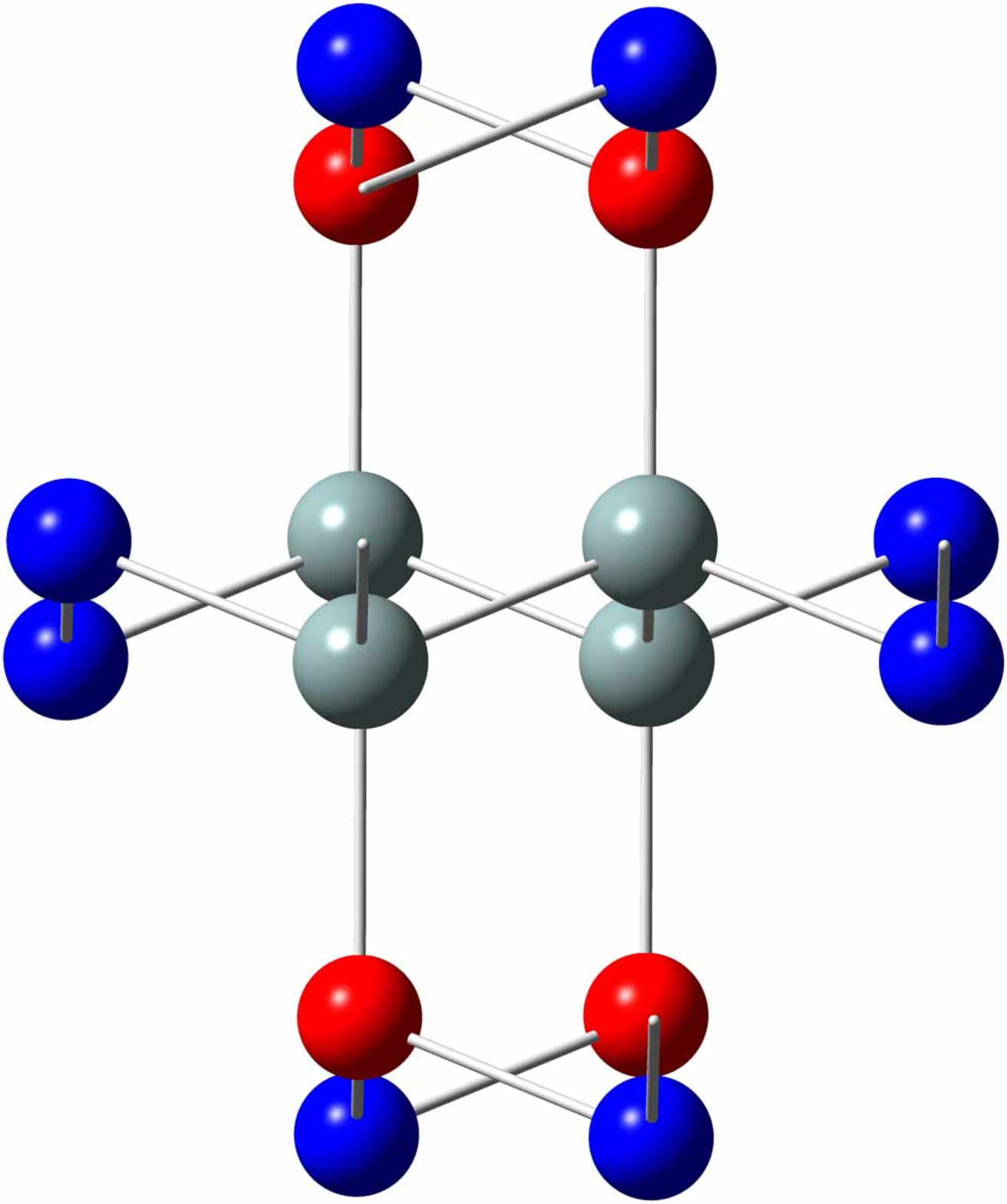}
\includegraphics[totalheight=0.087\textheight]{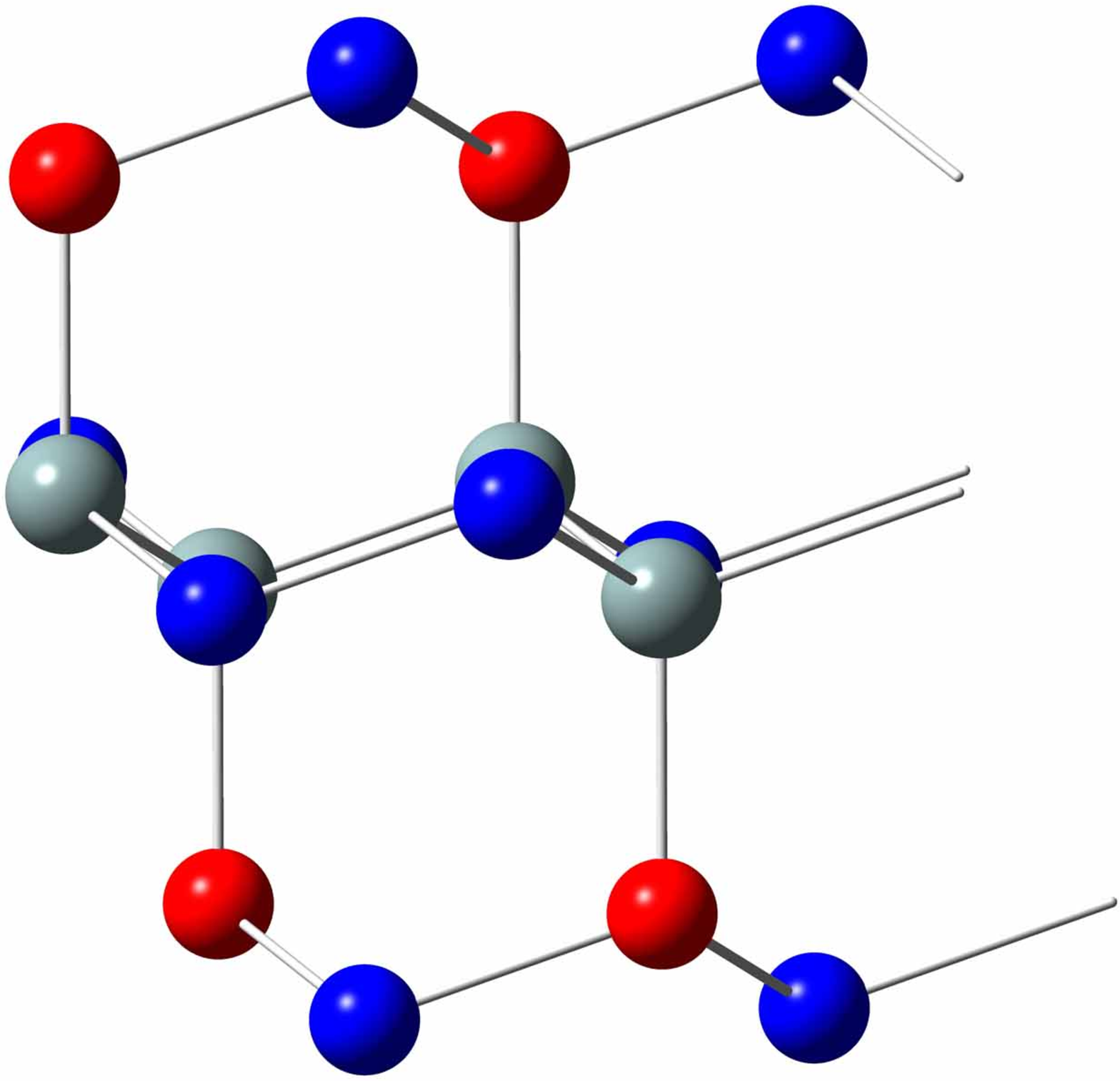}\hfill
\includegraphics[totalheight=0.106\textheight]{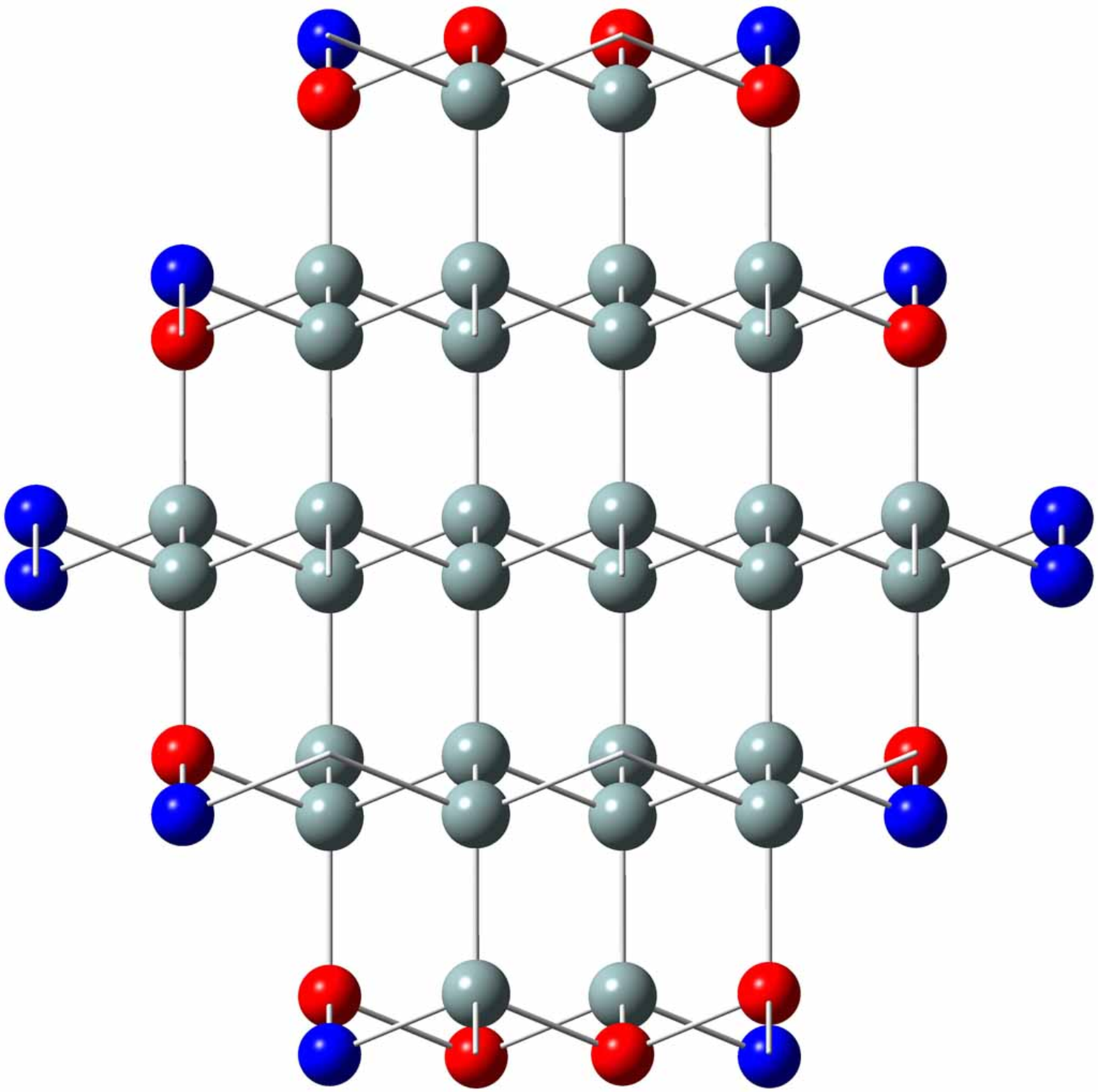}
\includegraphics[totalheight=0.106\textheight]{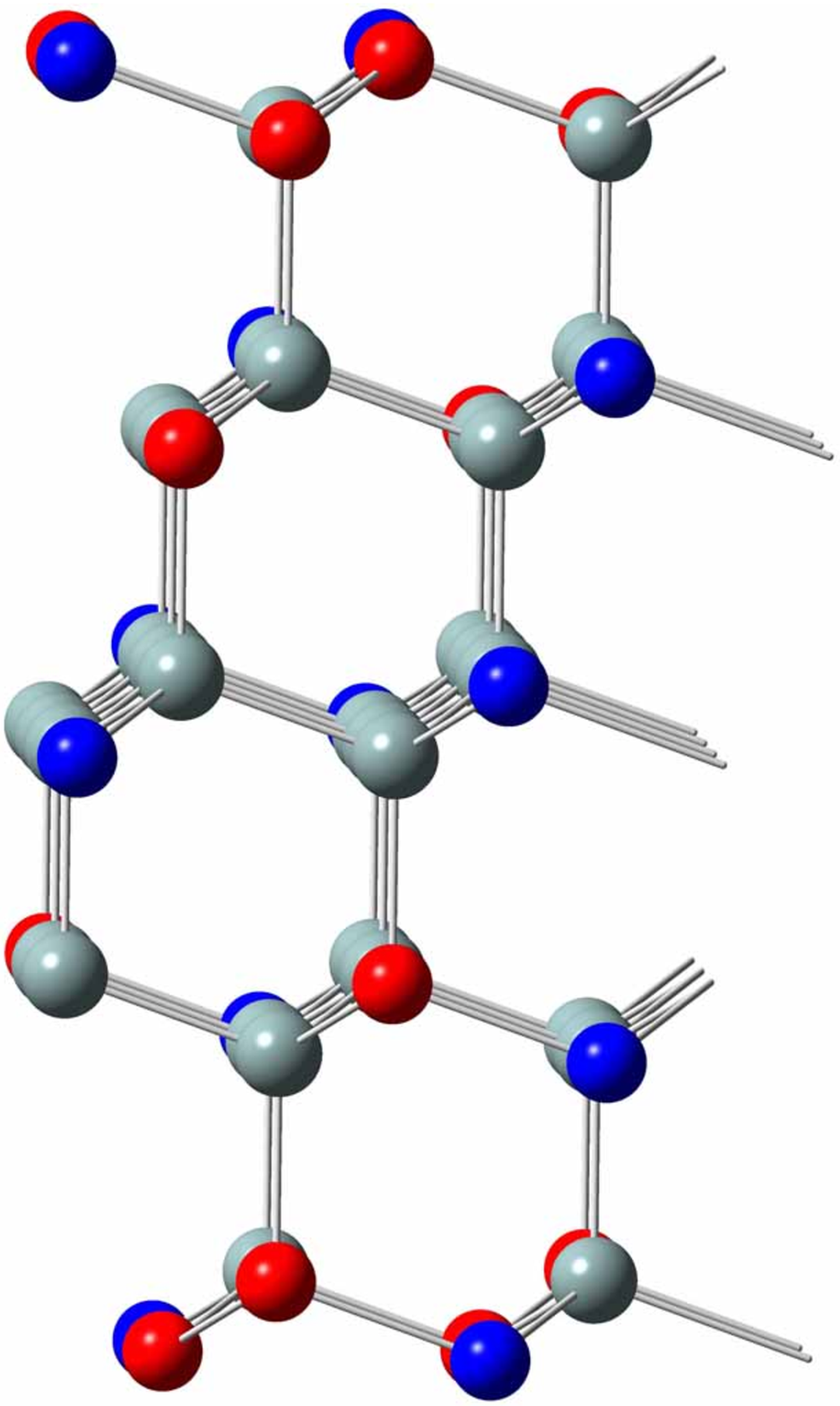}\\[-0.1cm]
\begin{picture}(0,0)
\hspace{-2.45cm}{\bf(b)}\hspace{3.91cm}{\bf (c)}
\end{picture}\\[0.2cm]
\includegraphics[totalheight=0.1136\textheight]{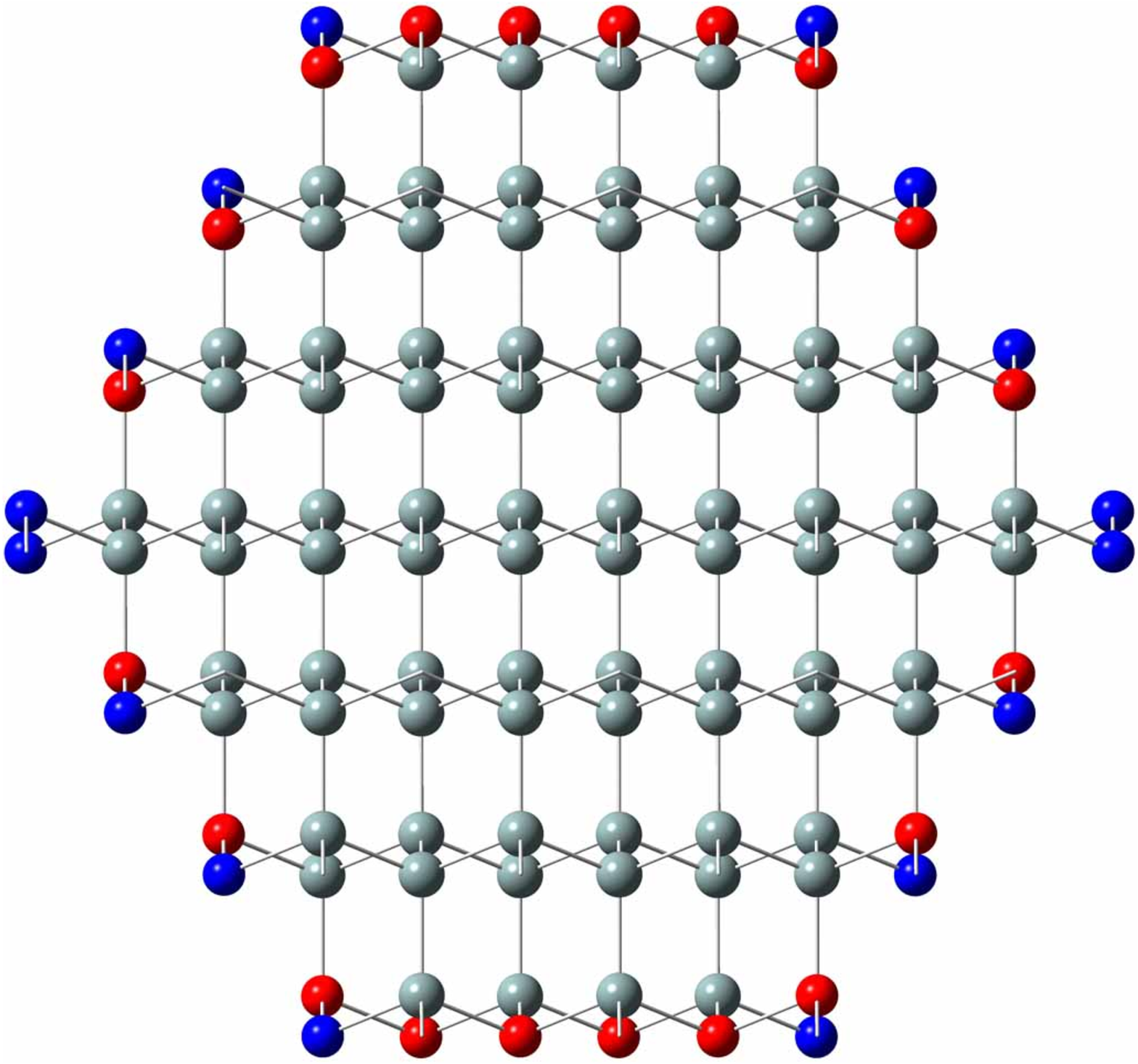}
\includegraphics[totalheight=0.1136\textheight]{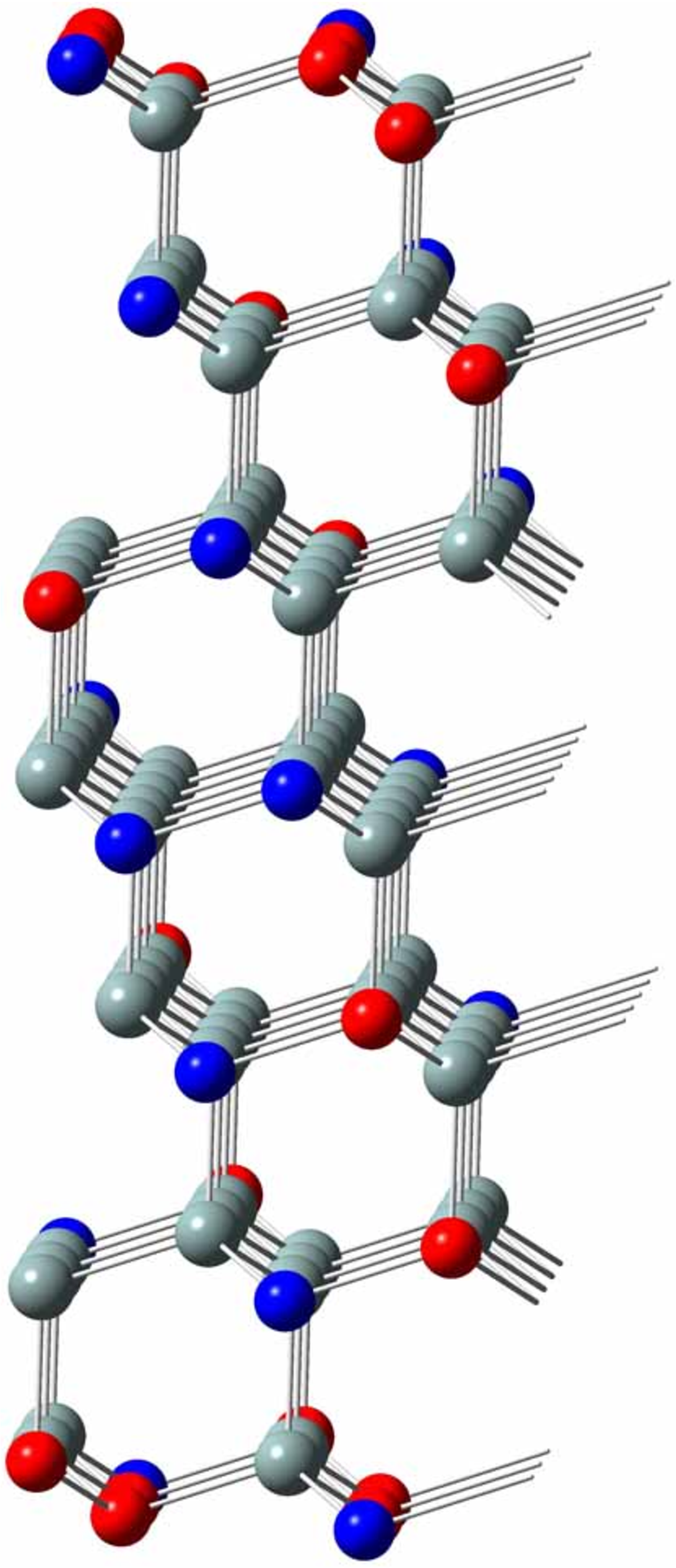}\hfill
\\[-0.1cm]
\begin{picture}(0,0)
\hspace{-2.45cm}{\bf(d)}
\end{picture}
\end{center}
\caption{\label{fig06}Definition of characteristic lengths for the zb-/diamond- lattice NWires with [11$\bar{2}$] growth axis, hexagonal cross section and four \{11$\bar{3}$\} plus two \{111\} interfaces shown by translucent black lines. Green markings show the assignment of interface atoms (red, blue) to the respective interface orientation in accord with number series for interface bonds $N_{\rm{IF}}^{11\bar{2}-\hexagon}[i]$. Top and side view of the first three members ($i=1$ to 3) of the \emph{even} series: X$_{16}$ (b), X$_{56}$ (c), X$_{120}$ (d). For atom colours see Figure \ref{fig03}.}
\end{figure}

\begin{eqnarray}\label{eqn-32}
N_{\rm{Wire,even}}^{11\bar{2}-\hexagon}[i]&=&4i(3i+1)
\end{eqnarray}
\begin{eqnarray}\label{eqn-33}
N_{\rm{bnd,even}}^{11\bar{2}-\hexagon}[i]&=&24i^2-2
\end{eqnarray}
\begin{eqnarray}\label{eqn-34}
N_{\rm{IF,even}}^{11\bar{2}-\hexagon}[i]&=&16\left(i+\frac{1}{4}\right)
\end{eqnarray}
The assignment of interface bonds to the \{111\} and \{1$\bar{3}$1\} planes and of characteristic lengths are shown in Fig. \ref{fig06}a, the cross section for the \emph{even} NWire for is shown in Figure \ref{fig06}b to d for $i=1$ to 3. With this assignment of interface atoms to \{111\}- and \{1$\bar{3}$1\}-planes we obtain
\begin{eqnarray}\label{eqn-35}
\frac{N_{\rm{1\bar{3}1-IF,even}}^{11\bar{2}-\hexagon}[i]}{N_{\rm{111-IF,even}}^{11\bar{2}-\hexagon}[i]}&=&\frac{4(3i+2)}{4(i-1)}\ =\ \frac{3i+2}{i-1}
\end{eqnarray}
As before, the center expression shows both number series in their explicit form, while the right side expression presents the simplied result of their ratio.
\begin{eqnarray}\label{eqn-37}
d_{\rm{111-IF,even}}^{11\bar{2}-\hexagon}[i]&=&a_{\rm{uc}}\sqrt{\frac{1}{8}}\,(2i-1)
\end{eqnarray}
\begin{eqnarray}\label{eqn-38}
d_{\rm{1\bar{3}1-IF,even}}^{11\bar{2}-\hexagon}[i]&=&a_{\rm{uc}}\sqrt{\frac{11}{24}}\,i
\end{eqnarray}
\begin{eqnarray}\label{eqn-39}
w_{\rm{even}}^{11\bar{2}-\hexagon}[i]&=&a_{\rm{uc}}\sqrt{\frac{1}{8}}(4i-1)
\end{eqnarray}
\begin{eqnarray}\label{eqn-40}
h_{\rm{even}}^{11\bar{2}-\hexagon}[i]&=&a_{\rm{uc}}\sqrt{\frac{4}{3}}\,i
\end{eqnarray}
\begin{eqnarray}\label{eqn-41}
A^{11\bar{2}-\hexagon}_{\rm{even}}[i]&=&\big(a_{\rm{uc}}\big)^2\sqrt{\frac{1}{6}}\,i(3i-1)
\end{eqnarray}

\begin{figure}[h!]
\begin{center}
\includegraphics[totalheight=0.087\textheight]{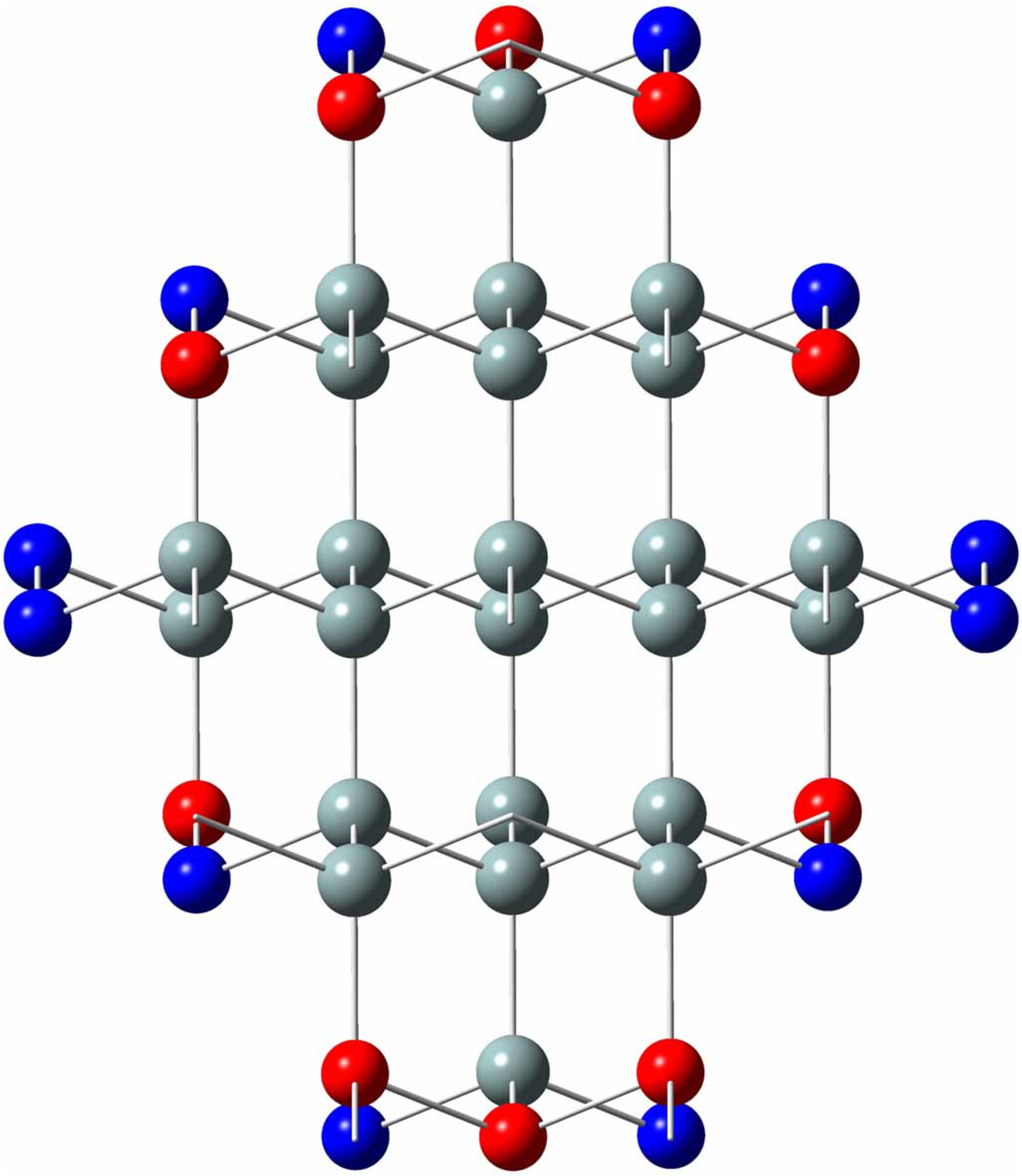}
\includegraphics[totalheight=0.087\textheight]{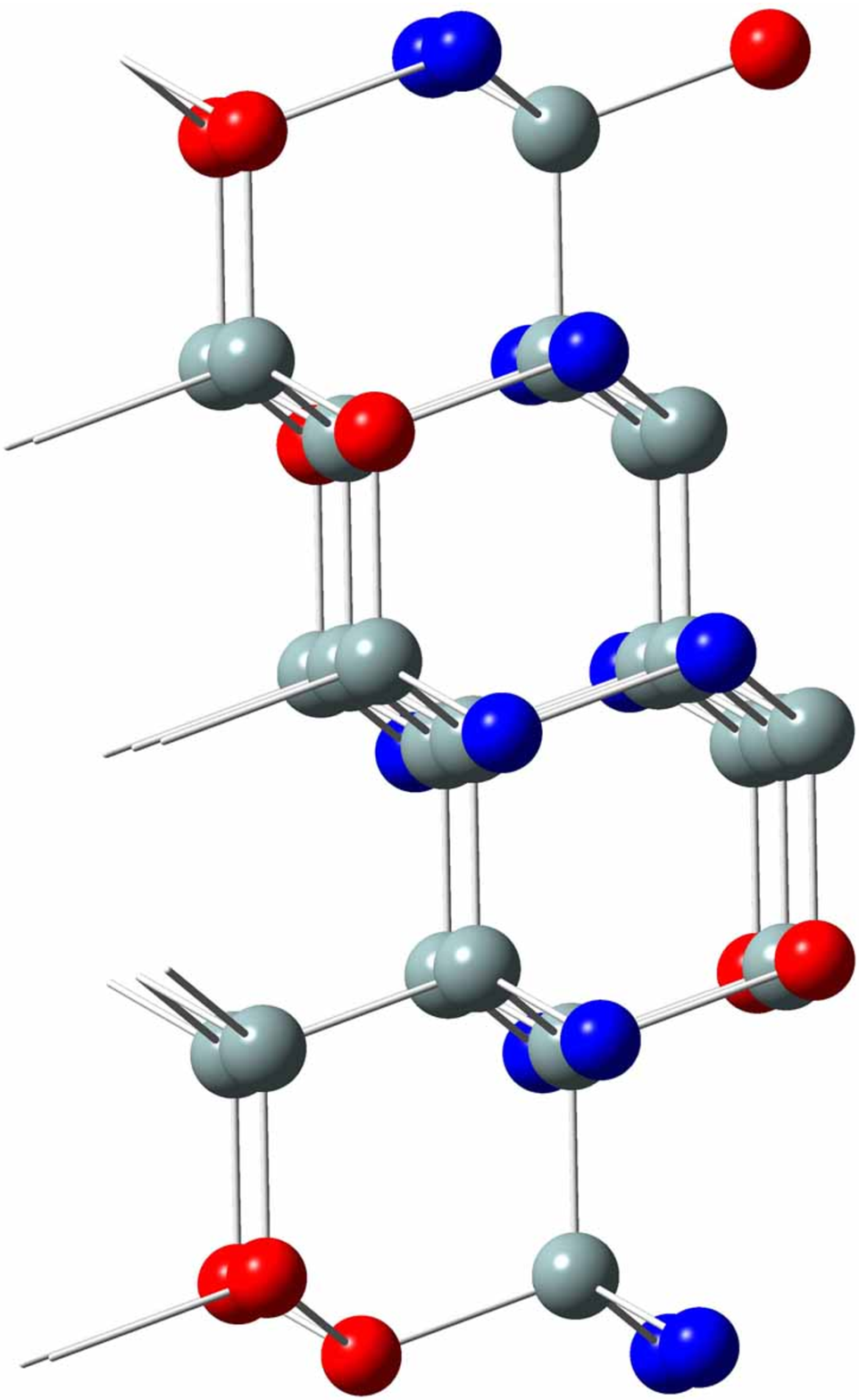}\hfill
\includegraphics[totalheight=0.1136\textheight]{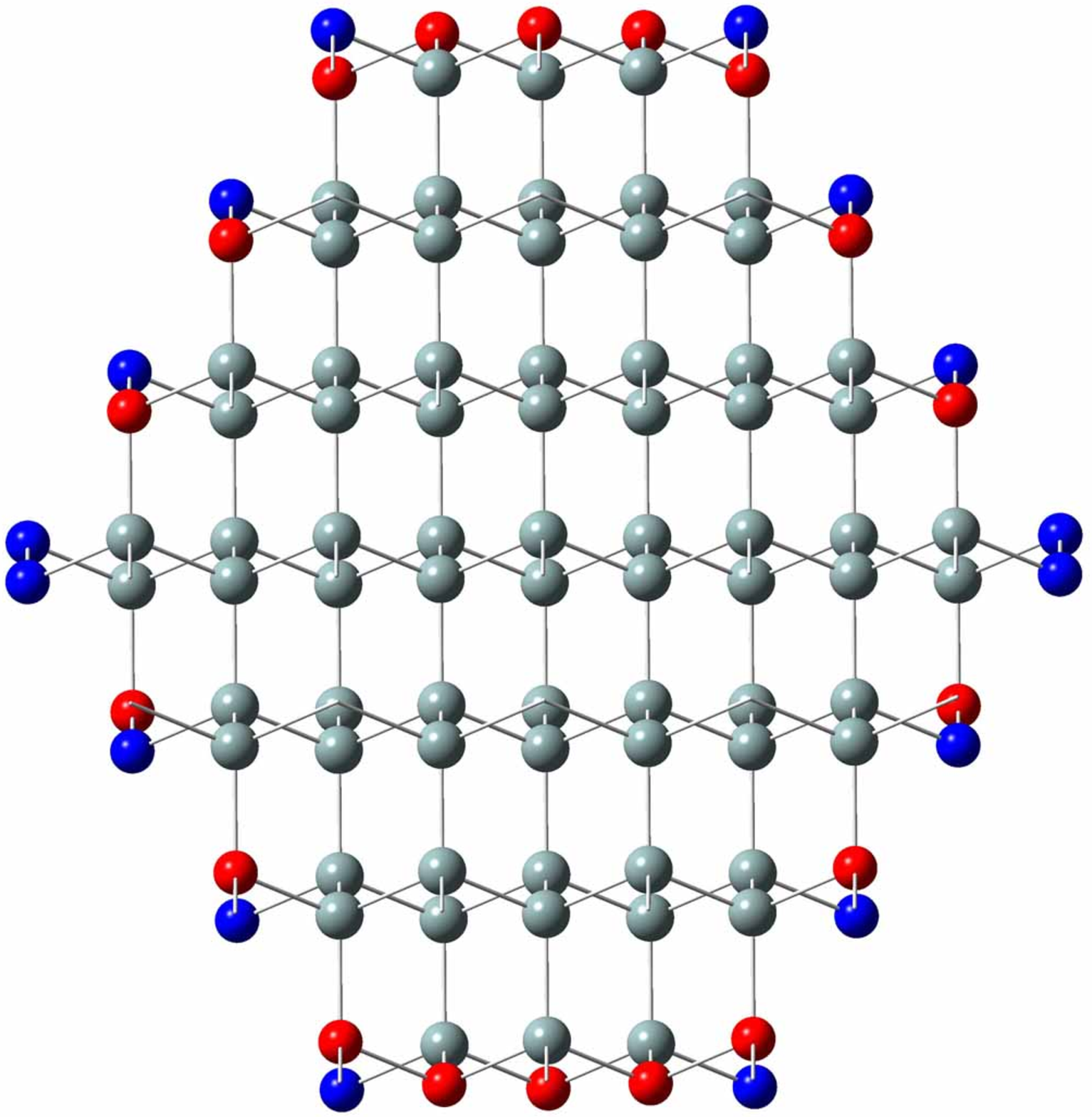}
\includegraphics[totalheight=0.1136\textheight]{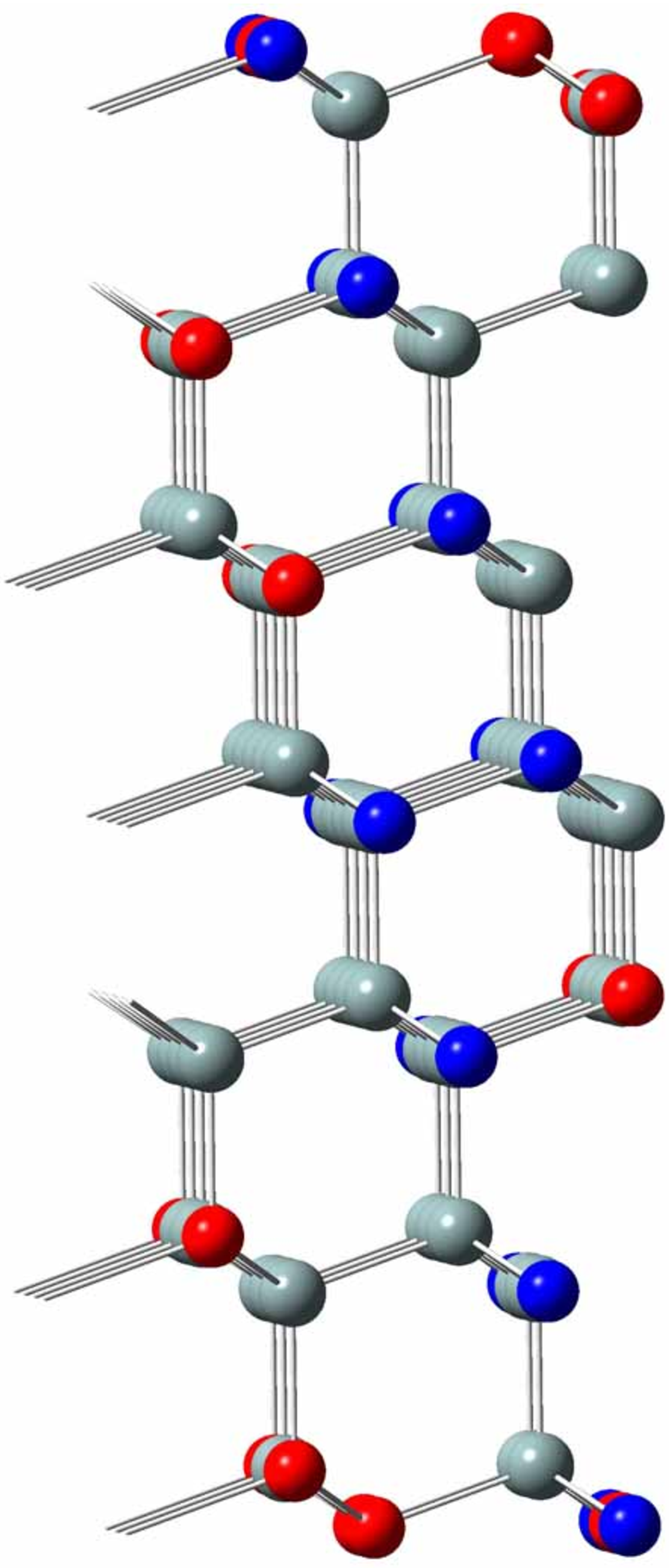}\\[-0.1cm]
\begin{picture}(0,0)
\hspace{-2.45cm}{\bf(a)}\hspace{3.91cm}{\bf (b)}
\end{picture}\\[0.2cm]
\includegraphics[totalheight=0.1204\textheight]{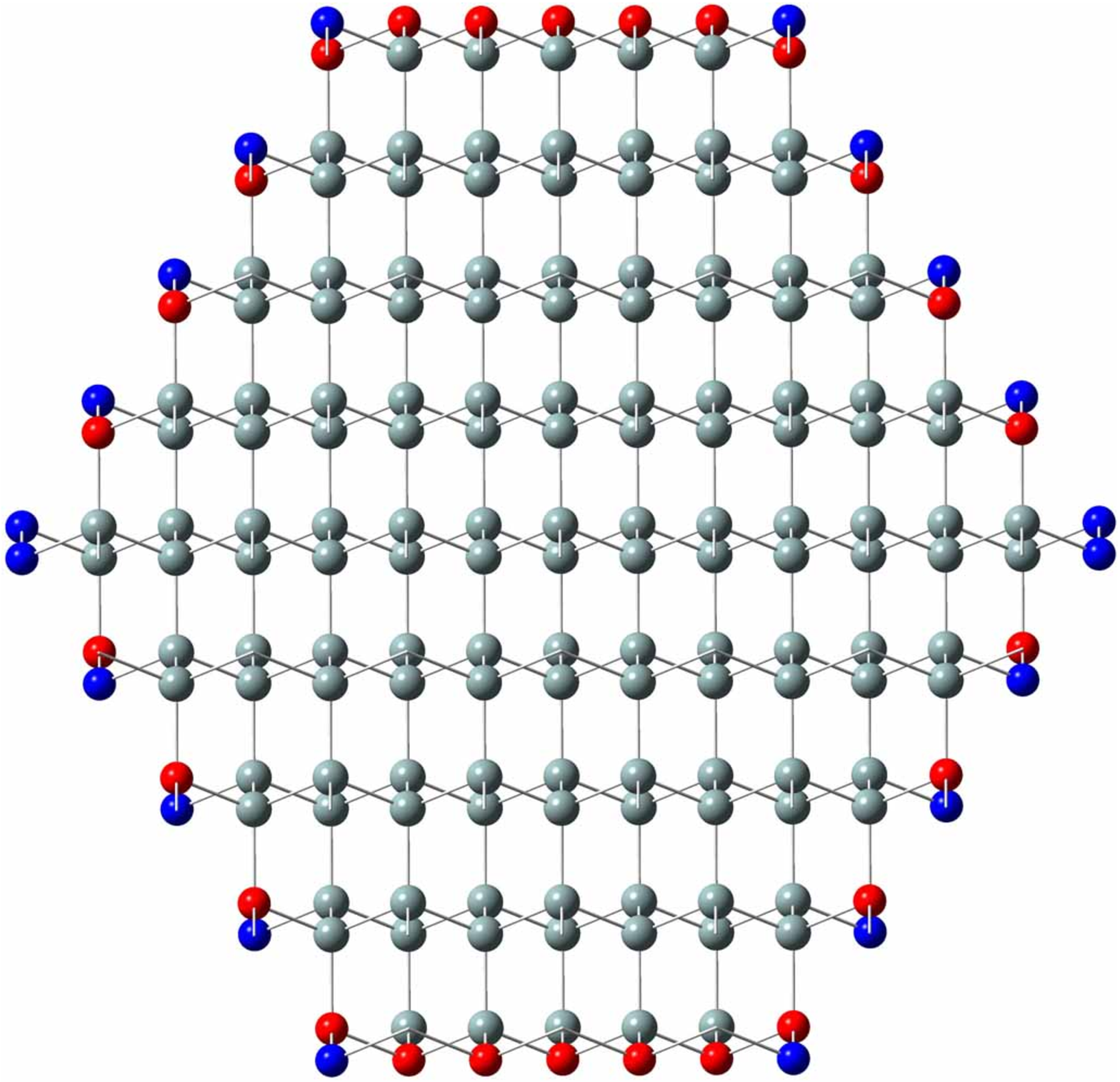}
\includegraphics[totalheight=0.1204\textheight]{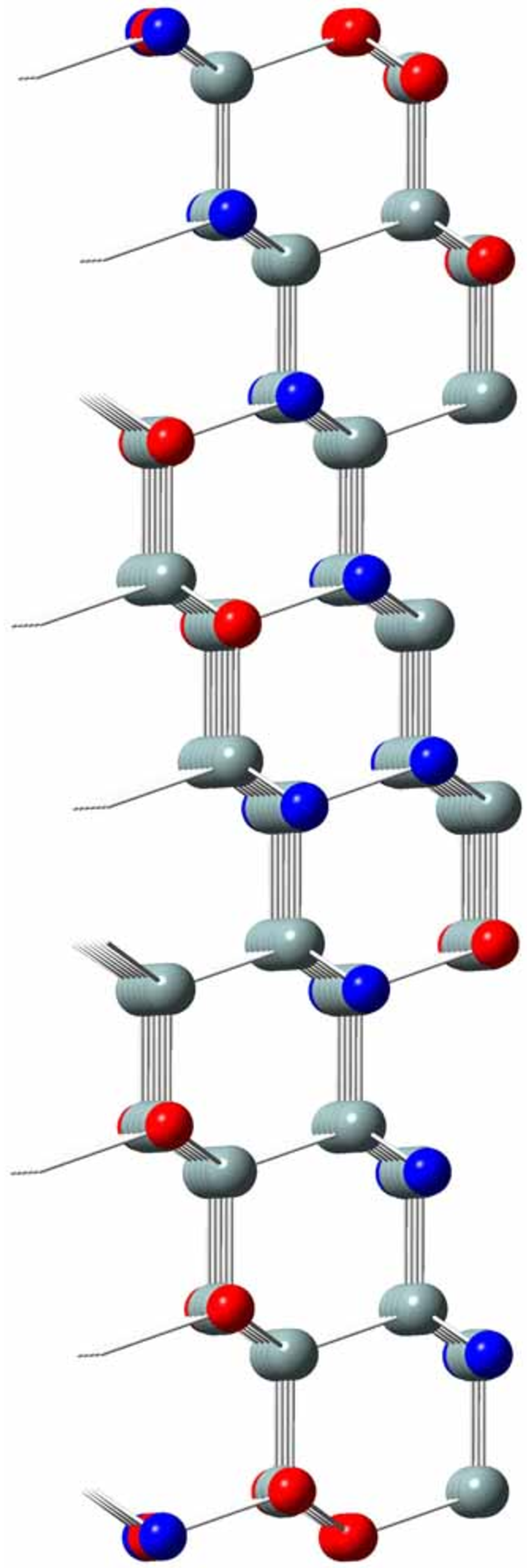}\hfill
\\[-0.1cm]
\begin{picture}(0,0)
\hspace{-2.45cm}{\bf(c)}
\end{picture}
\end{center}
\caption{\label{fig07}Cross section and side view of zb-/diamond-lattice NWires with [11$\bar{2}$] growth axis, hexagonal cross section and four \{11$\bar{3}$\} plus two \{111\} interfaces, \emph{odd} series, for run index $i=1$ to 3: X$_{46}$ (a), X$_{106}$ (b), X$_{190}$ (c). For atom colours and definition of characteristic lengths and interfaces, see Figure \ref{fig03} and Figure \ref{fig06}, respectively.}
\end{figure}

For the odd series we get
\begin{eqnarray}\label{eqn-42}
N_{\rm{Wire,odd}}^{11\bar{2}-\hexagon}[i]&=&12i(i+2)+10
\end{eqnarray}
\begin{eqnarray}\label{eqn-43}
N_{\rm{bnd,odd}}^{11\bar{2}-\hexagon}[i]&=&2(6i+1)(2i+3)+5
\end{eqnarray}
\begin{eqnarray}\label{eqn-44}
N_{\rm{IF,odd}}^{11\bar{2}-\hexagon}[i]&=&2(8i+9)
\end{eqnarray}
\begin{eqnarray}\label{eqn-45}
\frac{N_{\rm{1\bar{3}1-IF,odd}}^{11\bar{2}-\hexagon}[i]}{N_{\rm{111-IF,odd}}^{11\bar{2}-\hexagon}[i]}&=&\frac{4(3i+5)}{2(2i-1)}\ =\ \frac{2(3i+5)}{2i-1}
\end{eqnarray}
Again, the center expression shows both number series in their explicit form, while the right side expression presents the simplied result of their ratio.
\begin{eqnarray}\label{eqn-47}
d_{\rm{111-IF,odd}}^{11\bar{2}-\hexagon}[i]&=&
a_{\rm{uc}}\sqrt{\frac{1}{2}}\,i
\end{eqnarray}
\begin{eqnarray}\label{eqn-48}
d_{\rm{1\bar{3}1-IF,odd}}^{11\bar{2}-\hexagon}[i]&=&
a_{\rm{uc}}\sqrt{\frac{11}{24}}\,(i+1)
\end{eqnarray}
\begin{eqnarray}\label{eqn-49}
w_{\rm{odd}}^{11\bar{2}-\hexagon}[i]&=&a_{\rm{uc}}\sqrt{\frac{1}{2}}(2i+1)
\end{eqnarray}
\begin{eqnarray}\label{eqn-50}
h_{\rm{odd}}^{11\bar{2}-\hexagon}[i]&=&a_{\rm{uc}}\sqrt{\frac{4}{3}}(i+1)
\end{eqnarray}
\begin{eqnarray}\label{eqn-51}
A^{11\bar{2}-\hexagon}_{\rm{odd}}[i]&=&
\big(a_{\rm{uc}}\big)^2\sqrt{\frac{1}{6}}\,(3i+1)(i+1)
\end{eqnarray}
The cross section for this \emph{odd} series NWire type is shown in Figure \ref{fig07} for $i=1$ to 3.

\subsection{\label{AnaNomiHexa111-IF110} NWires growing along [111] Direction with Hexagonal Cross Section and Six \{110\} Interfaces}
\begin{figure}[h!]
\begin{center}
\includegraphics[totalheight=0.167\textheight]{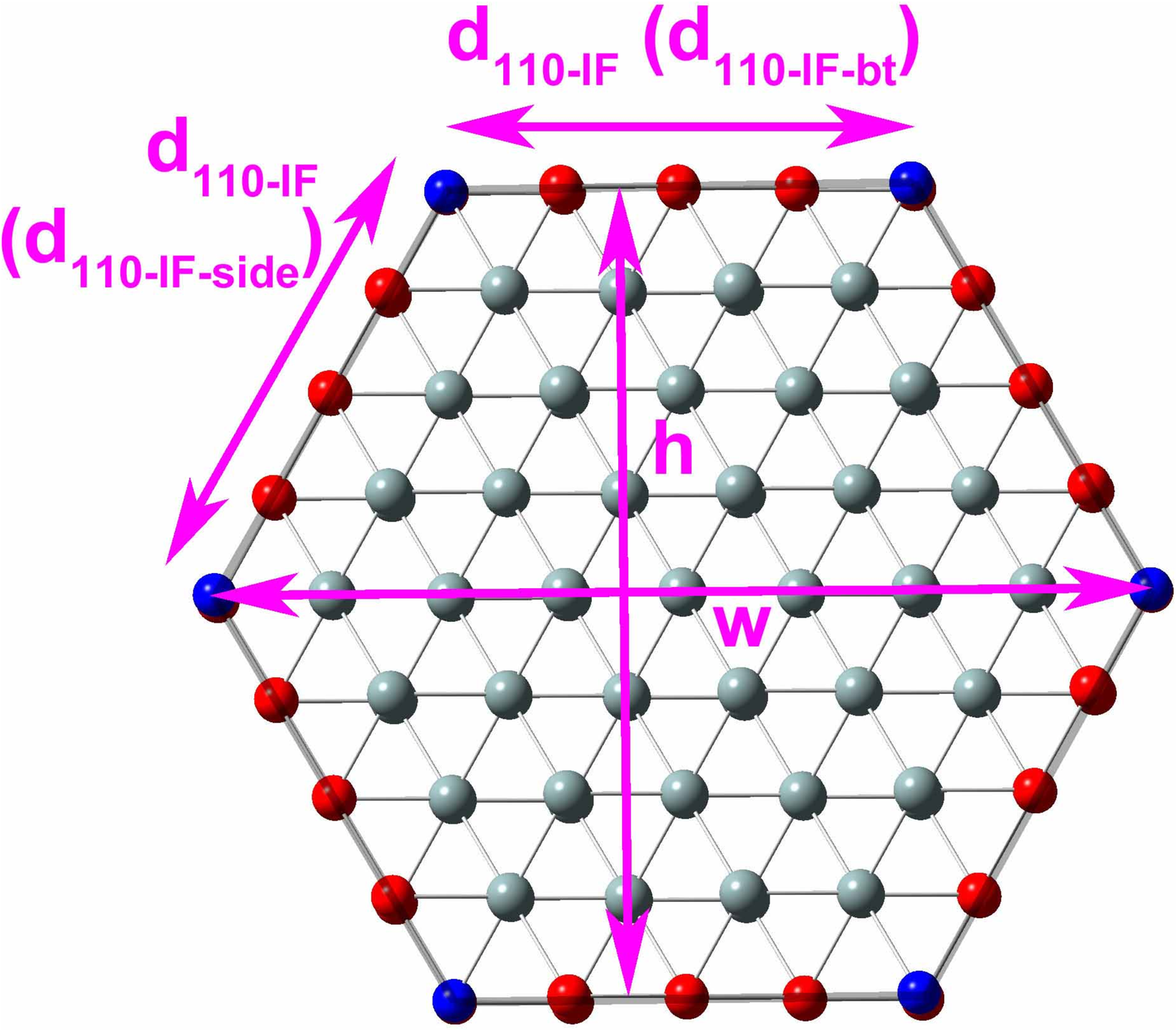}\\[-0.1cm]
\begin{picture}(0,0)
{\bf(a)}
\end{picture}\\[0.2cm]
\includegraphics[totalheight=0.0527\textheight]{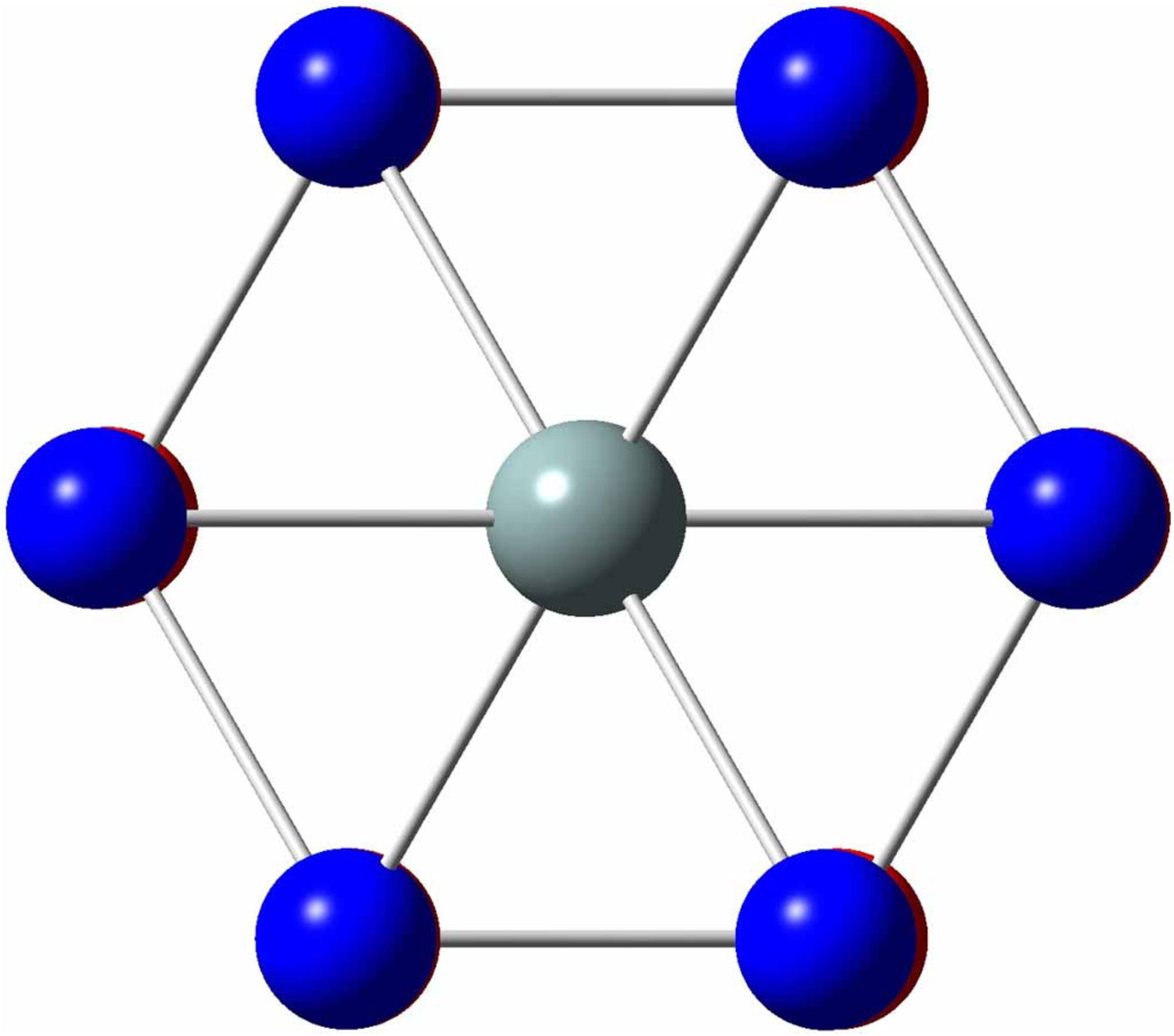}
\includegraphics[totalheight=0.0527\textheight]{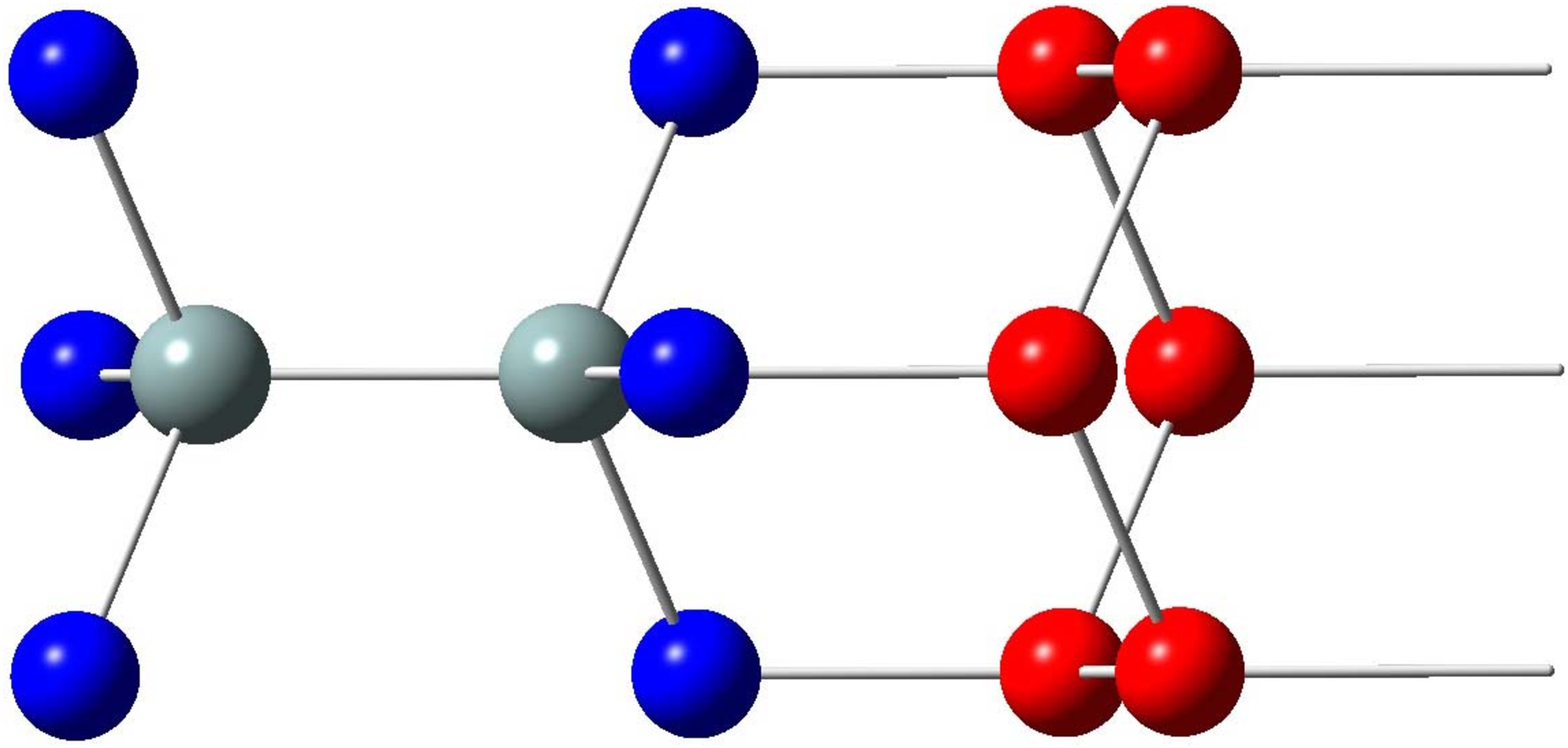}\hfill
\includegraphics[totalheight=0.0744\textheight]{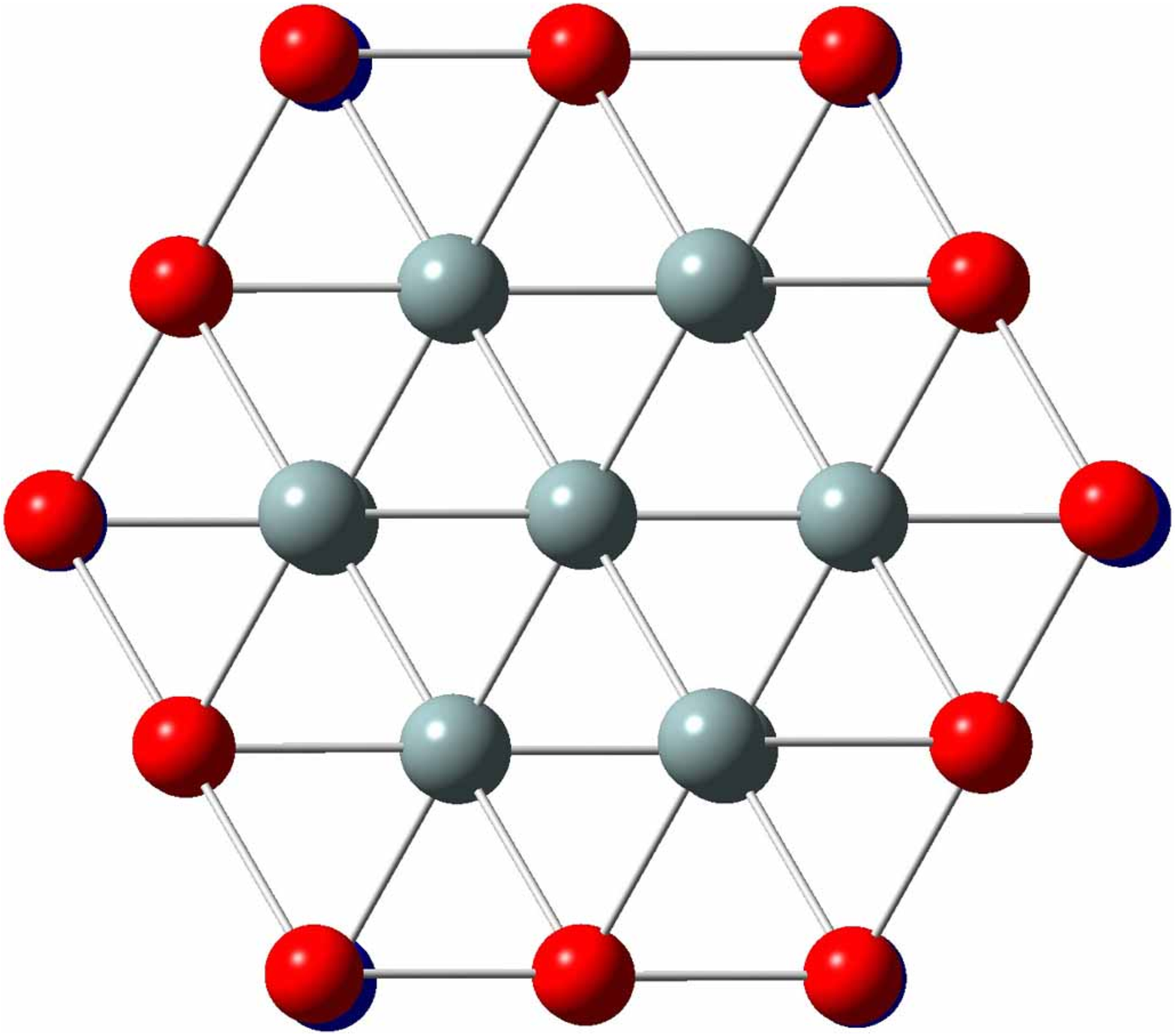}
\includegraphics[totalheight=0.0744\textheight]{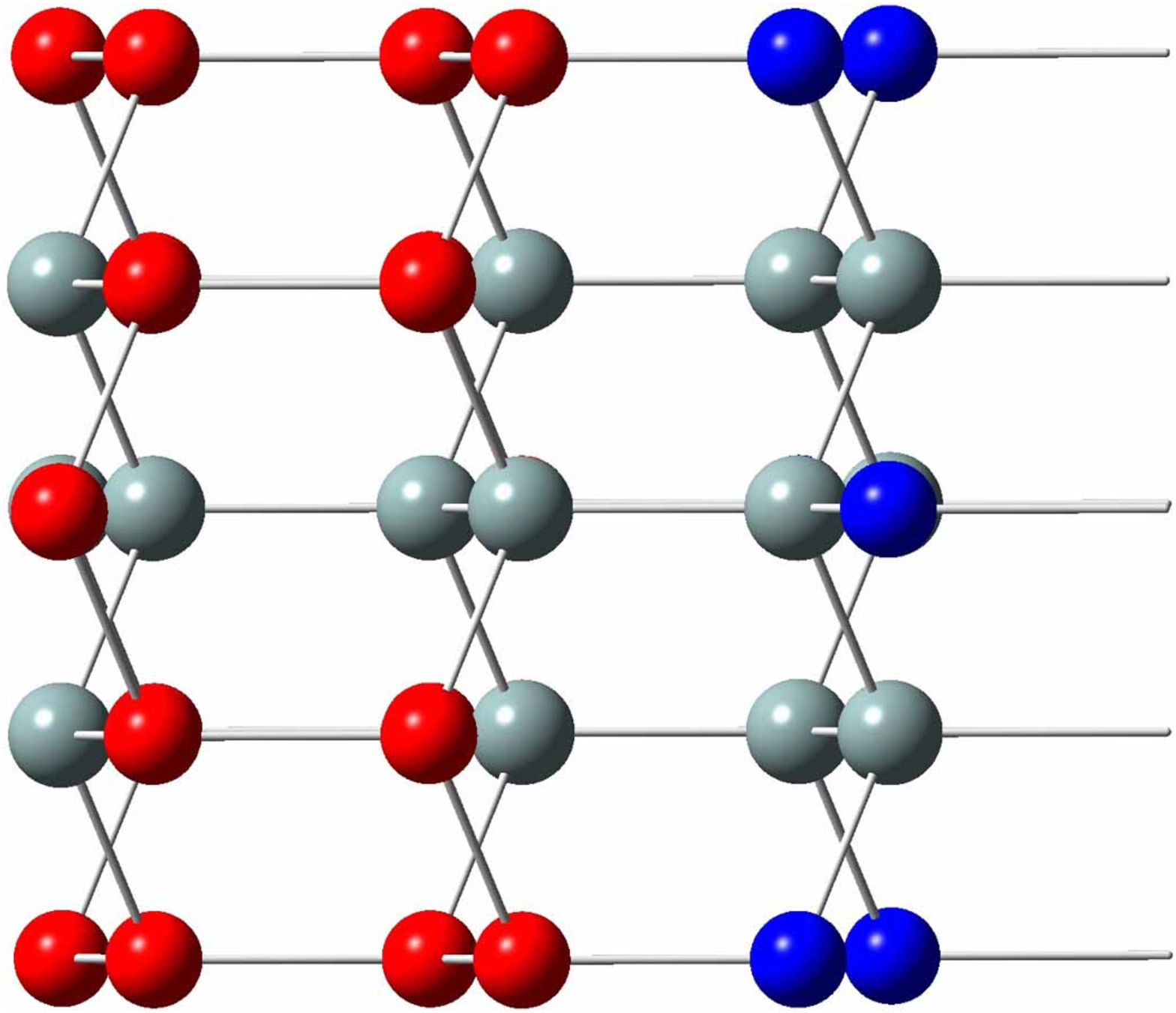}\\[-0.1cm]
\begin{picture}(0,0)
\hspace{-2.45cm}{\bf(b)}\hspace{3.91cm}{\bf (c)}
\end{picture}\\[0.2cm]
\includegraphics[totalheight=0.0847\textheight]{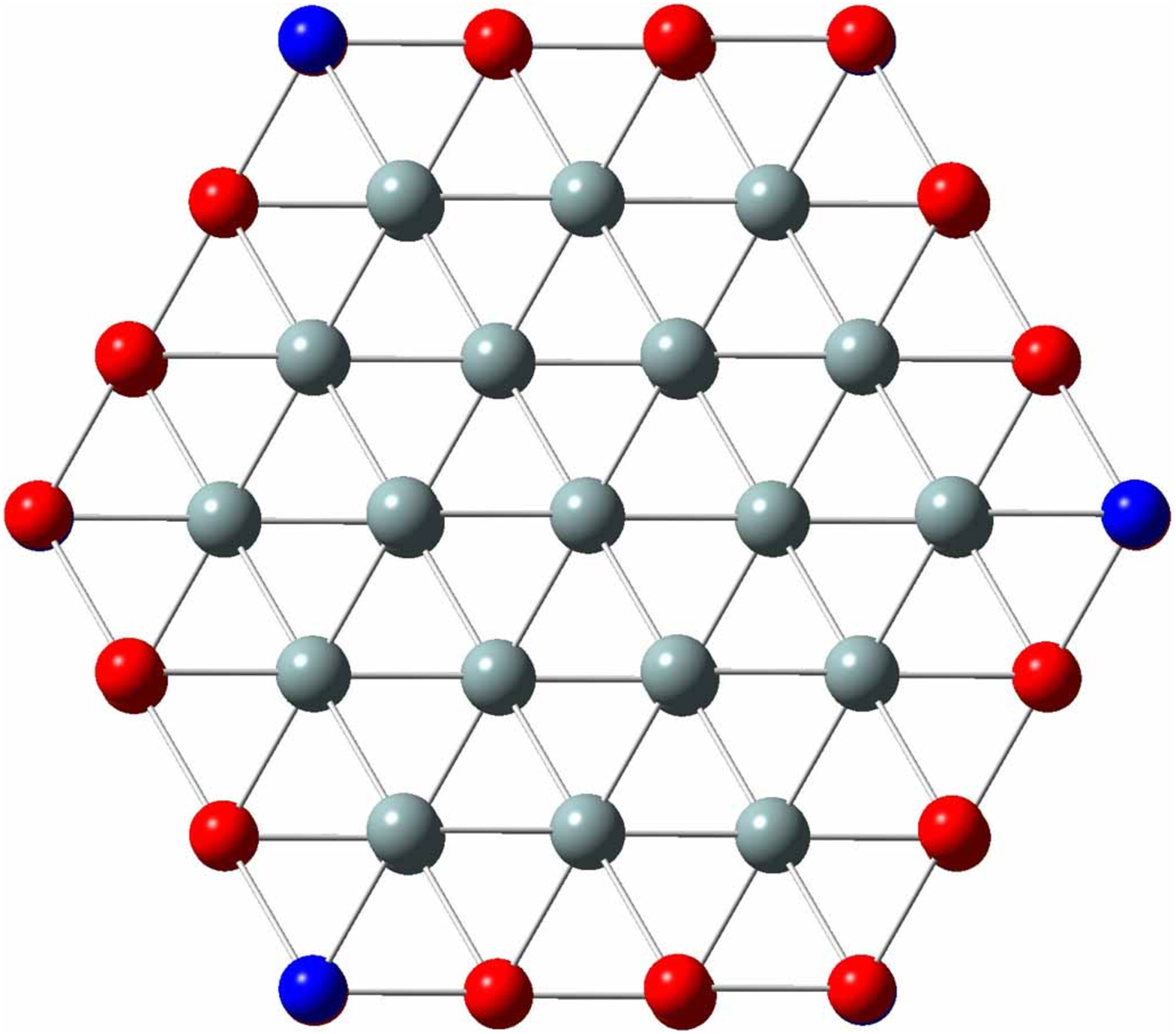}
\includegraphics[totalheight=0.0847\textheight]{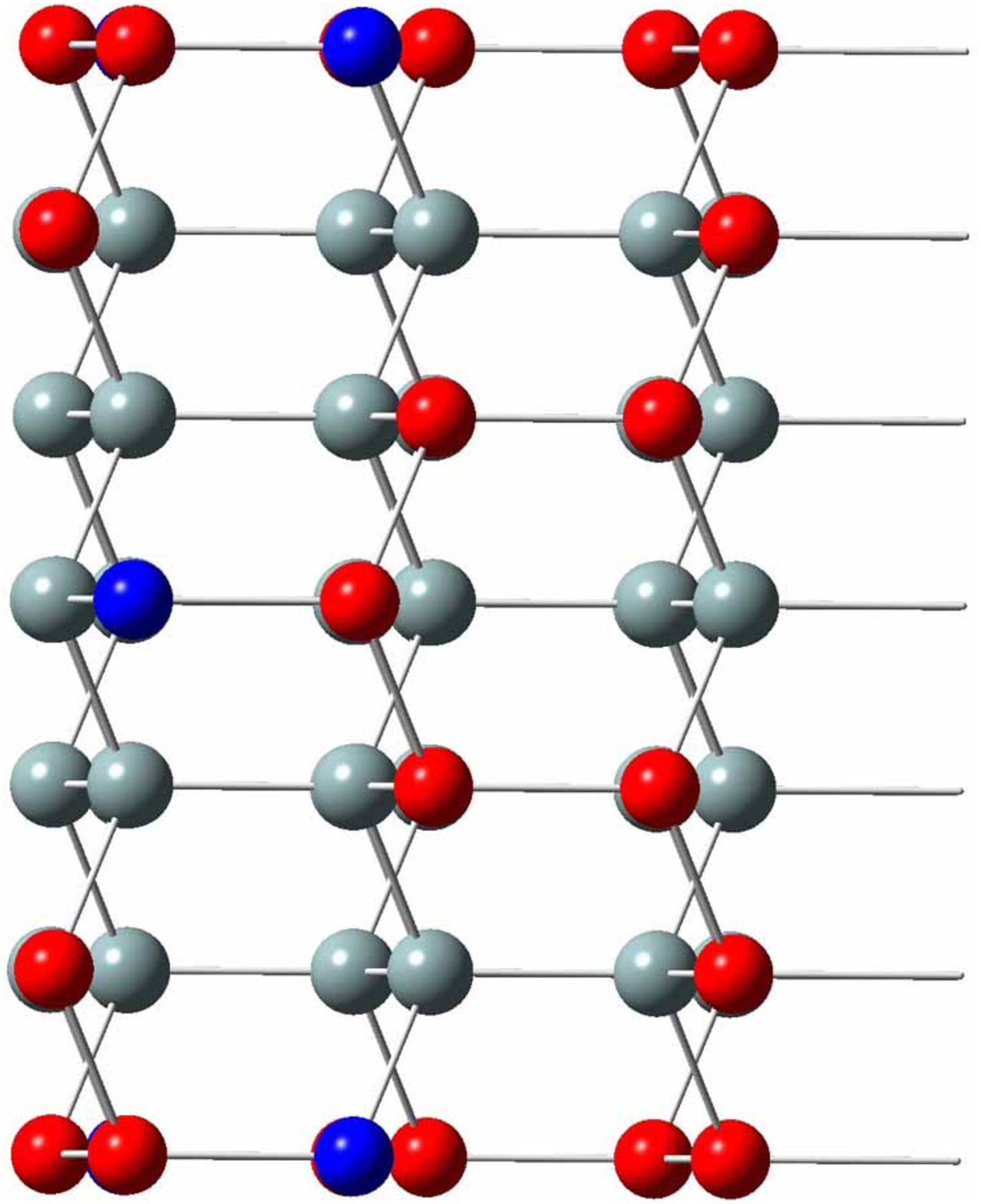}\hfill
\\[-0.1cm]
\begin{picture}(0,0)
\hspace{-2.45cm}{\bf(d)}
\end{picture}
\end{center}
\caption{Definition of characteristic lengths for the zb-/diamond-lattice NWires with [111] growth axis, hexagonal cross section and \{110\}-oriented interfaces shown by translucent black lines. Parameters in brackets refer to the \emph{odd} cross section series of where interface lenghts are not equal, see Figure \ref{fig09}. Top and side view of the first three members ($i=1$ to 3) of the \emph{even} series: X$_{14}$ (b), X$_{38}$ (c), X$_{74}$ (d). For atom colours see Figure \ref{fig03}.}
\label{fig08}
\end{figure}

\begin{eqnarray}\label{eqn-52}
N_{\rm{Wire,even}}^{111-\hexagon|110}[i]&=&6i(i+1)+2
\end{eqnarray}
\begin{eqnarray}\label{eqn-53}
N_{\rm{bnd,even}}^{111-\hexagon|110}[i]&=&6i(2i+1)+1
\end{eqnarray}
\begin{eqnarray}\label{eqn-54}
N_{\rm{IF,even}}^{111-\hexagon|110}[i]&=&6\left(2i+1\right)
\end{eqnarray} 
\begin{eqnarray}\label{eqn-55}
d_{\rm{IF,even}}^{111-\hexagon|110}[i]&=&a_{\rm{uc}}\,\frac{\sqrt{11}}{8}\ i
\end{eqnarray}
\begin{eqnarray}\label{eqn-56}
w_{\rm{even}}^{111-\hexagon|110}[i]&=&a_{\rm{uc}}\frac{\sqrt{11}}{4}\,i
\end{eqnarray}
As for a regular hexagon with $60^{\circ}$ rotational symmetry, we get $w_{\rm{even}}^{111-\hexagon|110}[i]= 2\,d_{\rm{IF,even}}^{111-\hexagon|110}[i]$.
\begin{eqnarray}\label{eqn-57}
h_{\rm{even}}^{111-\hexagon|110}[i]&=&a_{\rm{uc}}\frac{1}{\sqrt{2}}\,i
\end{eqnarray}
\begin{eqnarray}\label{eqn-58}
A^{111-\hexagon|110}_{\rm{even}}[i]&=&
\big(a_{\rm{uc}}\big)^2\ \frac{3}{8}\sqrt{\frac{11}{8}}\ i^2
\end{eqnarray}
Figure \ref{fig08} shows the cross section of this \emph{even} NWire type for $i=1$ to 3 together with the definition of characteristic lengths.

\begin{figure}[h!]
\begin{center}
\includegraphics[totalheight=0.0814\textheight]{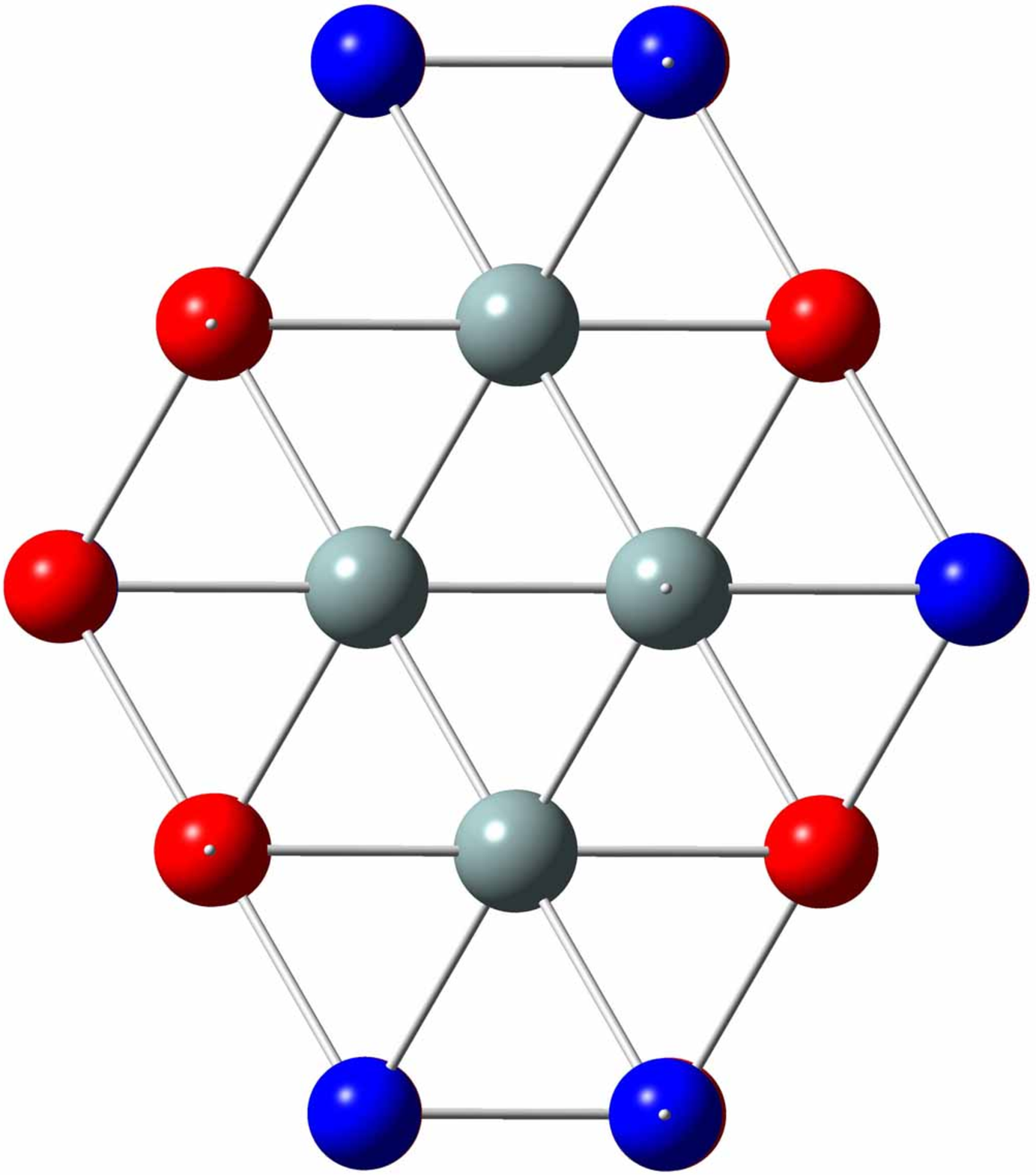}
\includegraphics[totalheight=0.0814\textheight]{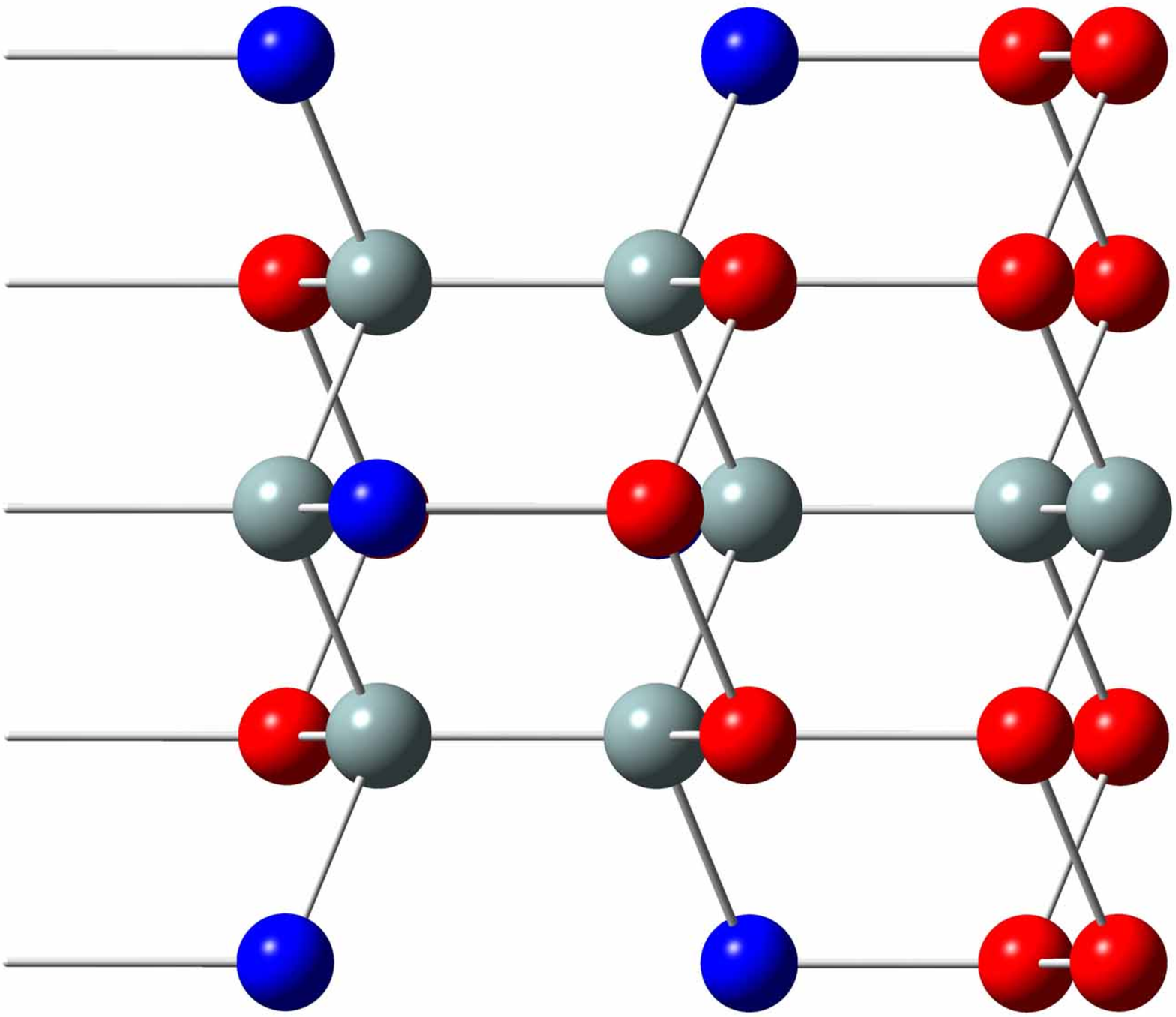}\hfill
\includegraphics[totalheight=0.0938\textheight]{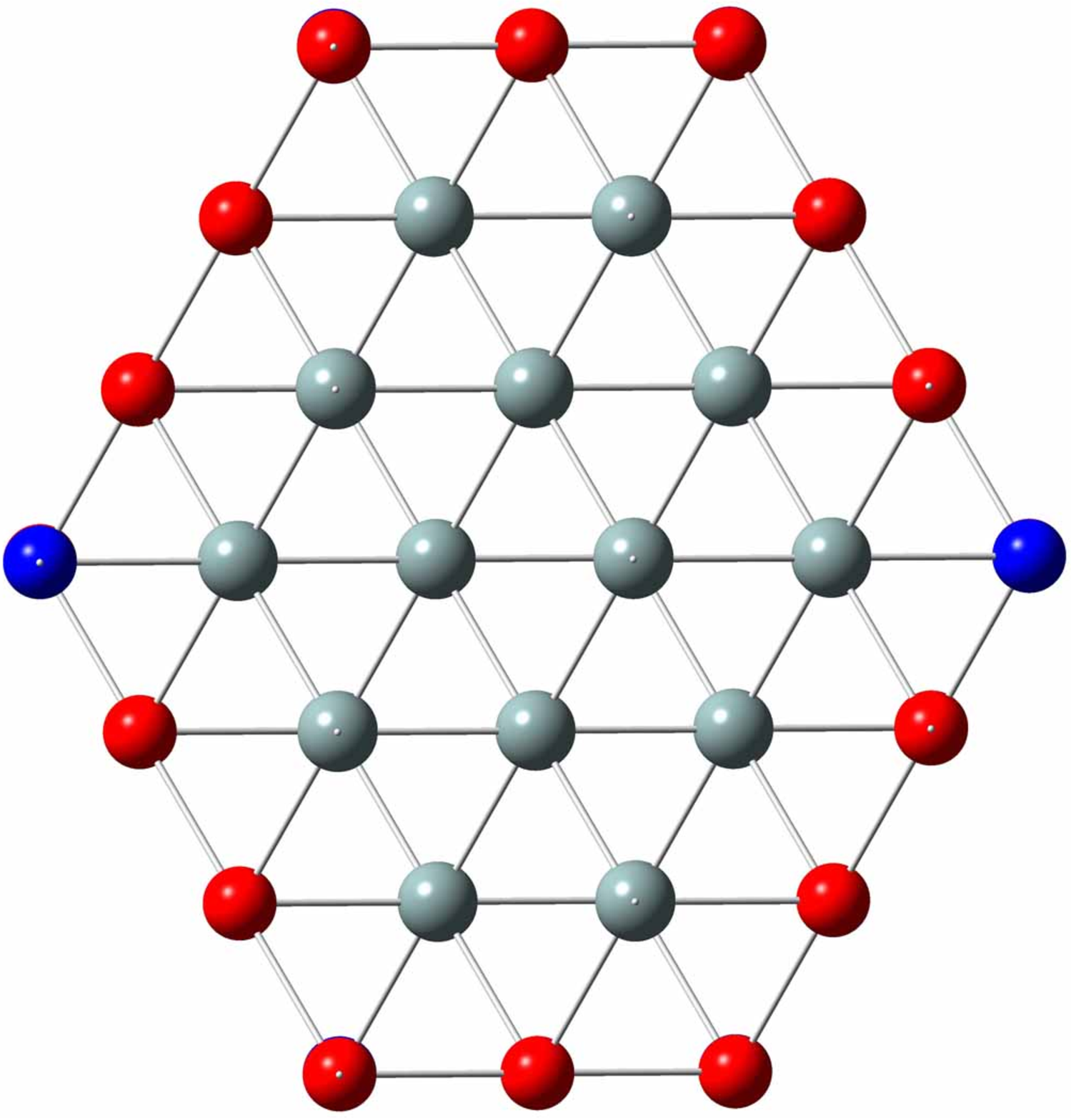}
\includegraphics[totalheight=0.0938\textheight]{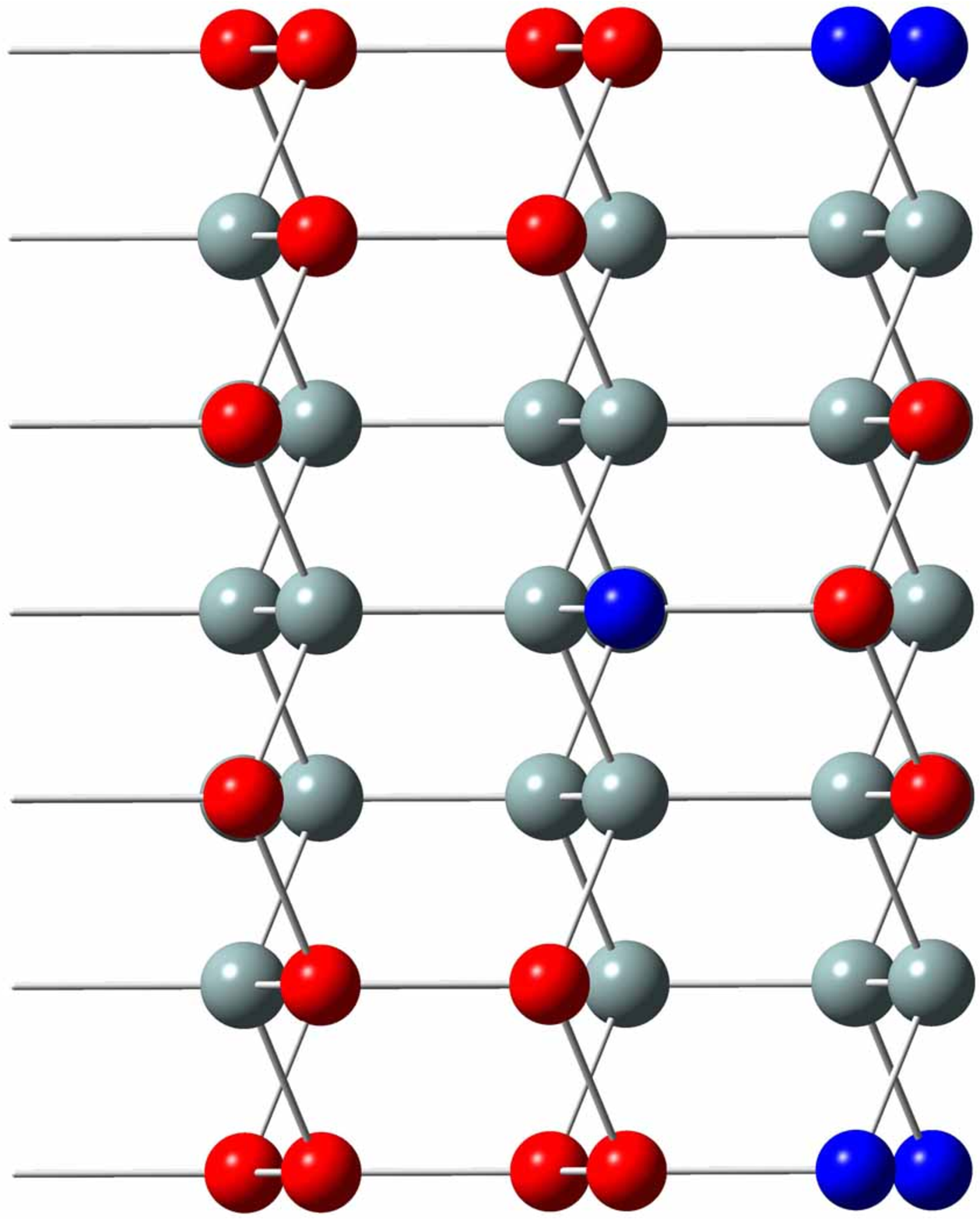}\\[-0.1cm]
\begin{picture}(0,0)
\hspace{-2.45cm}{\bf(a)}\hspace{3.91cm}{\bf (b)}
\end{picture}\\[0.2cm]
\includegraphics[totalheight=0.1048\textheight]{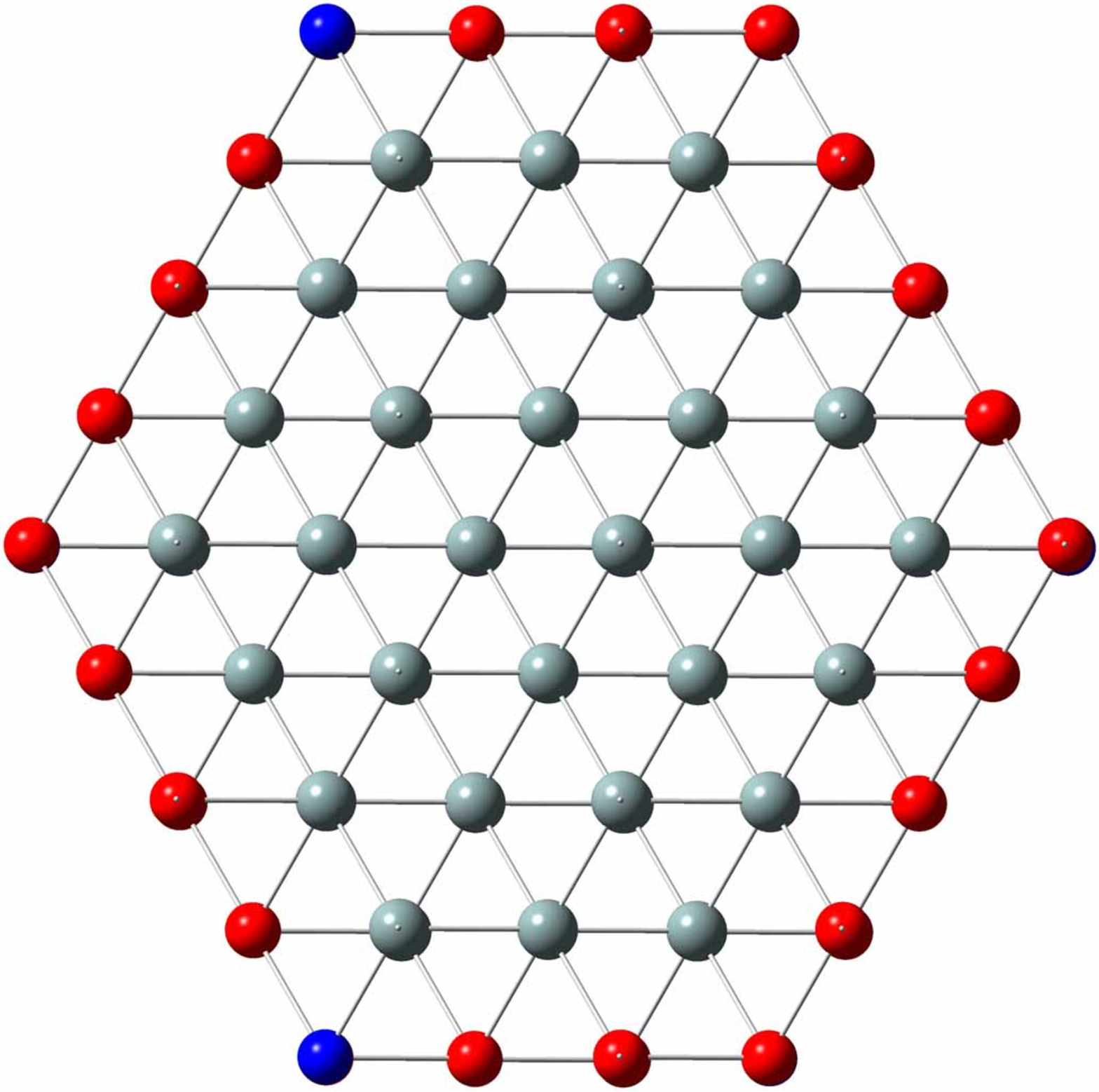}
\includegraphics[totalheight=0.1048\textheight]{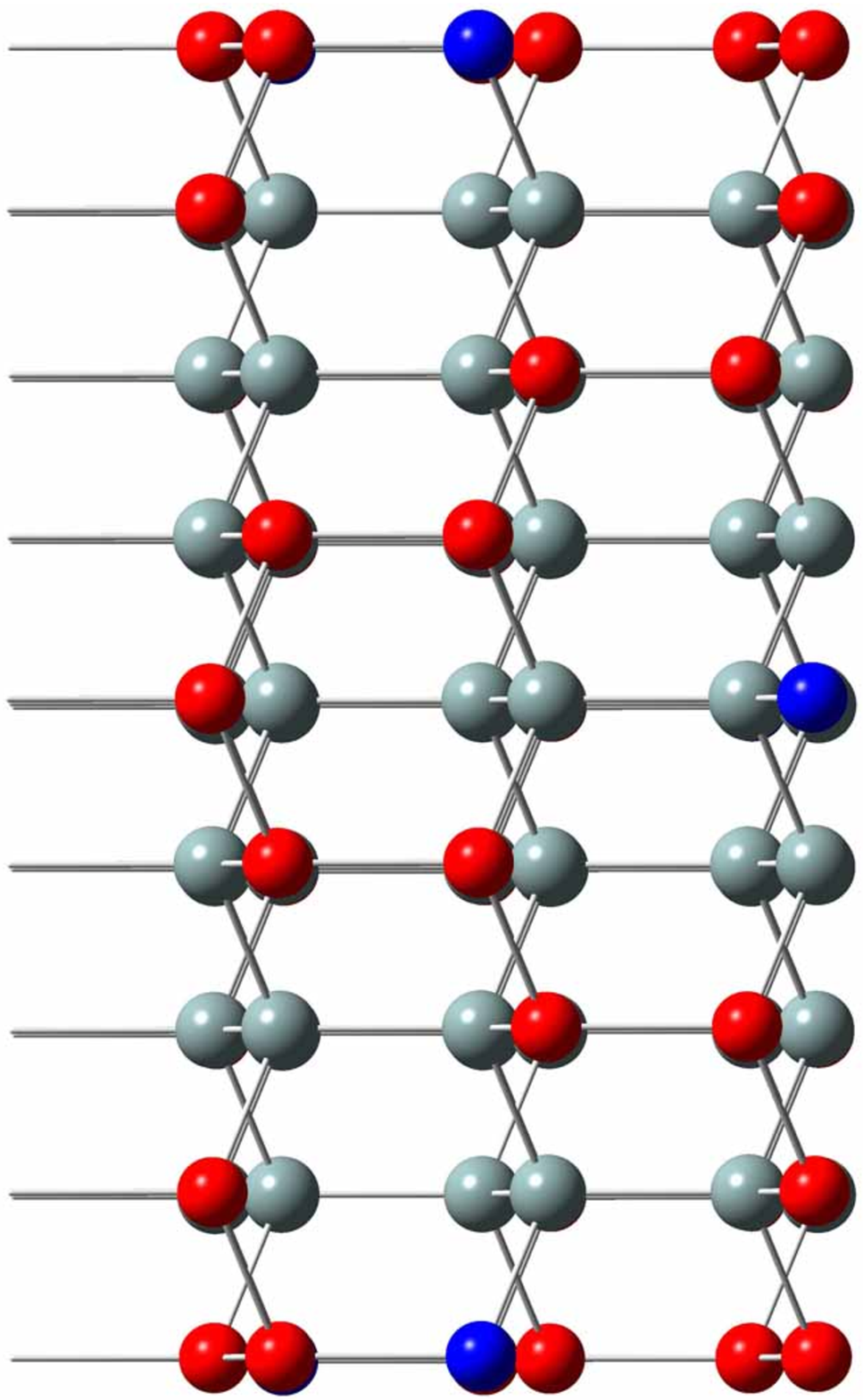}\hfill
\\[-0.1cm]
\begin{picture}(0,0)
\hspace{-2.45cm}{\bf(c)}
\end{picture}
\end{center}
\caption{Cross section and side view of zb-/diamond-lattice NWires with [111] growth axis, hexagonal cross section and \{110\}-oriented interfaces, \emph{odd} series, for run index $i=1$ to 3: X$_{28}$ (a), X$_{60}$ (b), X$_{104}$ (c). For atom colours and definition of characteristic lengths, see Figure \ref{fig03} and Figure \ref{fig08}, respectively.}
\label{fig09}
\end{figure}

For the odd series, we obtain
\begin{eqnarray}\label{eqn-59}
N_{\rm{Wire,odd}}^{111-\hexagon|110}[i]&=&2i(3i+7)+8
\end{eqnarray}
\begin{eqnarray}\label{eqn-60}
N_{\rm{bnd,odd}}^{111-\hexagon|110}[i]&=&2i(6i+11)+9
\end{eqnarray}
\begin{eqnarray}\label{eqn-61}
N_{\rm{IF,odd}}^{111-\hexagon|110}[i]&=&2(6i+7)
\end{eqnarray} 
\begin{eqnarray}\label{eqn-62}
d_{\rm{IF,odd,tb}}^{111-\hexagon|110}[i]&=&a_{\rm{uc}}\,\frac{\sqrt{11}}{8}\ i
\end{eqnarray}
This result is equal to $d_{\rm{IF,even}}^{111-\hexagon|110}[i]$.
\begin{eqnarray}\label{eqn-63}
d_{\rm{IF,odd,side}}^{111-\hexagon|110}[i]&=&
a_{\rm{uc}}\,\frac{\sqrt{11}}{8}\ (i+1)
\end{eqnarray}
We see that $d_{\rm{IF,odd,side}}^{111-\hexagon|110}[i]=d_{\rm{IF,even}}^{111-\hexagon|110}[i+1]$.
\begin{eqnarray}\label{eqn-64}
w_{\rm{odd}}^{111-\hexagon|110}[i]&=&
a_{\rm{uc}}\frac{\sqrt{11}}{4}\,\left(i+\frac{1}{2}\right)
\end{eqnarray}
Here, the relation to the \emph{even} series is $w_{\rm{odd}}^{111-\hexagon|110}[i]=w_{\rm{even}}^{111-\hexagon|110}[i+\frac{1}{2}]$.
\begin{eqnarray}\label{eqn-65}
h_{\rm{odd}}^{111-\hexagon|110}[i]&=&a_{\rm{uc}}\frac{1}{\sqrt{2}}\,(i+1)
\end{eqnarray}
Again, this result relates to the \emph{even} series via $h_{\rm{odd}}^{111-\hexagon|110}[i]=h_{\rm{even}}^{111-\hexagon|110}[i+1]$.
\begin{eqnarray}\label{eqn-66}
A^{111-\hexagon|110}_{\rm{odd}}[i]&=&\big(a_{\rm{uc}}\big)^2\ \frac{1}{8}\sqrt{\frac{11}{8}}\,(3i+1)(i+1)
\end{eqnarray}
The cross section of this \emph{odd} NWire type is shown in Figure \ref{fig09} for $i=1$ to 3.

\begin{figure}[h!]
\begin{center}
\includegraphics[totalheight=0.158\textheight]{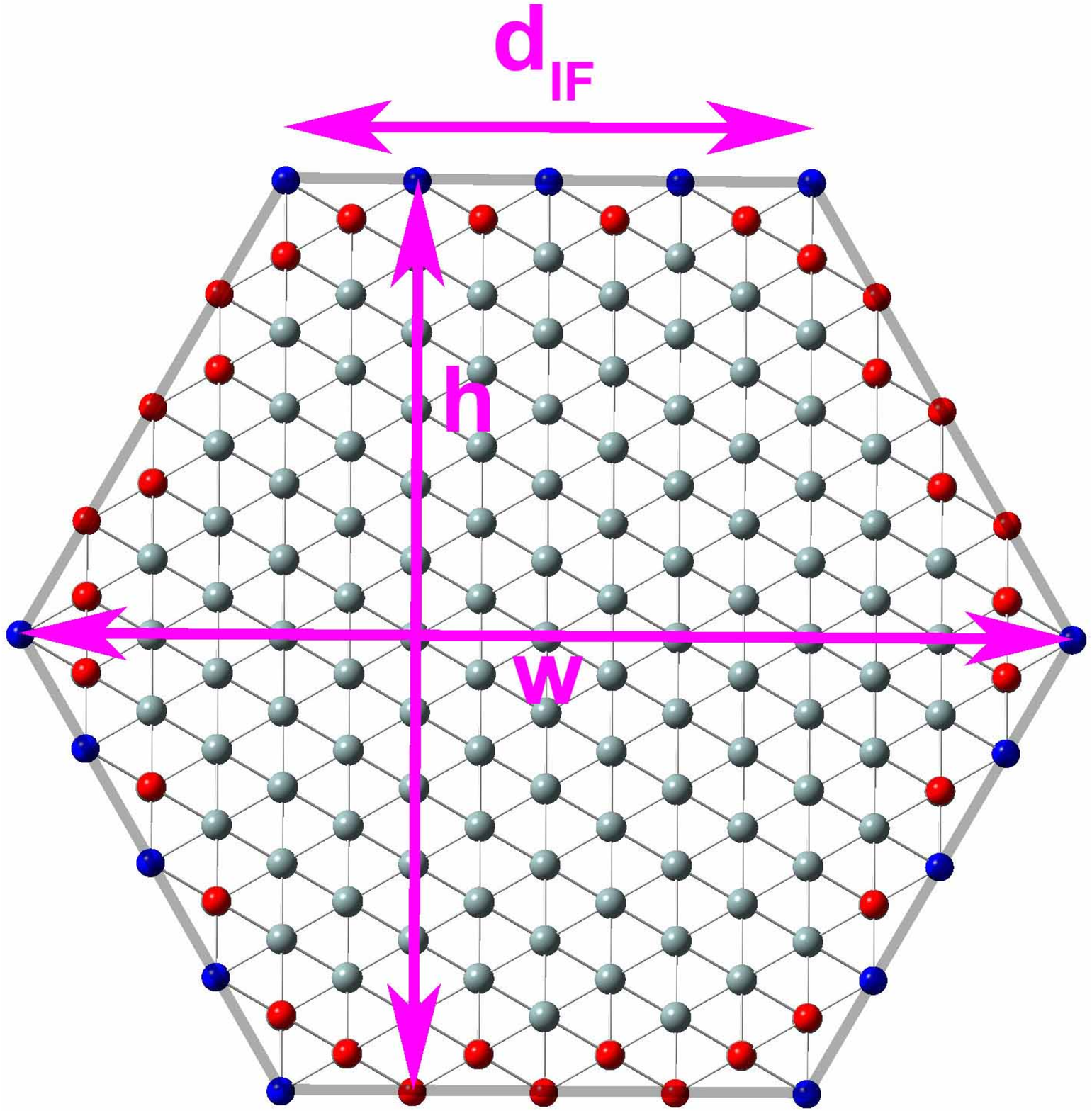}\\[-0.1cm]
\begin{picture}(0,0)
{\bf(a)}
\end{picture}\\[0.2cm]
\includegraphics[totalheight=0.0695\textheight]{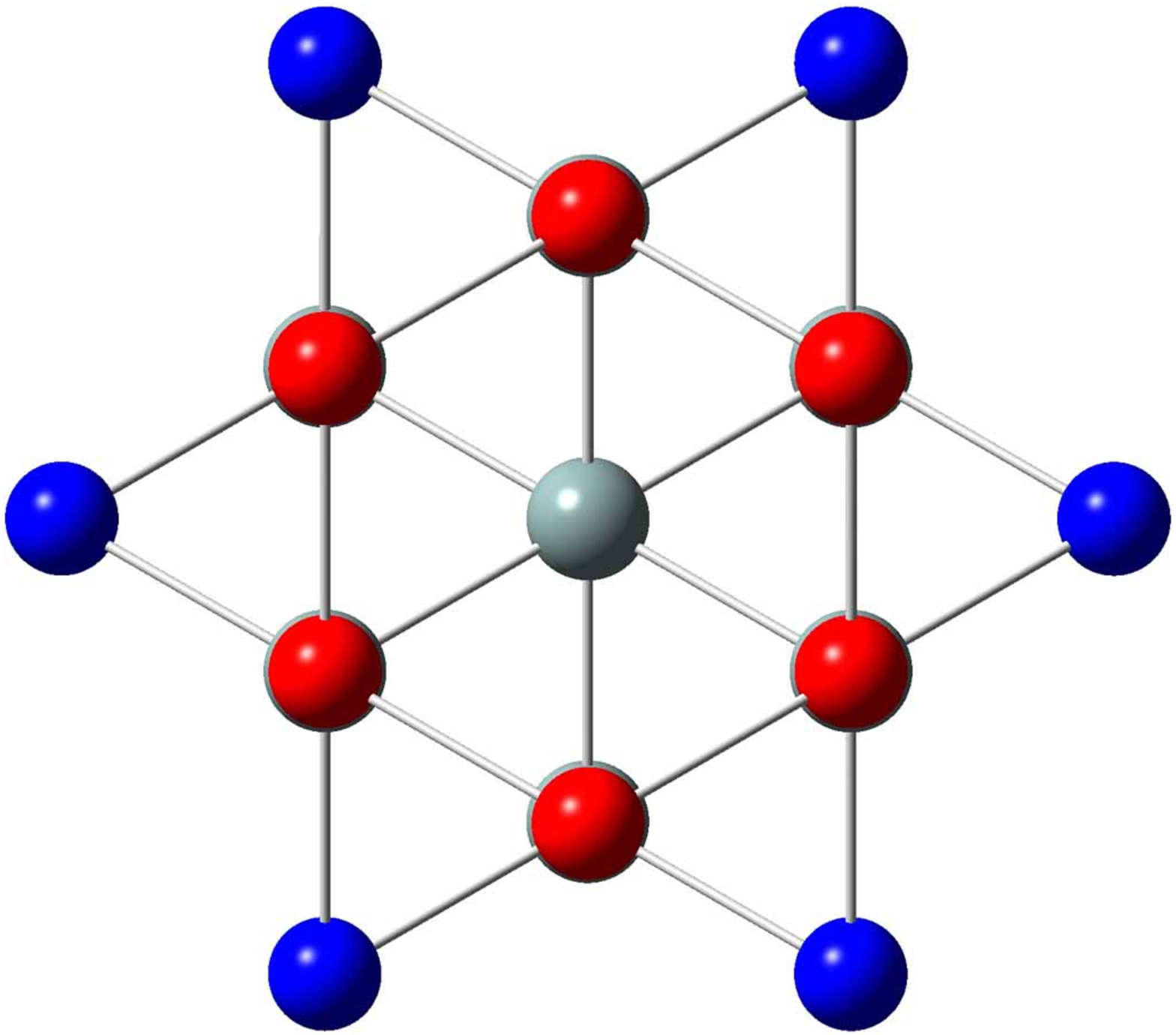}
\includegraphics[totalheight=0.0695\textheight]{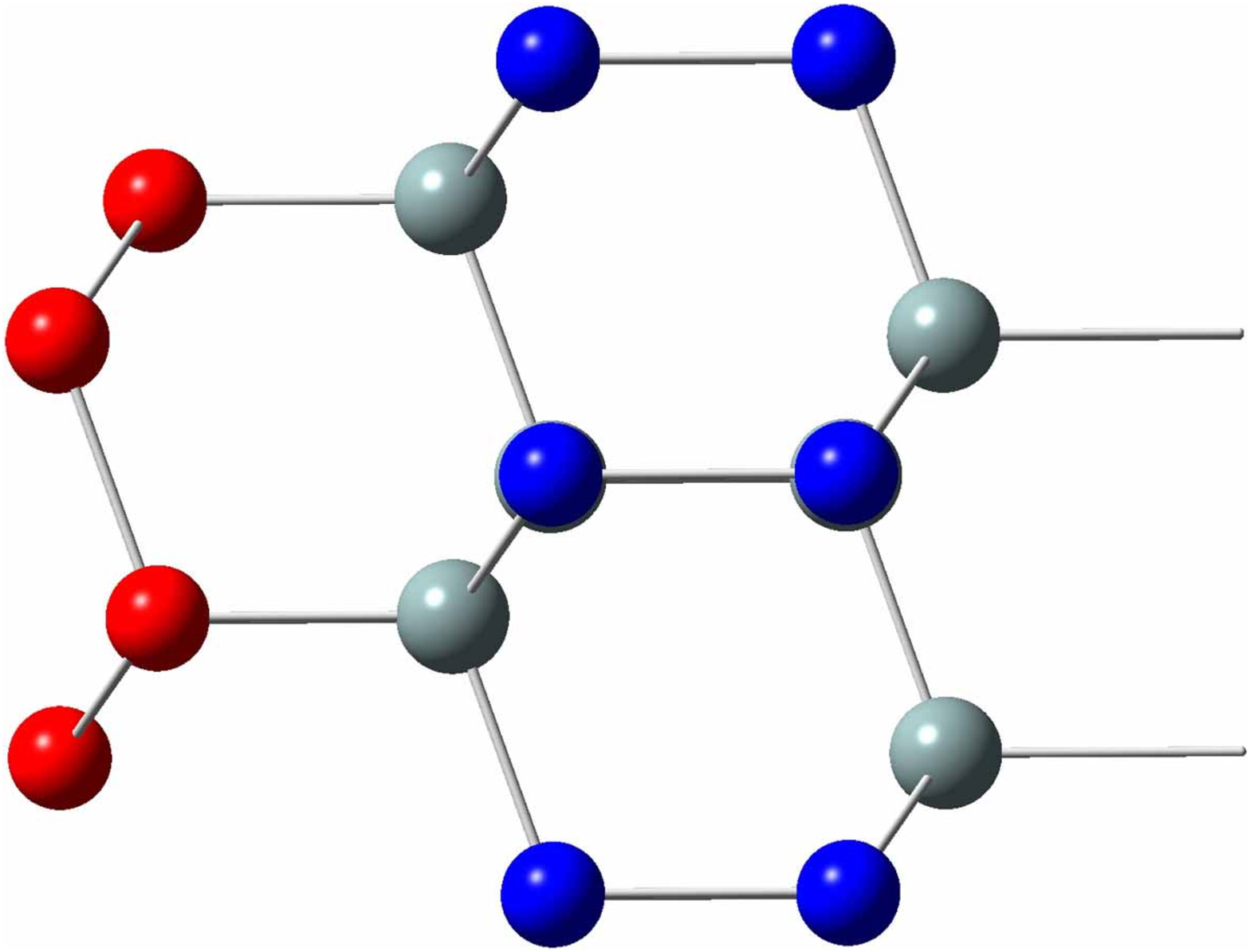}\hfill
\includegraphics[totalheight=0.0929\textheight]{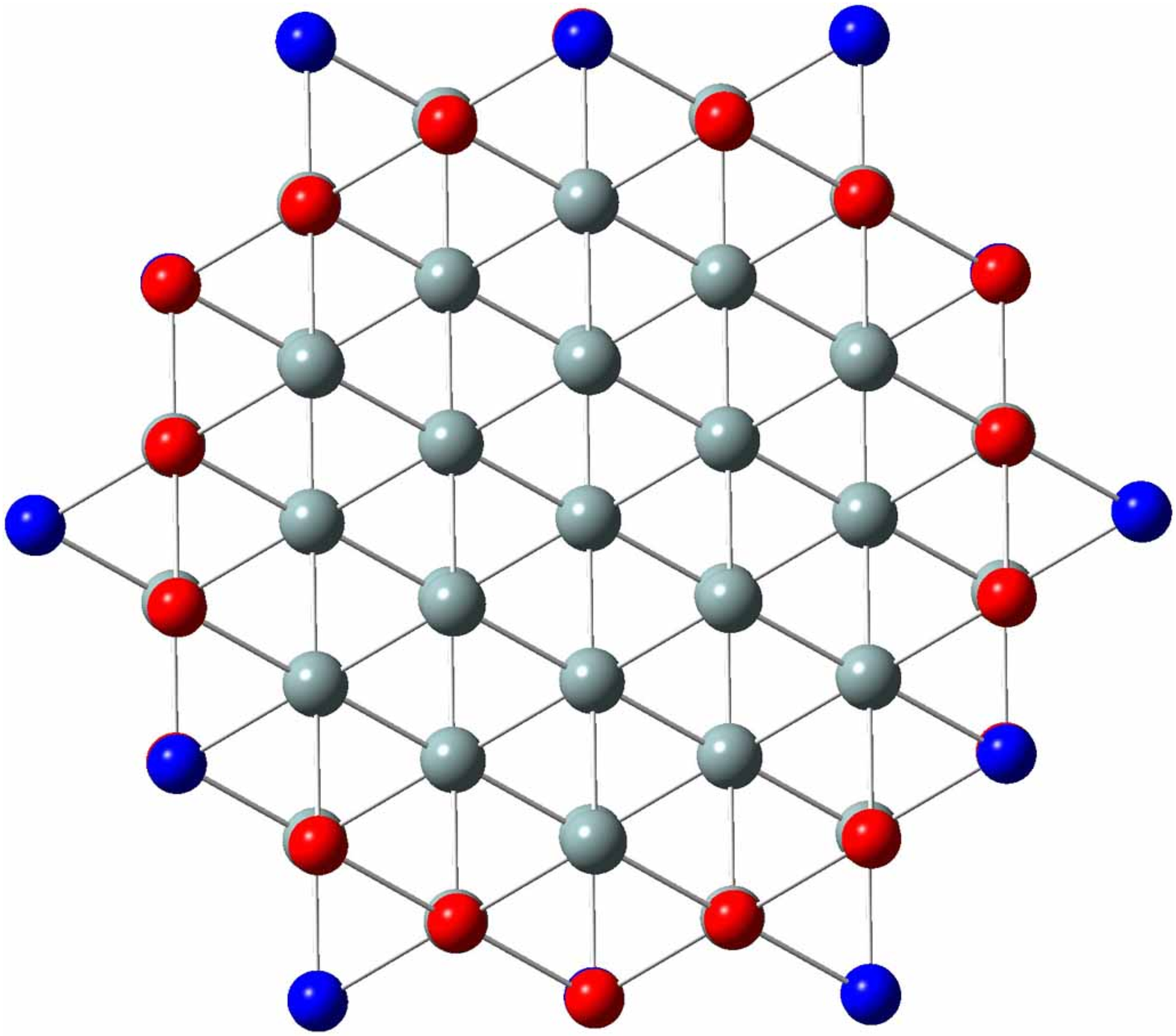}
\includegraphics[totalheight=0.0929\textheight]{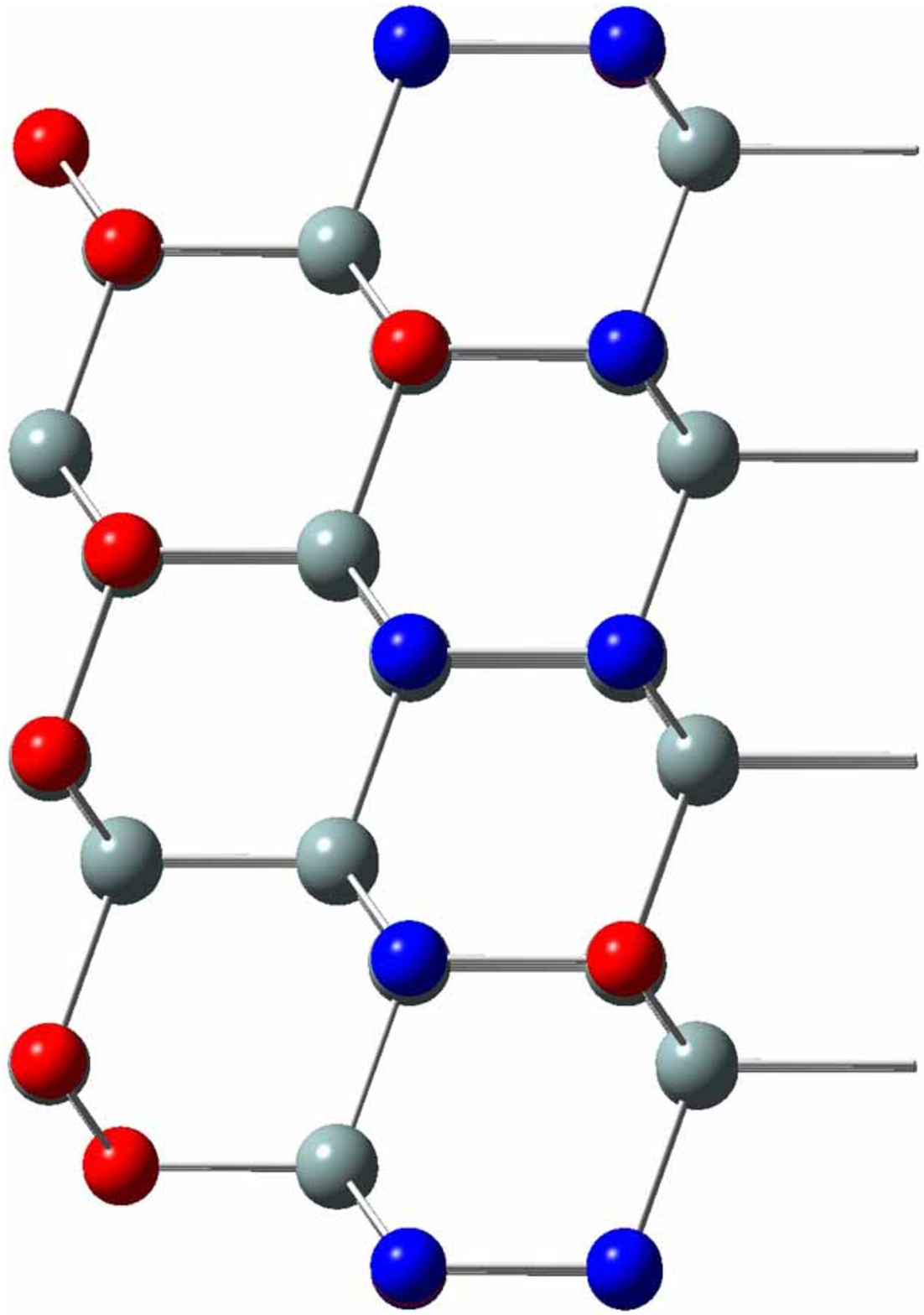}\\[-0.1cm]
\begin{picture}(0,0)
\hspace{-2.45cm}{\bf(b)}\hspace{3.91cm}{\bf (c)}
\end{picture}\\[0.2cm]
\includegraphics[totalheight=0.1041\textheight]{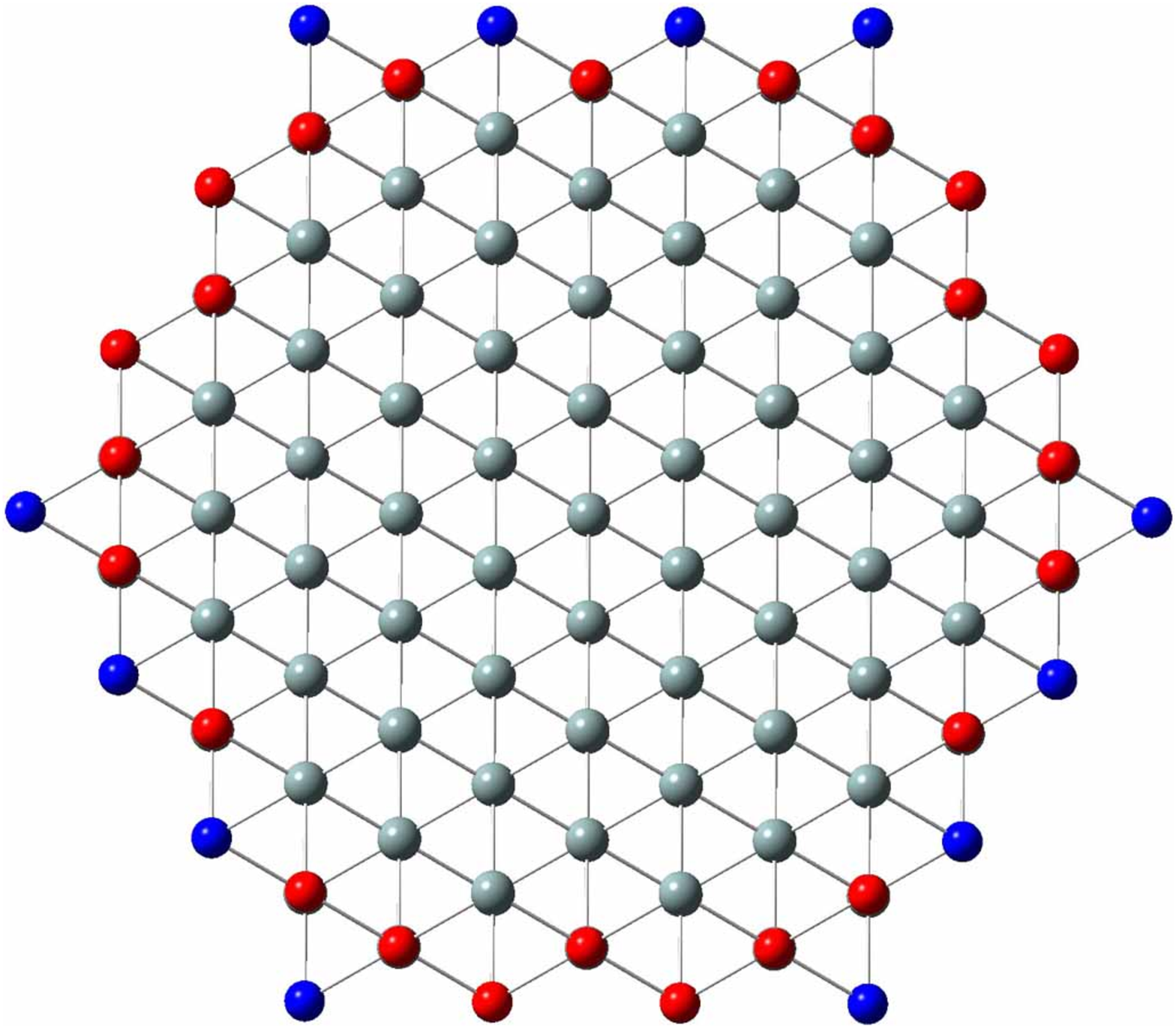}
\includegraphics[totalheight=0.1041\textheight]{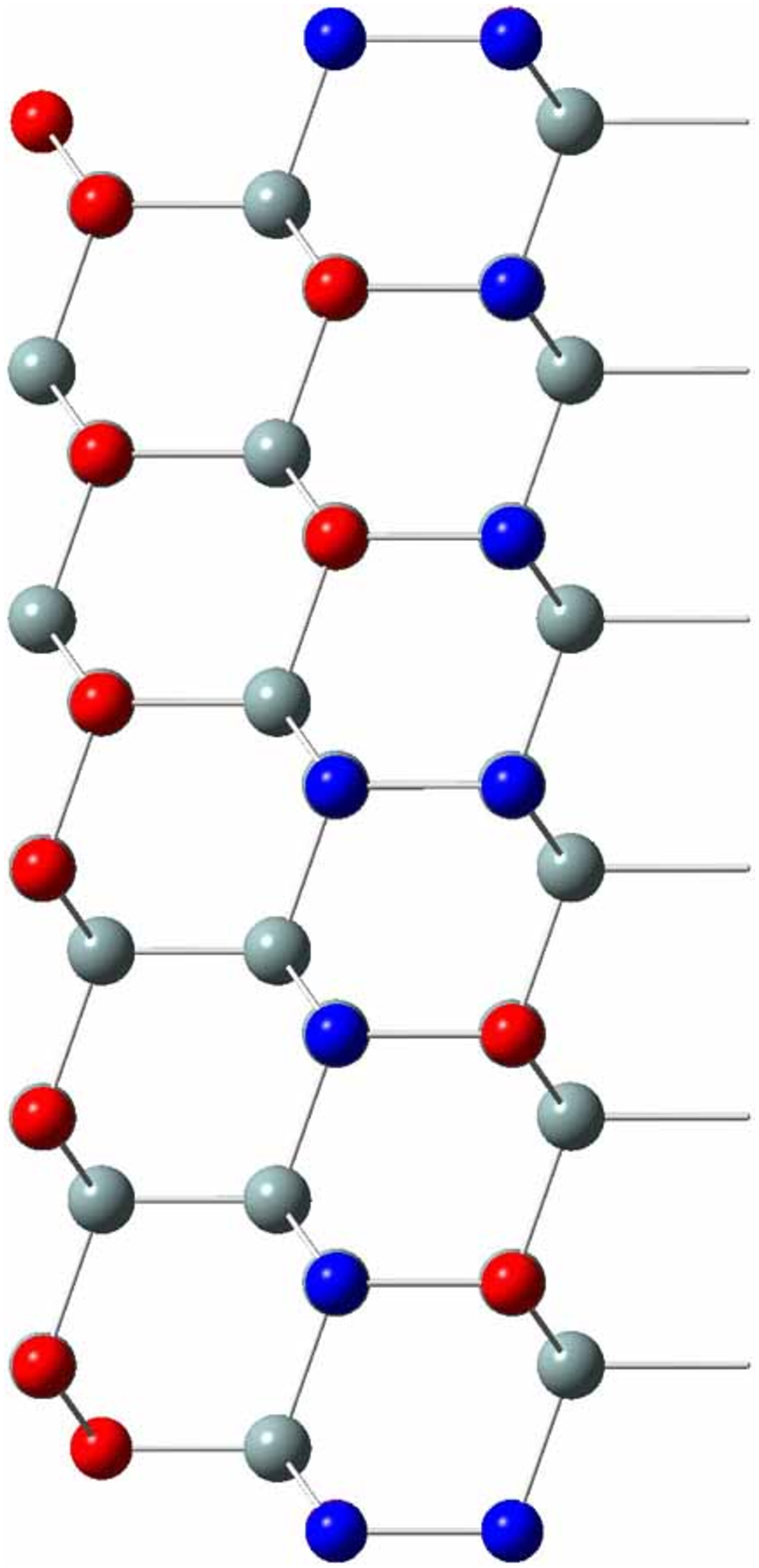}\hfill
\\[-0.1cm]
\begin{picture}(0,0)
\hspace{-2.45cm}{\bf(d)}
\end{picture}
\end{center}
\caption{Definition of characteristic lengths for the \emph{even} series of zb-/diamond-lattice NWires with [111] growth axis and hexagonal cross section. All interfaces have \{11$\bar{2}$\} orientation, shown by translucent black lines. For definitions of the \emph{odd} series see Figure \ref{fig11}a. Top and side view of the first three members ($i=1$ to 3) of the \emph{even} series: X$_{26}$ (b), X$_{86}$ (c), X$_{182}$ (d). For atom colours see Figure \ref{fig03}.}
\label{fig10}
\end{figure}

\subsection{\label{AnaNomiHexa111_IF112} NWires growing along [111] Direction with Hexagonal Cross Section and Six \{11$\mathbf{\bar{2}}$\} Interfaces}

\begin{eqnarray}\label{eqn-67}
N_{\rm{Wire,even}}^{111-\hexagon|11\bar{2}}[i]&=&6i(3i+1)+2
\end{eqnarray}
\begin{eqnarray}\label{eqn-68}
N_{\rm{bnd,even}}^{111-\hexagon|11\bar{2}}[i]&=&(6i)^2+1
\end{eqnarray}
\begin{eqnarray}\label{eqn-69}
N_{\rm{IF,even}}^{111-\hexagon|11\bar{2}}[i]&=&6\left(4i+1\right)
\end{eqnarray} 
\begin{eqnarray}\label{eqn-70}
d_{\rm{IF,even}}^{111-\hexagon|11\bar{2}}[i]&=&
a_{\rm{uc}}\,\frac{1}{\sqrt{2}}\ i
\end{eqnarray}
\begin{eqnarray}\label{eqn-71}
w_{\rm{even}}^{111-\hexagon|11\bar{2}}[i]&=&a_{\rm{uc}}\sqrt{2}\,i
\end{eqnarray}
As for a regular hexagon with $60^{\circ}$ rotational symmetry, we get $w_{\rm{even}}^{111-\hexagon|11\bar{2}}[i]= 2\,d_{\rm{IF,even}}^{111-\hexagon|11\bar{2}}[i]$.
\begin{eqnarray}\label{eqn-72}
h_{\rm{even}}^{111-\hexagon|11\bar{2}}[i]&=&
a_{\rm{uc}}\,\frac{3}{8}\sqrt{11}\ i
\end{eqnarray}
\begin{eqnarray}\label{eqn-73}
A^{111-\hexagon|11\bar{2}}_{\rm{even}}[i]&=&
\big(a_{\rm{uc}}\big)^2\,\left(\frac{3}{4}\right)^2\sqrt{\frac{11}{2}}\ i^2
\end{eqnarray}
The cross section of this \emph{even} NWire type is shown in Figure \ref{fig10} for $i=1$ to 3 together with the definition of characteristic lengths.

\begin{figure}[h!]
\begin{center}
\includegraphics[totalheight=0.2068\textheight]{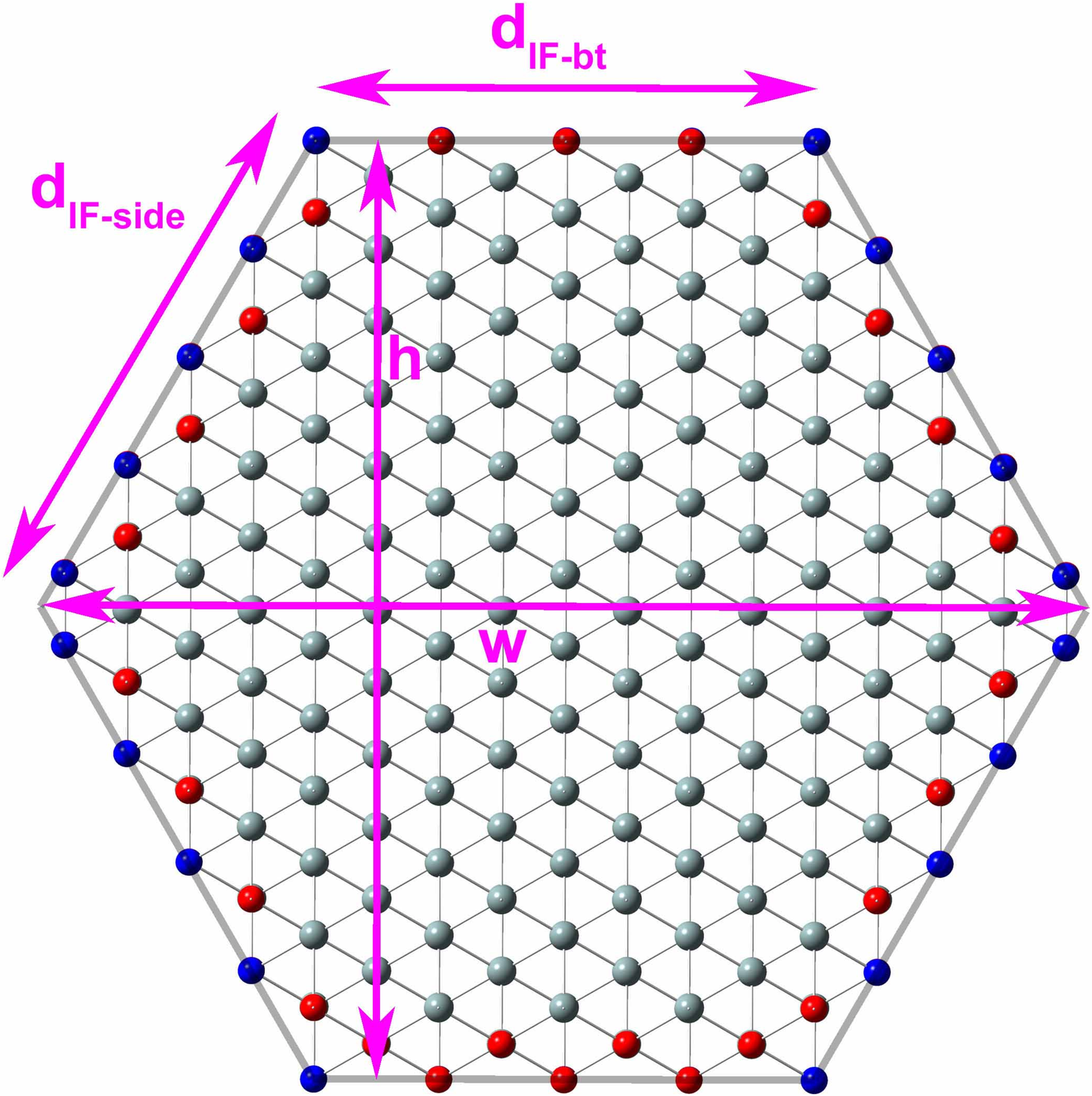}\\[-0.1cm]
\begin{picture}(0,0)
{\bf(a)}
\end{picture}\\[0.2cm]
\includegraphics[totalheight=0.0856\textheight]{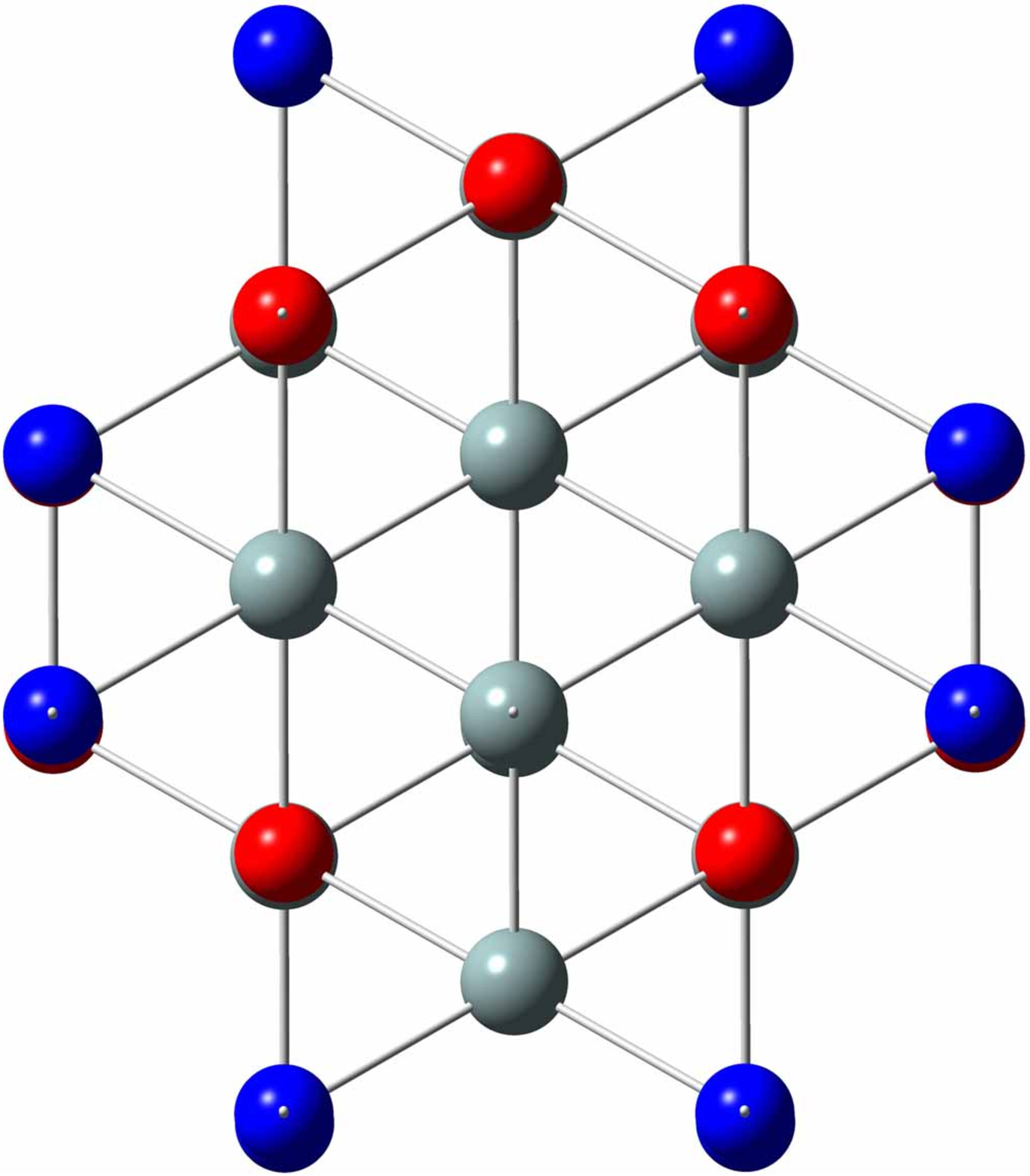}
\includegraphics[totalheight=0.0856\textheight]{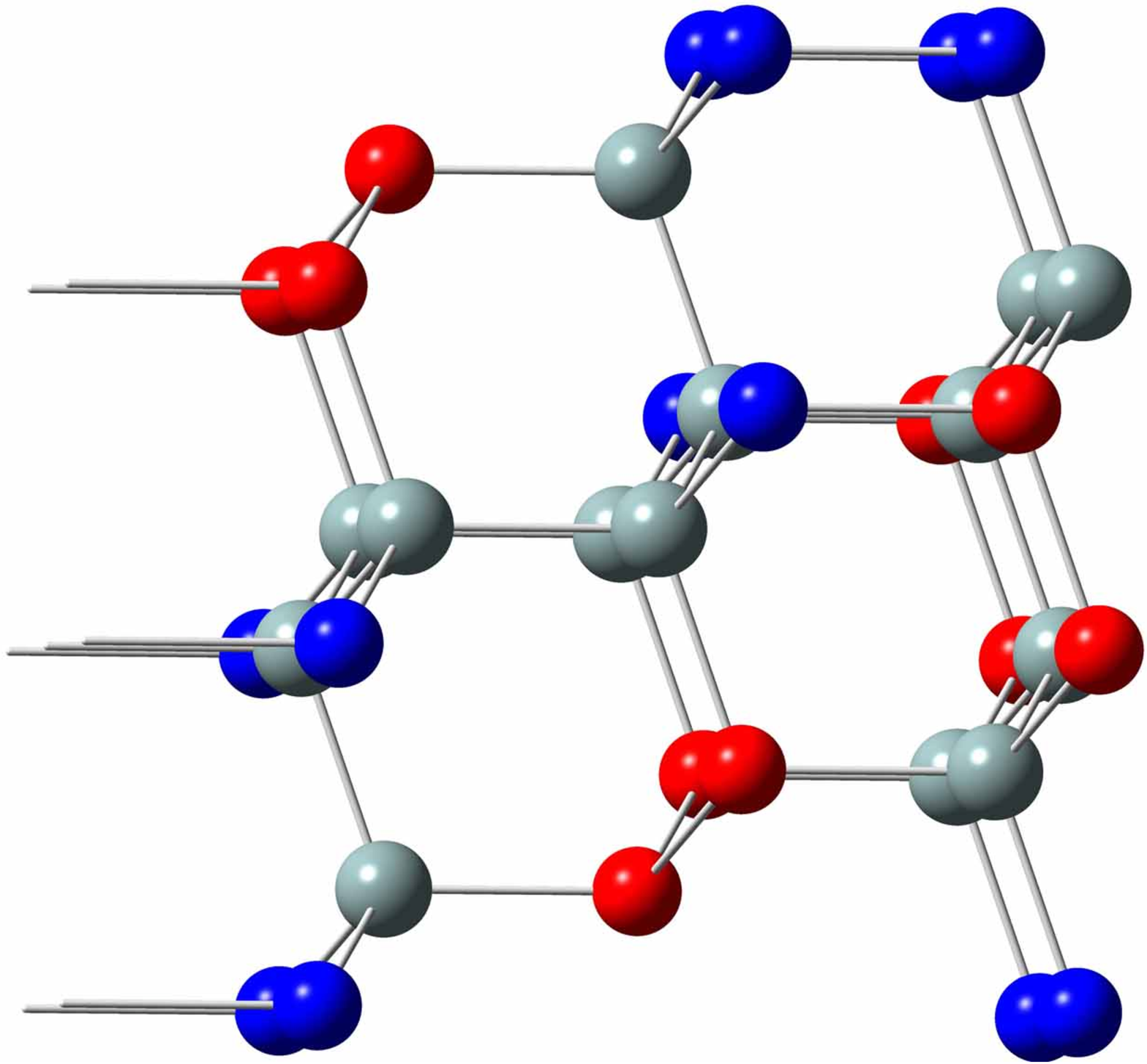}\hfill
\includegraphics[totalheight=0.1064\textheight]{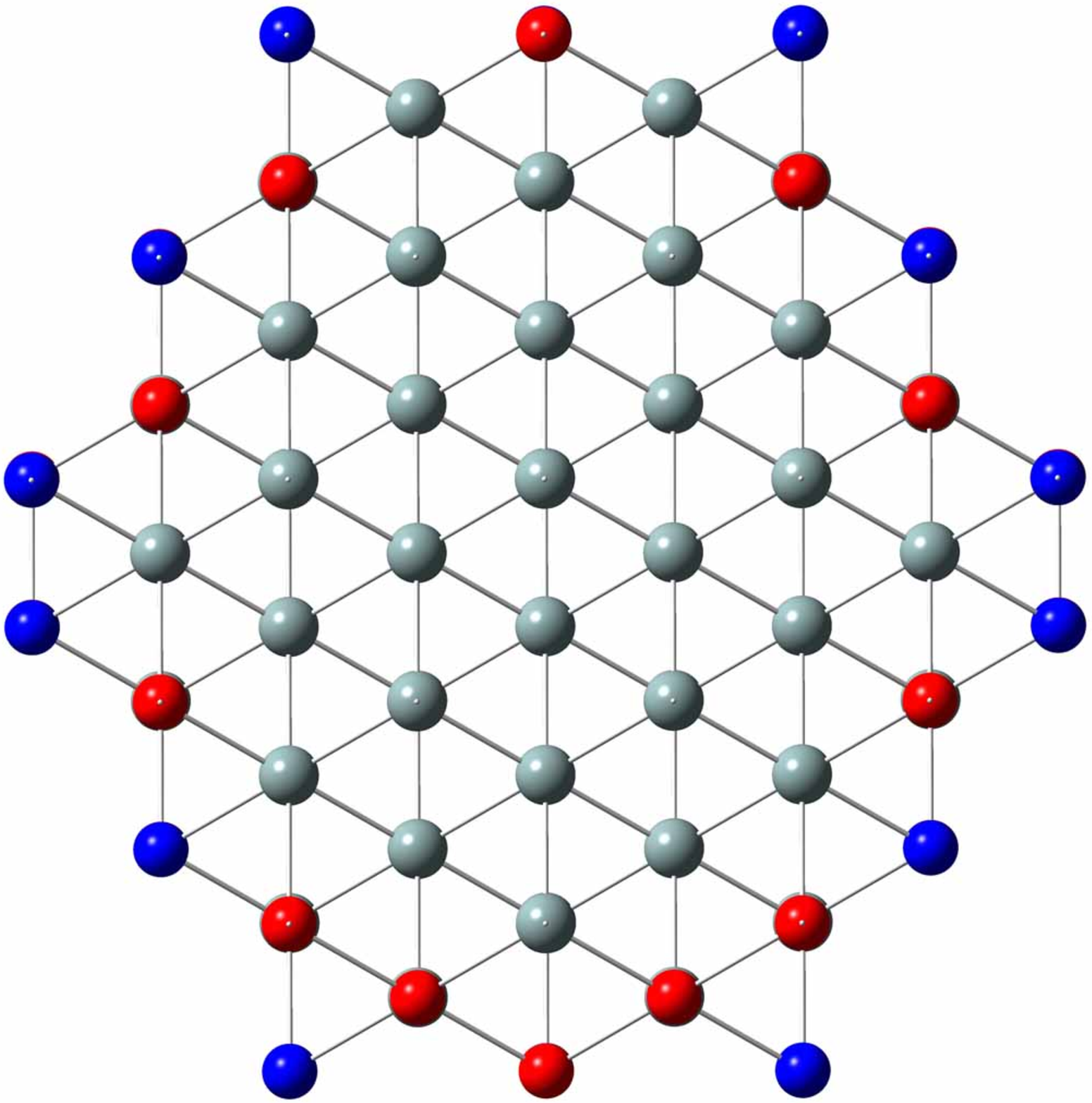}
\includegraphics[totalheight=0.1064\textheight]{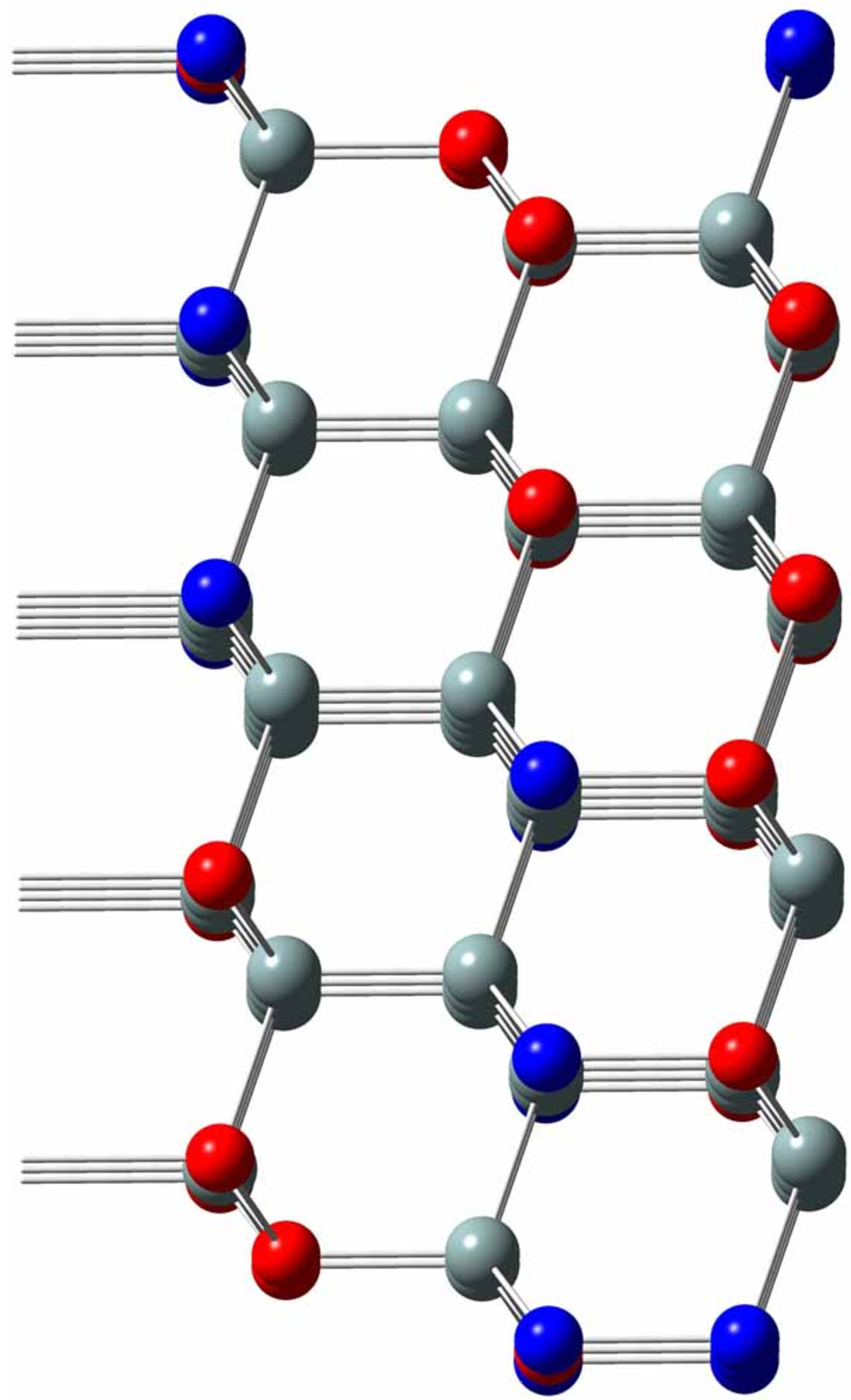}\\[-0.1cm]
\begin{picture}(0,0)
\hspace{-2.45cm}{\bf(b)}\hspace{3.91cm}{\bf (c)}
\end{picture}\\[0.2cm]
\includegraphics[totalheight=0.1139\textheight]{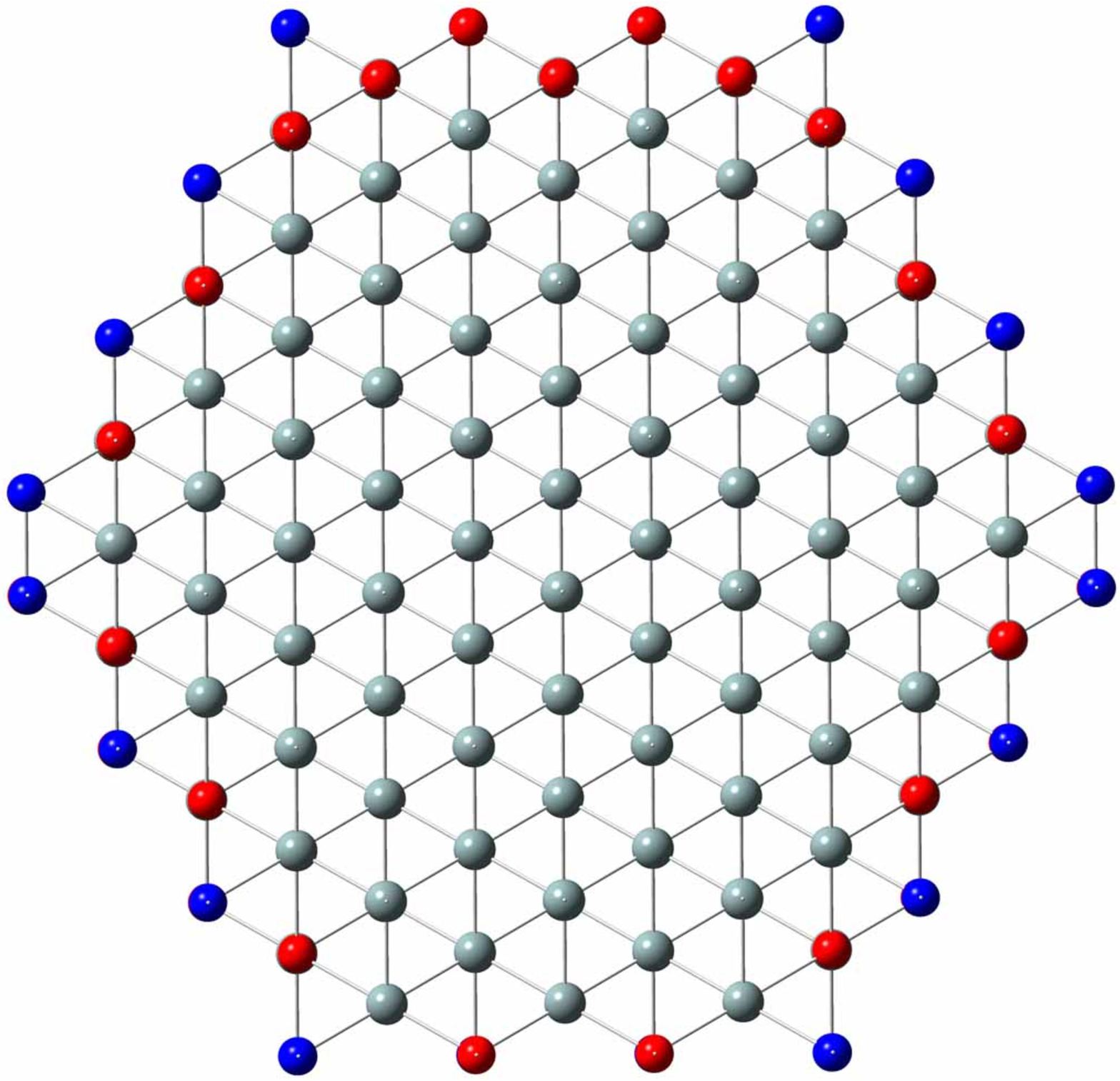}
\includegraphics[totalheight=0.1139\textheight]{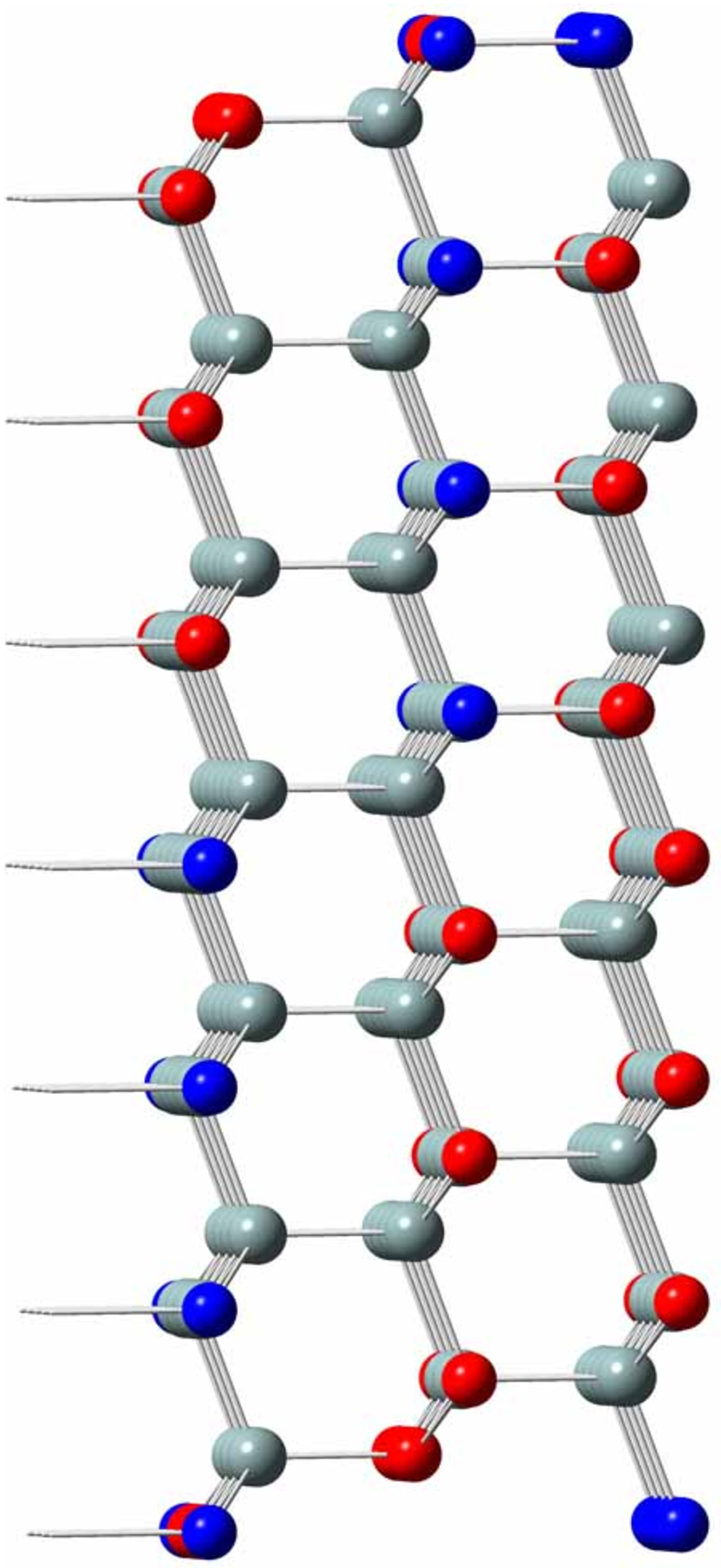}\hfill
\\[-0.1cm]
\begin{picture}(0,0)
\hspace{-2.45cm}{\bf(d)}
\end{picture}
\end{center}
\caption{Definition of characteristic lengths for the \emph{odd} series of zb-/diamond-lattice NWires growing along [111] axis with hexagonal cross section and all interfaces with \{11$\bar{2}$\} orientation, shown by translucent black lines. Top and side view of the first three members ($i=1$ to 3) of the \emph{odd} series: X$_{36}$ (b), X$_{104}$ (c), X$_{208}$ (d), X$_{348}$ (e). For atom colours see Figure \ref{fig03}.}
\label{fig11}
\end{figure}

For the odd series, we get
\begin{eqnarray}\label{eqn-74}
N_{\rm{Wire,odd}}^{111-\hexagon|11\bar{2}}[i]&=&2i(9i+7)+4
\end{eqnarray}
\begin{eqnarray}\label{eqn-75}
N_{\rm{bnd,odd}}^{111-\hexagon|11\bar{2}}[i]&=&4i(9i+4)+3
\end{eqnarray}
\begin{eqnarray}\label{eqn-76}
N_{\rm{IF,odd}}^{111-\hexagon|11\bar{2}}[i]&=&2\left(12i+5\right)
\end{eqnarray} 
\begin{eqnarray}\label{eqn-77}
d_{\rm{IF,odd,tb}}^{111-\hexagon|11\bar{2}}[i]&=&
a_{\rm{uc}}\,\frac{1}{\sqrt{2}}\ i
\end{eqnarray}
Comparison to Equation \ref{eqn-70} yields  $d_{\rm{IF,odd,tb}}^{111-\hexagon|11\bar{2}}[i]=d_{\rm{IF,even}}^{111-\hexagon|11\bar{2}}[i]$.
\begin{eqnarray}\label{eqn-78}
d_{\rm{IF,odd,side}}^{111-\hexagon|11\bar{2}}[i]&=&
a_{\rm{uc}}\,\frac{1}{\sqrt{2}}\ \left(i+\sqrt{\frac{11}{96}}\right)
\end{eqnarray}
This result differs from $d_{\rm{IF,odd,tb}}^{111-\hexagon|11\bar{2}}[i]$ by the offset of $\sqrt{\frac{11}{96}}$ to $i$ due to the corners between ${11\bar{2}}$ interfaces, see Figure 11.
\begin{eqnarray}\label{eqn-79}
w_{\rm{odd}}^{111-\hexagon|11\bar{2}}[i]&=&
a_{\rm{uc}}\sqrt{2}\,\left(i+\frac{1}{2}\sqrt{\frac{11}{96}}\,\right)
\end{eqnarray}
The corners between ${11\bar{2}}$ interfaces are reflected in an offset of the run index as compared to the \emph{even} series so that $w_{\rm{odd}}^{111-\hexagon|11\bar{2}}[i]=w_{\rm{even}}^{111-\hexagon|11\bar{2}}[i+\frac{1}{2}\sqrt{\frac{11}{96}}]$.
\begin{eqnarray}\label{eqn-80}
h_{\rm{odd}}^{111-\hexagon|11\bar{2}}[i]&=&
a_{\rm{uc}}\frac{\sqrt{11}}{8}\,(3i+1)
\end{eqnarray}
\begin{eqnarray}\label{eqn-81}\\
&&\big(a_{\rm{uc}}\big)^2\,\frac{9}{8}\sqrt{\frac{11}{8}}\,\left(i+\frac{\sqrt{33}}{48}\right)\left(i+\frac{1}{3}\right)\nonumber
\end{eqnarray}
The cross section of this \emph{odd} NWire type is shown in Figure \ref{fig11} for $i=1$ to 3 together with the definition of characteristic lengths.

\section{\label{ApplExamples}Application Examples of Nanowire Cross Sections} 
\subsection{\label{Example_35Grwth}Morphologic Description of III-V Nanowires}
\begin{figure*}[t!]
\begin{center}
\includegraphics[totalheight=0.24\textheight]{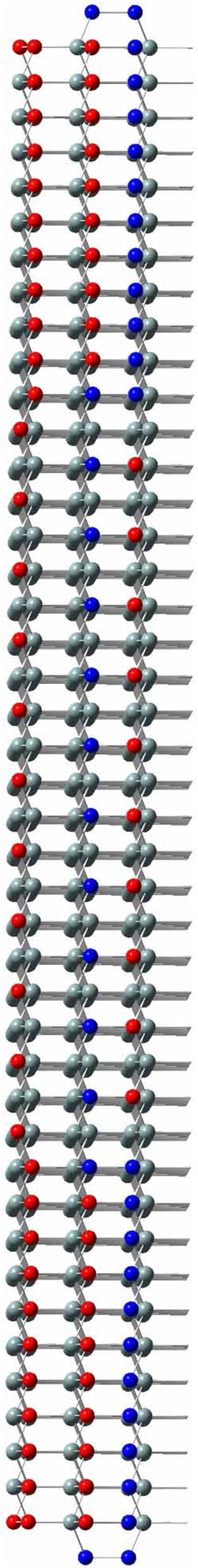}
\includegraphics[totalheight=0.24\textheight]{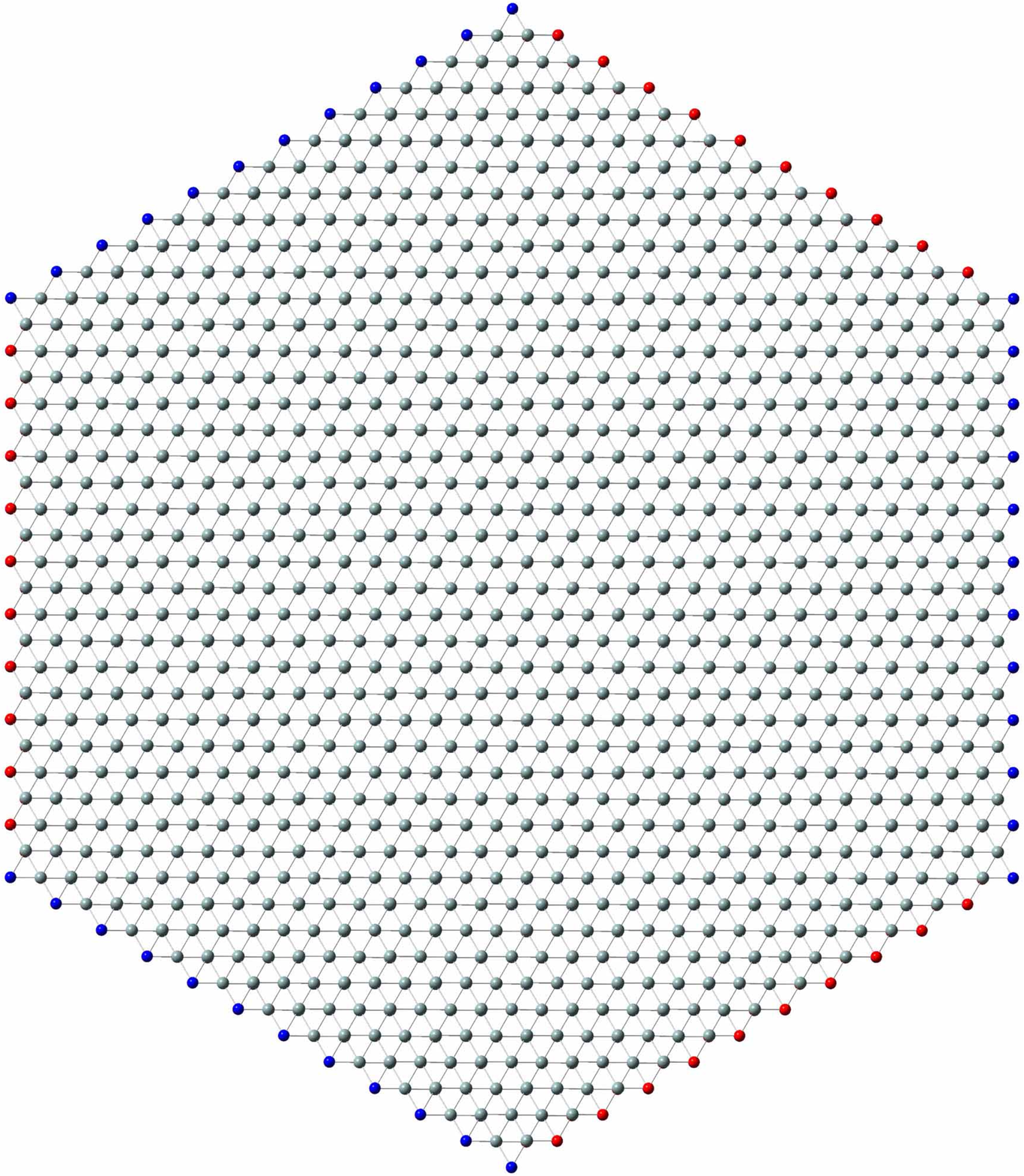}\hfill
\begin{picture}(2.5,12)
 \put(-10.73,31.14){\framebox(25,25){\shortstack[c]{{\large\bf (b)}\\ \ \\see Fig. 1b,\\Nano Lett.\\{\bf6}, 2743 (2006)}}}
  \put(-10.73,0){\framebox(25,25){\shortstack[c]{{\large\bf (c)}\\ \ \\see Fig. 1c,\\Nano Lett.\\{\bf6}, 2743 (2006)}}}
\end{picture}
\hfill
\includegraphics[totalheight=0.24\textheight]{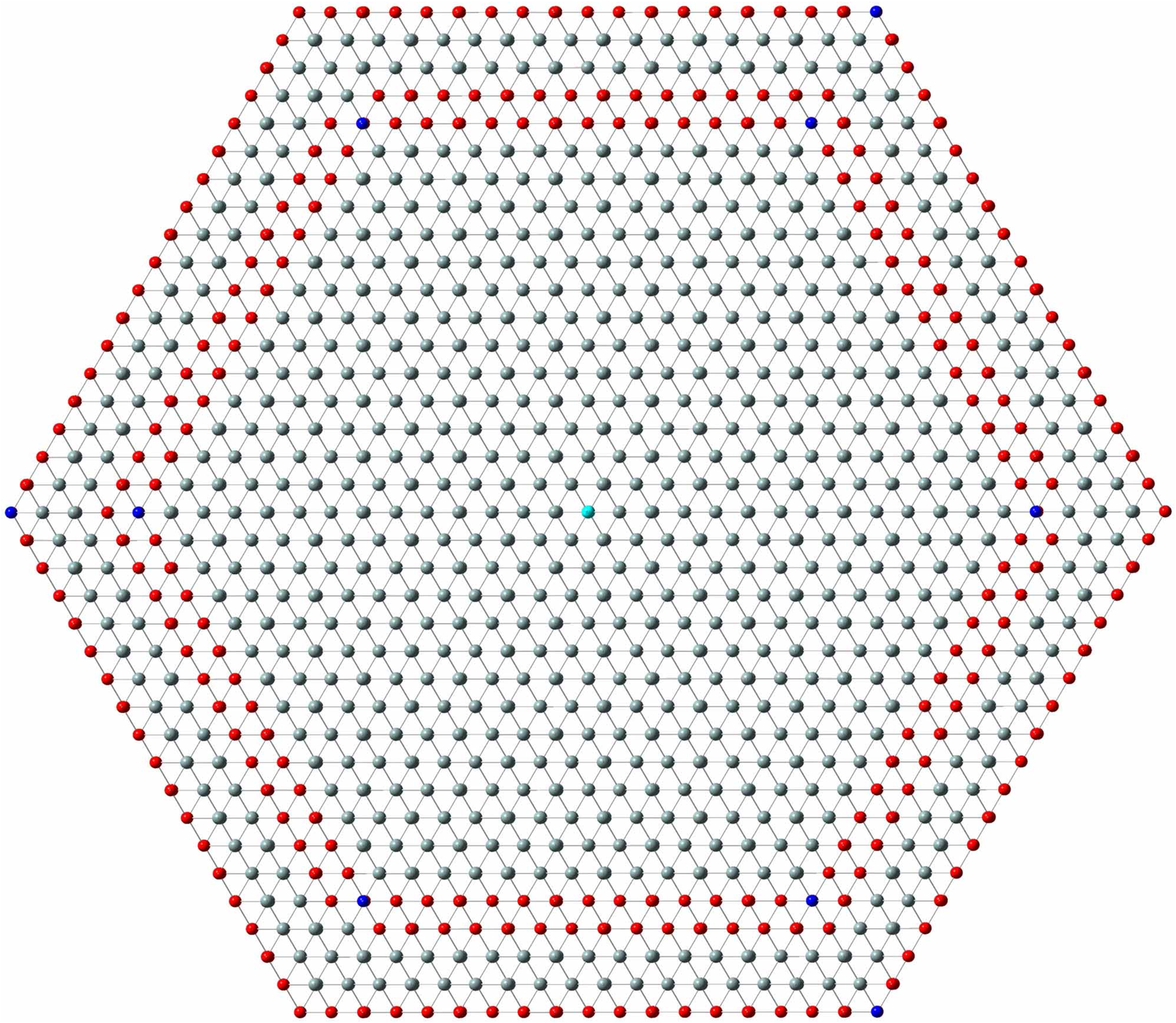}
\includegraphics[totalheight=0.24\textheight]{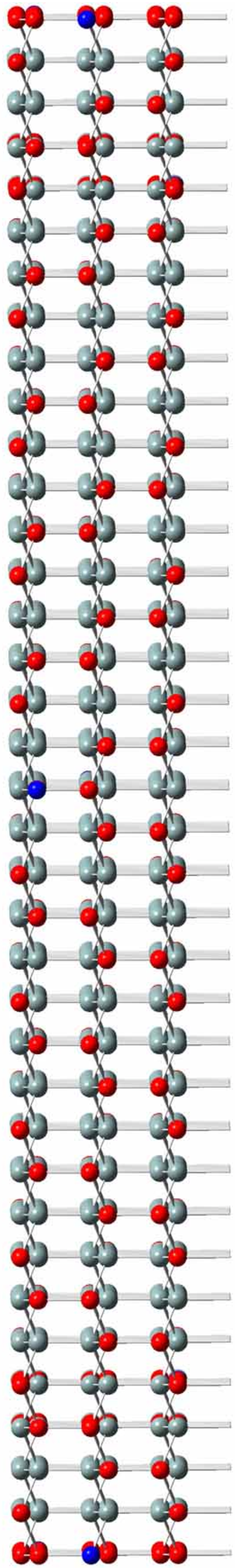}\\
\begin{picture}(18,0)
\hspace*{-5.399cm}{\bf (a)}\hspace*{10.875cm}{\bf (d)}
\end{picture}
\end{center}
\caption{\label{fig12}Side and cross sectional view of X$_{2246}$ ($i=11$) hexagonal NWire with [111] growth axis and \{11$\bar{2}$\} interfaces (a). Cross sectional STEM image of same NWire type consisting of GaAs (b) \cite{Skol06}. Cross sectional STEM image of a NWire with [111] growth axis, GaAs core and AlInP shell visible as bright rim (c) \cite{Skol06}. Graphs not shown in accord with copyright policy of \emph{ArXiV}. Scale bars in graphs (b) and (c) correspond to 100 \,nm. Side and cross sectional view of a hexagonal NWire -- X$_{2054}$ ($i=18$) -- in analogy to graph (c), growing along [111] axis with \{110\} interfaces and internal \{110\} interface separating a core from a shell (d). The symmetry center is marked by cyan atom. For other atom colours see Figure \ref{fig03}.}
\end{figure*}

III-V NWires are often found to grow along the [111] axis which requires the least energy (see Table \ref{tab_1}) with hexagonal cross sections \cite{Joyc11,Treu15}. Such NWires have experimental cross sections of \{11$\bar{2}$\} interfaces (Figures \ref{fig12}a and \ref{fig12}b, see \cite{Skol06} for the latter) or \{110\} interfaces (Figures \ref{fig12}d and \ref{fig12}c, see \cite{Skol06} for the latter). Below, we will show the structural parameters we can derive using equations \ref{eqn-70} and \ref{eqn-55} to match the side lengths and obtain the respective run index $i$. 

Figure \ref{fig12}c Figure 10c shows a GaAs core with hexagonal cross section, surrounded by a Al$_{0.5}$In$_{0.5}$P shell visible as bright hexagonal ring. For now, we focus on the GaAs partition of both NWires. The scale bars in those images present 100 nm and allow to estimate the side length of the hexagonal cross sections which is 72 nm for the $11\bar{2}$ interfaces and 92.4 nm for the $110$ interfaces. With $a_{\rm{uc}}({\rm GaAs}) = 0.56533$\,nm \cite{BoerI}, these side lengths can be found to match $d_{\rm{IF,even}}^{111-\hexagon|11\bar{2}}[i=180]=a_{\rm{uc}}\,180/\sqrt{2}=$ 71.95\,nm and
$d_{\rm{IF,even}}^{111-\hexagon|110}[i=394]$ $=a_{\rm{uc}}\,394\sqrt{11}/8$ $=92.34$\,nm. As \emph{even} and \emph{odd} series converge in their parameters for large $i$, the use of latter series does not yield more accurate information unless lattice information around the centerpoint of the cross section is available; see Figure 13 for an example. Data obtained from $i=180$ for hexagonal cross sections with [111] growth and $11\bar{2}$ interfaces and $i=394$ for hexagonal cross sections with [111] growth and \{110\} interfaces are listed in \ref{tab_3}.

Analytic number series of the same cross section type can be combined in any superposition seen fit to describe experimental data. We turn our attention to the NWire in Figure 10c again to evaluate its complete morphology, i.e. GaAs core with Al$_{0.5}$In$_{0.5}$P shell, using our findings presented in Section \ref{AnaNomiHexa111-IF110}. The shell has the same orientation as the GaAs core, growing in [111] direction with \{110\} interfaces. With the morphology of the entire core-shell NWire having a side length of 108 nm and monolithic lattice growth \cite{Skol06} ($a_{\rm{uc}}({\rm Al_{0.5}In_{0.4}P})\approx a_{\rm{uc}}({\rm GaAs})$), we arrive at $d_{\rm{IF,even}}^{111-\hexagon|110}[i=461]$, from which it is straightforward to get all parameters as for the GaAs core, see Table 5. The parameters of the Al$_{0.5}$In$_{0.5}$P shell can be calculated by substracting parameters of the inner GaAs core from the parameters of the entire core-shell cross section. Due to the finite thickness of the interface layer, we choose the GaAs core parameters for the run index matching this core plus one ($i=394+1$) to calculate the cross section area of the Al$_{0.5}$In$_{0.5}$P shell: $A_{\rm{IF}}=A^{111-\hexagon|110}_{\rm{even}}[i=395]-A^{111-\hexagon|110}_{\rm{even}}[i=394]$. Along the same lines, we calculate the number of shell-internal bonds which is $N_{\rm{bnd,even}}^{111-\hexagon|110}[i=461]-N_{\rm{bnd,even}}^{111-\hexagon|110}[i=395]$ and cross ection area of the shell which is $A=A^{111-\hexagon|110}_{\rm{even}}[i=461]-A^{111-\hexagon|110}_{\rm{even}}[i=395]$. The total number of interface bonds for the Al$_{0.5}$In$_{0.5}$P shell is the sum of its outer and internal interface to the GaAs core, yielding $N_{\rm{IF}}(\rm{shell})=N_{\rm{IF,even}}^{111-\hexagon|110}[i=394]+N_{\rm{IF,even}}^{111-\hexagon|110}[i=461]$. The NWire cross section is a regular hexagon. Thereby, simple trigonometry yields the thickness of the Al$_{0.5}$In$_{0.5}$P shell: $w_{\rm{shell}}=\Delta d_{\rm{IF}}/(2\tan[30^{\circ}])=a_{\rm{uc}}\sqrt{33}/16\,(461-395)$, with $\Delta d_{\rm{IF}}$ as the difference in side lengths of inner and outer hexagons defining the shell. All values are listed in \ref{tab_3}.
\begin{table*}[t!]
\caption{\label{tab_3}Structural parameters for the \emph{even} series of hexagonal cross sections with [111] growth and ${11\bar{2}}$ interfaces with $i=180$, \emph{cf.} Figure 10, and for the \emph{even} series of hexagonal cross sections with [111] growth and \{110\} interfaces with $i=394$, \emph{cf.} Figure 8. The slab thickness of the NWire cross section is $d_{\rm{slab}}=a_{\rm{uc}}\sqrt{3}$ for all cases.}
\begin{ruledtabular}
\begin{tabular}{l|c|c|c|c}
parameter&$111-\hexagon|11\bar{2}$&$111-\hexagon|110$&$111-\hexagon|110$&$111-\hexagon|110$\\
&$[i=180]$&$[i=394]$&$[i=461]$&Al$_{0.5}$In$_{0.5}$P shell\\
\hline
$N_{\rm{Wire}}/d_{\rm{slab}}$&584,282&933,782&1,277,894&344,112\\
$N_{\rm{bnd}}/d_{\rm{slab}}$&1,166,401&1,865,197&2,553,019&678,348\\
$N_{\rm{IF}}/d_{\rm{slab}}$&4,326&4,734&5,538&10,272\\
$d_{\rm{IF}}$\,[nm]&71.95&92.34&108.0& \\
$w$\,[nm]&143.9&184.7&216.1&\\
$w_{\rm{shell}}$\,[nm]&&&&13.40\\
$h$\,[nm]&126.6&157.5&184.3&\\
$A$\,[nm$^2$]&13,660&21,816&29,867&7939.7\\
$A_{\rm{IF}}$\,[nm$^2$]&&&&110.88\\ \hline
$N_{\rm{bnd}}/N_{\rm{Wire}}$&1.9963&1.9975&1.9978&1.9713\\
$N_{\rm{IF}}/N_{\rm{Wire}}$&$7.404\times 10^{-3}$&$5.070\times 10^{-3}$&$4.334\times 10^{-3}$&$2.985\times 10^{-2}$\\
$N_{\rm{IF}}/N_{\rm{bnd}}$&$3.709\times 10^{-3}$&$2.538\times 10^{-3}$&$2.169\times 10^{-3}$&$1.514\times 10^{-2}$\\
\end{tabular}
\end{ruledtabular}
\end{table*}

We briefly look at some of the outcomes which can be derived from the values in Table \ref{tab_3}. As will be discussed in Section \ref{Applicat}, the ratio of NWire-internal bonds per NWire atom $N_{\rm{bnd}}/N_{\rm{Wire}}$ is a sensitive stress indicator since less bonds per NWire atom are available to compensate external or internal stress. A good example of the latter is substitutional doping where the dopant usually foreign to the NWire lattice causes significant amounts of strain. The values in Table \ref{tab_3} show that it should be either notably more difficult to dope the Al$_{0.5}$In$_{0.5}$P shell as compared to the GaAs core, or that we can expect a significant rate of lattice defects introduced by dopants. A quantitative statement will be given Section \ref{Applicat} for Si NWires where experimental data for the increasing failure of impurity doping is available.

While the case above delivered valuable structural data, the full capacity of analytical NWire cross section metrology comes to light when structural information on an atomic base is available, e.g. by HR-TEM. We will demonstrate these capabilities in Section \ref{Example_Si-Ox-Etch}.

\subsection{\label{Example_Si-Ox-Etch}Morphologic Description of Si Nanowires}
\begin{figure}[t!]
\begin{center}
\begin{picture}(86,26)
  \put(18,0){\framebox(50,25){\shortstack[c]{{\large\bf (a, b)}\\ \ \\see Fig. 1b,\\Nano Lett.\\{\bf11}, 5465 (2011)}}}
\end{picture}
\includegraphics[totalheight=0.1371\textheight]{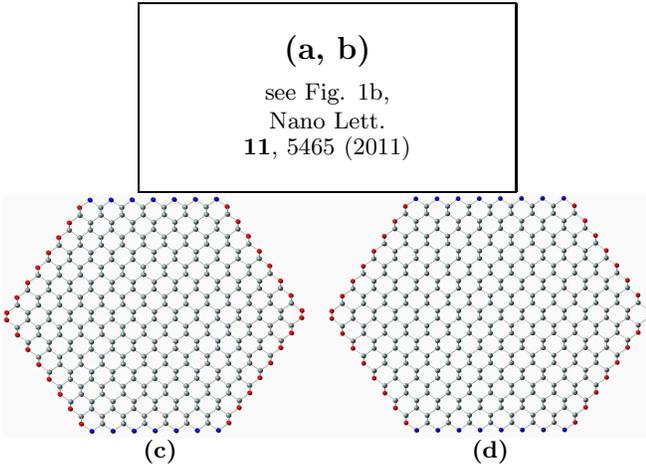}\\[-0.1cm]
\begin{picture}(0,0)
\hspace{-2.45cm}{\bf(c)}\hspace{3.91cm}{\bf (d)}
\end{picture}
\end{center}
\caption{\label{fig13}Ultrathin Si nanowire with [110] axis embedded in SiO$_2$ constituting NWire FET structure \cite{Yi11}; graph not shown in accord with copyright policy of \emph{ArXiV}. Graph (a) refers to original image, (b) includes the contours of the two analyic NWire cross sections; not included due to copyright, see graph (a). The red semi-translucent outline shows the \emph{odd} series of NWire growing along [110] axis with \{110\} and \{001\} interfaces ($i=7;\,{\rm X}_{704}$) as shown in graph (c). The blue semi-translucent outline shows the same NWire type, but using its \emph{even} series ($i=8;\,{\rm X}_{768}$) as shown in graph (d). For atom colours see Figure \ref{fig03}. Details of the specific NWire cross sections shown in graphs (c) and (d) are explained in section \ref{AnaNomiHexa110}. Additional structural information of both cross sections is listed in Table \ref{tab_4}.}
\end{figure}

Si NWires are key candidates for future VLSI devices in sub-14 nm technology nodes \cite{Webe17}. At such device dimensions, NWires have diameters significantly below 10 nm which allows to image their entire cross sections at atomic resolution. The latter provides the information to pick the right analytic description from Section \ref{nomiWire} in terms of \emph{odd} or \emph{even} series. Often, such NWires follow a top-down approach by thermal oxidation combined with SiO$_2$ back-etching \cite{Webe17}. Figure \ref{fig13}a shows a HR-TEM image of such NWire with [110] axis orientation, \{001\} interfaces on top and bottom and \{111\} side interfaces \cite{Yi11}. This cross section type was described in Section \ref{AnaNomiHexa110}, the nearest members of the respective \emph{odd} and \emph{even} series to fit the HR-TEM image are shown in Figure \ref{fig13}c and d, with their cross section superimposed onto the TEM image in Figure \ref{fig13}b. All structural parameters derived from the number series are listed in Table \ref{tab_4}.

With these data, accurate structural metrology statements can be made for any characterisation which has a structural dependence. The exact cross section area allows to calculate current densities. In contrast to absolute current values which are used in most works at present, current densities help to interpret electronic phenomena such as impact ionization or inelastic carrier-carrier scattering. Thereby, device characteristics of NWire-based electronic devices \cite{Tomi11,Webe17} can be assessed and interpreted in more detail.

\begin{table}[t!]
\caption{\label{tab_4}Structural parameters for hexagonal cross sections with [110] growth axis, \{111\} top and bottom interfaces and \{001\} side interfaces of the \emph{odd} series with $i=7$, and of the \emph{even} series with $i=8$, \emph{cf.} section \ref{AnaNomiHexa110} and Figure \ref{fig13}. The slab thickness of NWire cross sections is $d_{\rm{slab}}=a_{\rm{uc}}\sqrt{2}$.}
\begin{ruledtabular}
\begin{tabular}{l|c|c}
parameter&$i=7$, \emph{odd}&$i=8$, \emph{even}\\
\hline
$N_{\rm{Wire}}/d_{\rm{slab}}$&704&768\\
$N_{\rm{bnd}}/d_{\rm{slab}}$&1348&1472\\
$N_{\rm{IF}}/d_{\rm{slab}}$&120&128\\
$d_{\rm{001-IF}}^{110-\hexagon}$\,[nm]&2.50&2.88\\
$d_{\rm{111-IF}}^{110-\hexagon}$\,[nm]&2.58&2.58\\
$w^{110-\hexagon}$\,[nm]&5.47&5.86\\
$h^{110-\hexagon}$\,[nm]&4.21&4.21\\
$A^{110-\hexagon}$\,[nm$^2$]&16.76&18.38\\ \hline
$N_{\rm{bnd}}/N_{\rm{Wire}}$&1.915&1.917\\
$N_{\rm{IF}}/N_{\rm{Wire}}$&0.170&0.167\\
$N_{\rm{IF}}/N_{\rm{bnd}}$&0.089&0.087\\
\end{tabular}
\end{ruledtabular}
\end{table}

\section{\label{Applicat}Usage of Number Series Ratios on Nanowire Cross Sections}
The ratio of intra-NWire bonds per NWire-atom converges against 2 for NWire cross sections approaching infinity via their run index $i$; $N_{\rm{bnd}}[i]/N_{\rm{Wire}}[i]\rightarrow 2$ for $i\rightarrow\infty$. This result is straightforward from the consideration of the zb- or diamond-lattice unit cell, see Figure \ref{fig01}. Every atom has four bonds to its 1-nn atoms and each of these four bonds is shared with one of the 1-nn atoms. 
Depending on the size of the NWire, its interface faceting and surface-to-volume ratio, $N_{\rm{bnd}}[i]/N_{\rm{Wire}}[i]$ can be significantly smaller than 2.

In impurity doping, dopants reside on substitutional lattice sites within the semiconductor such as Si. This process requires a certain activation energy of which a significant part is taken up by neighboring Si---Si bonds, thereby inducing local stress. At a very high dopant densities, their solubility limit is reached where many local doping events build up a global stress which either induces dopant clustering and interstitials or the disintegration of the semiconductor lattice into its amorphous form. For nano-Si systems such as NWires or NCs, the ratio $N_{\rm{bnd}}[i]/N_{\rm{Wire}}[i]$ decreasing with system size reflects a decreasing ability of the nano-Si system to cope with local stress as induced by subsitutional doping. 

\begin{figure}[t!]
\begin{center}
\includegraphics[totalheight=0.255\textheight]{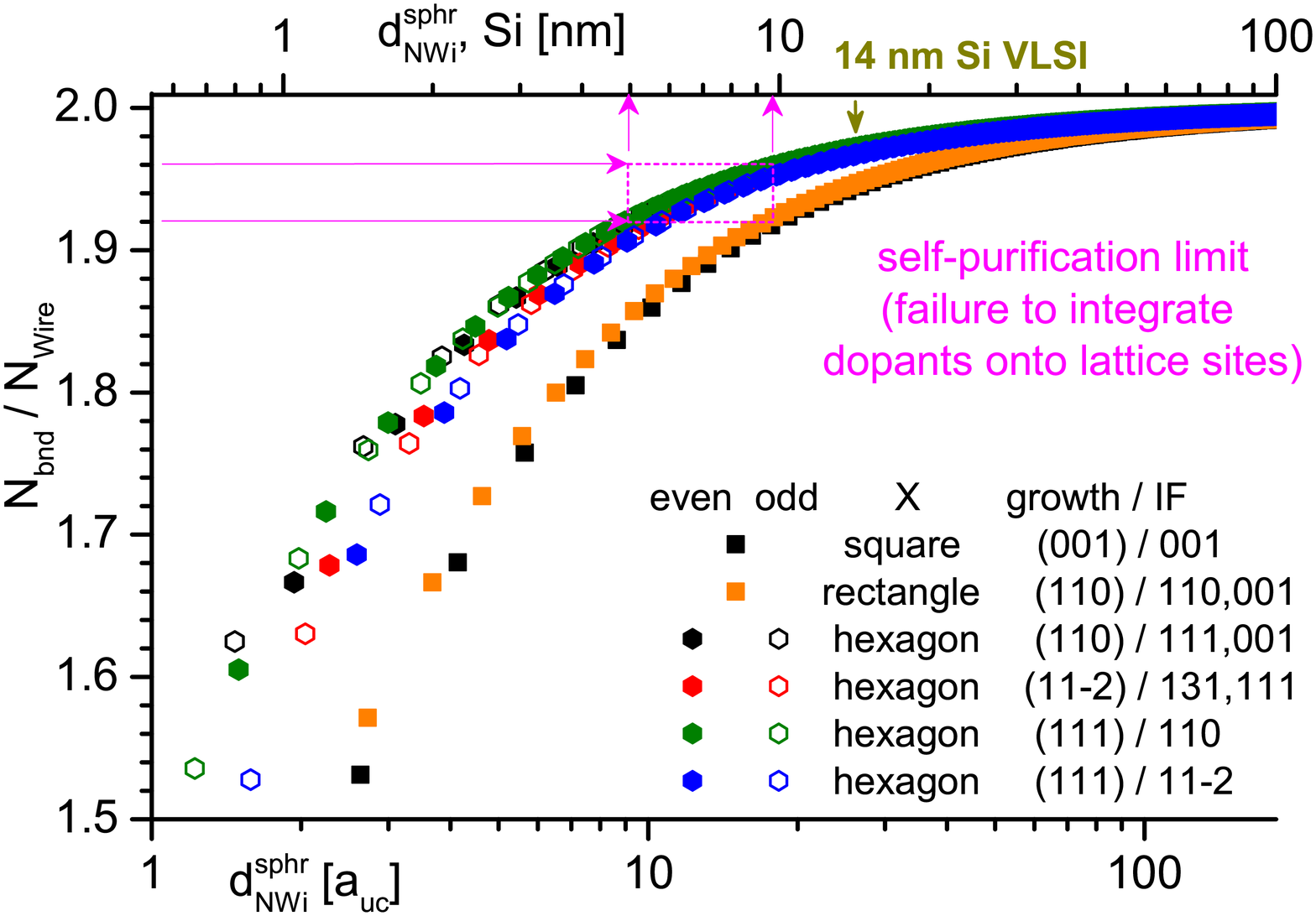}
\\[-0.1cm]
\begin{picture}(0,0)
{\bf(a)}
\end{picture}\\[0.2cm]
\begin{picture}(86,26)
  \framebox(35,25){\shortstack[c]{{\large\bf (b)}\\ \ \\see Fig. 4d,\\Nature Nanotech.\\{\bf4}, 103 (2009)}}
  \put(15.75,0){\framebox(35,25){\shortstack[c]{{\large\bf (c)}\\ \ \\see Fig. 7,\\Phys. Rev. B\\{\bf80}, 165326 (2009)}}}
\end{picture}
\end{center}
\caption{\label{fig14}Ratio of internal bonds per NWire atom $N\mathrm{_{bnd}}/N\mathrm{_{Wire}}$ as a function of cross section area expressed by the spherical NWire diameter $d_{\rm{Wire}}^{\,\rm{sphr}}$ in units of $a_{\rm{uc}}$ and absolute values for Si (upper x-axis) (a). Magenta marks show the area where self-purification sets in for bottom-up (grown) Si NWires. Relative increase of Si NWire resistivity over bulk value as a function of NWire radius (b); graph not shown in accord with copyright policy of \emph{ArXiV}. The magenta slab shows the hard limit where dopant failure occurs. Donors with unpaired electron within Si nanocrystals (c) \cite{Steg09}; graphs not shown in accord with copyright policy of \emph{ArXiV}. Round symbols refer to donors did not getter Si dangling bonds and potentially can donate an electron at room temperature. Green symbols show region of interest, magenta lines the doping probability dropping by a factor of ca. 400.}
\end{figure}

This finding is reflected in the strong resistivity increase of Si NWires with shrinking diameter, see Figure \ref{fig14}b. The argument of the authors in \cite{Bjoe09} that dielectric mismatch with the enviroment prevents dopant ionization is certainly correct. However, dielectric mismatch is not the only cause of dopant failure. For Si NCs, their rejection of dopants to be located on substitutional lattice sites by self-purification was discussed in detail \cite{Dalp06,Dalp08,Koe15,Hill17a,Hill17b}. Similar limits exist for Si NWires as discussed below. Even for doping densities of $N_{\rm{D}}=1.5\times 10^{20}$ cm$^{-3}$ which present the semiconductor to semi-metal limit for Si \cite{Pear49}, substitutional doping fails for NWire diameters in the range of 7 to 10 nm \cite{Bjoe09}, see Figure \ref{fig14}b. At this diameter, we have a ratio of $N_{\rm{bnd}}/N_{\rm{Wire}}=1.94\pm0.01$ for Si NWires with hexagonal cross section \cite{Schm08} depending on interface orientation and growth axis. Values are significantly lower for square and rectangular cross sections.
For Si NCs, probabilities of built-in dopants dropped by 2$^{\,1\!}/_2$ order of magnitude when decreasing NC diameters from 20 to 10 nm \cite{Steg09}, see Figure \ref{fig14}c. For this size range, a ratio of NC-internal bonds per NC atom of $N_{\rm{bnd}}/N_{\rm{NC}}=1.94\pm0.02$ was found where the same analytical metrology presented here was applied to zb- and diamond lattice NCs \cite{Koe16}. Using independent experimental results from two different nano-Si systems (NWires and NCs), our metrology shows that $N_{\rm{bnd}}/N_{\rm{NC}}=1.94\pm0.02$ is a general limit below which self-purification prevents effective substitutional doping. 

\begin{figure}[t!]
\begin{center}
\includegraphics[totalheight=0.255\textheight]{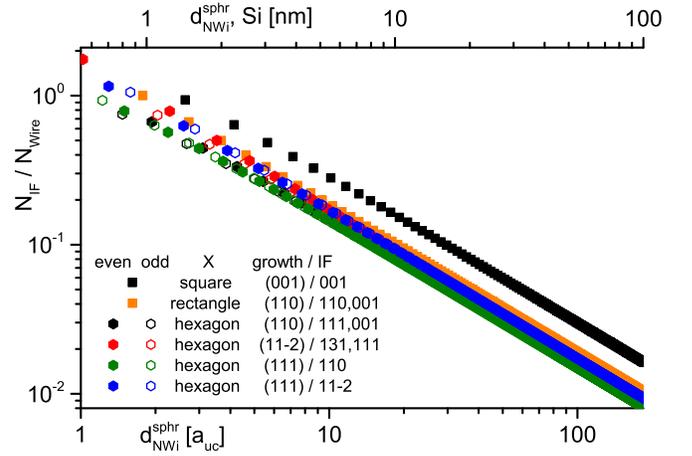}
\end{center}
\caption{ \label{fig15}Ratio of interface bonds per NWire atom $N\mathrm{_{IF}}/N\mathrm{_{Wire}}$ as function of cross section area expressed by the spherical NWire diameter $d_{\rm{Wire}}^{\,\rm{sphr}}$ in units of $a_{\rm{uc}}$ and absolute values for Si (upper x-axis) shown for all NWires of section \ref{nomiWire}.}
\end{figure}

The ratio $N\mathrm{_{IF}}[i]/N_{\mathrm{Wire}}[i]$ yields the number of interface bonds per NWire atom, see Figure \ref{fig15}. This key parameter quantifies electronic phenomena occurring across NWire interfaces. It follows the opposite trend as discussed for $N\mathrm{_{bnd}}[i]/N_{\mathrm{Wire}}[i]$: Any bond not available for connecting NWire atoms occurs at an interface. NWire cross sections with high aspect ratio have a higher value of $N\mathrm{_{IF}}/N\mathrm{_{Wire}}$ which results in bigger $d_{\mathrm{Wire}}$ values up to which the embedding dielectric dominates electronic and optical NWire properties. This finding is important for \{001\}-terminated Si NWires with square cross section which can be morphed to a high aspect ratio as mentioned at the end of Section \ref{WrapUp}. Such \{001\}-terminated cross sections are encountered in fin-FETs and can be used to exploit nanoscopic phenomena. As an example, the electronic impact of ultrathin SiO$_2$ and Si$_3$N$_4$ coatings onto Si-NWires with thicknesses as shown in Figure \ref{fig13} could replace conventional doping while maintaining CMOS-compatibility for VLSI devices \cite{Koe14,Koe18a}.

The quantity $N\mathrm{_{IF}}/N\mathrm{_{bnd}}$ serves as gauge for the stress balance between NWires and an embedding matrix or coating interacting via interface bonds such as GaAs NWires grown with Al$_{0.5}$In$_{0.5}$P shell \cite{Skol06}, see Section \ref{Example_35Grwth}. 
There may exist a minimum of $N\mathrm{_{IF}}/N\mathrm{_{bnd}}$ below which the NWire structure is unstable and may switch to a different gross section or even symmetry group like intermittent zb-W\"urtzite phases of InAs NWires \cite{Park15}. It is therefore likely that there are certain limits of minimum NWire diameters per cross section type below which a phase change occurs. This phenomenon was observed for Si NCs in SiO$_2$ which have strong preference for an octahedral shape with \{111\} interfaces for NC sizes of $\leq3$ nm \cite{Gode08,Koe16}. Such limits are a function of $N\mathrm{_{IF}}/N\mathrm{_{bnd}}$, the Young's moduli of the semiconductor and its embedding environment and can be derived for any NWire with appropriate experimental input.
\begin{figure}[t!]
\begin{center}
\includegraphics[totalheight=0.255\textheight]{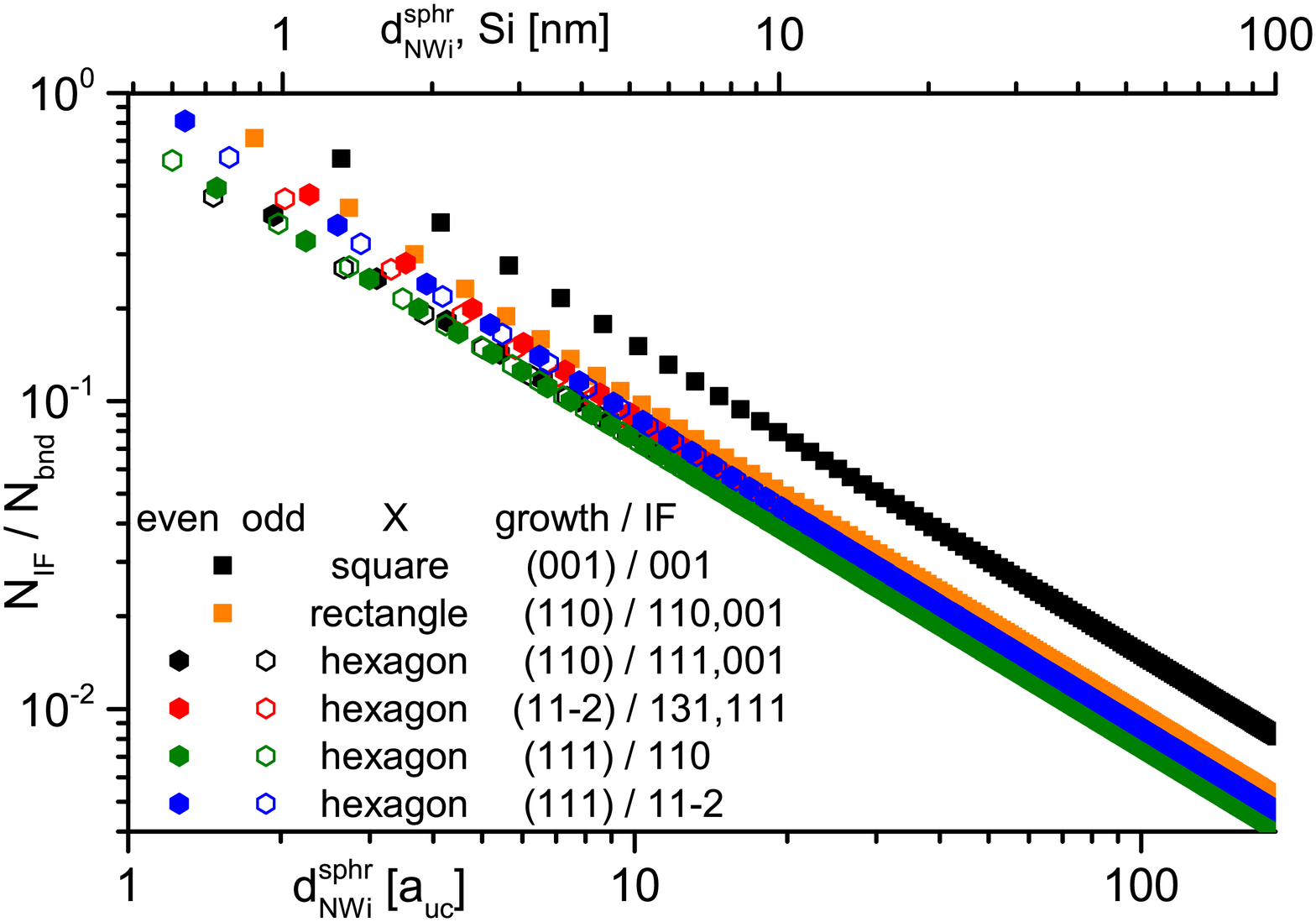}
\end{center}
\caption{\label{fig16}Ratio of nterface to NWire-internal bonds $N\mathrm{_{IF}}/N\mathrm{_{bnd}}$ as function of cross section area expressed by the spherical NWire diameter $d_{\rm{Wire}}^{\,\rm{sphr}}$ in units of $a_{\rm{uc}}$ and absolute values for Si (upper x-axis)  shown for all NWires of section \ref{nomiWire}.}
\end{figure}

\section{\label{Outlook}Outlook}
Work is in progress for the NWire cross sections shown here to develop lateral number series which allow asymmetric cross section morphing to any rectangular aspect ratio for square and rectangular NWires and from low-aspect hexagonal up to full rhombic cross section for hexagonal NWires. As an example, square NWire cross sections with exclusive \{001\} interface could be morphed to large aspect rectangles which describe Si fin-FETs in 14 nm VLSI technology. Morphed cross sections possess a ratio of interface-specific bonds changing with the shape of cross section. As an example, the ratio of \{001\} to \{111\} interface bonds can be exploited for the interface-specific Si dangling bond (DB) defects labeled P$_{\rm{b0}}$ and P$_{\rm{b1}}$, thereby enabling non-destructive shape detection of NWire cross sections. Such characterisation was already applied to Si NCs by Stesmans \emph{et al.} \cite{Stes08} using EPR and recently underpinned by crystallographic metrology \cite{Koe16}.

\section{\label{WrapUp} Conclusions}
We deduced analytical number series for zb- and diamond-lattice NWires as a function of diameter, shape and interface faceting for the following NWire cross sections: 
Square, \{001\} growth axis and interfaces; rectangular, \{110\} growth axis and \{110\} plus \{001\} interfaces; hexagonal, \{110\} growth axis and \{001\} plus \{111\} interfaces; hexagonal, ${11\bar{2}}$ growth axis and \{111\} plus ${1\bar{3}1}$ interfaces; hexagonal, \{111\} growth axis and \{110\} interfaces; hexagonal, \{111\} growth axis and ${11\bar{2}}$ interfaces. 
All hexagonal cross sections were presented in an \emph{even} and an \emph{odd} scheme to facilitate matching to different symmetry centers encountered in experiment. Calculated parameters are the number of NWire atoms $N_{\mathrm{Wire}}[i]$, the number of bonds between such atoms $N\mathrm{_{bnd}}[i]$ and the number of NCWire interface bonds $N\mathrm{_{IF}}[i]$, interface lengths $d_{\rm{IF}}[i]$, cross section widths $w[i]$, heights $h[i]$ and total cross section areas $A[i]$. All expressions are linked to NWire spherical diameters $d_{\mathrm{Wire}}[i]$ to enable direct parameter comparison between different morphologies.

Use of the analytic metrology was shown on III-V core-shell and Si NWire STEM/TEM images. The available atomic resolution of the latter allowed exact parameter description down to the atom/bond which opens a new avenue to interpret any experimental spectroscopic data of zb- and diamond-lattice NWire cross sections.

The ratio $N\mathrm{_{bnd}}/N\mathrm{_{Wire}}$ is useful to gauge internal stress of NWires which is key to evaluate self-purification and dopant segregation as encountered in impurity doping, and the general stress response of NWires to an external force. Both, $N\mathrm{_{IF}}/N\mathrm{_{bnd}}$ and $N\mathrm{_{bnd}}/N\mathrm{_{Wire}}$, can be applied to optical spectroscopy methods such as FT-IR, Raman, PL or EL to interprete and deconvolute spectra into NWire-immanent (internal) and matrix/shell (external) components. The ratio $N\mathrm{_{IF}}/N_{\mathrm{Wire}}$ describes the electronic interaction of NWires with the embedding matrix or ligands to gauge the impact of interface dipoles or interface charge transfer onto the NC electronic structure. 

The analytic metrology of zb- and diamond-lattice NWire cross sections can provide major advancements in experimental data interpretation and understanding of III-V, II-VI and group-IV based NWires. 
The number series allow for a deconvolution of experimental data into environment-exerted, interface-related and NC-internal phenomena. The predictive power of our metrology can render it to be an essential tool to predict NWire cross sections and to tune process conditions for tayloring NWires towards desired shape and interface properties.\\

\begin{acknowledgments}
D.~K. acknowledges funding by the 2015 UNSW Blue Sky Research Grant, funding by 2012, 2014 and 2016 DAAD-Go8 joint research cooperation schemes and the 2018 Theodore-von-K{\`a}rm{\`a}n Fellowship of RWTH Aachen University, Germany.\\
\end{acknowledgments}

\end{document}